\documentclass[twocolumn,astrosymb]{aastex7}
\usepackage{color}
\usepackage{soul}
\usepackage[normalem]{ulem}
\usepackage{cancel}

\begin{document}

    \title{Quantifying the Dominance and Emergence of Brightest Cluster Galaxies In SPT and Planck Clusters: A Novel Application of the Gini Index}

\author[0000-0003-0298-4106]{Katiya Fosdick}
\affiliation{Department of Physics, Massachusetts Institute of Technology, Cambridge, MA 02139, USA}
\ifdefined\iflatexml
  \email{kfosdick@mit.edu}
\else
  \email[show]{kfosdick@mit.edu}
\fi

\author[0000-0001-5226-8349]{Michael McDonald} 
%\altaffiliation{Las Campanas Observatory}
\affiliation{Department of Physics, Massachusetts Institute of Technology, Cambridge, MA 02139, USA}
\affiliation{MIT Kavli Institute for Astrophysics and Space Research, Massachusetts Institute of Technology, Cambridge, MA 02139, USA}
\email{mcdonald@space.mit.edu}

\author[0000-0002-3475-7648]{Gourav Khullar}
\affiliation{Department of Astronomy, University of Washington, Physics-Astronomy Building, Box 351580, Seattle, WA 98195-1700, USA}
\affiliation{eScience Institute, University of Washington, Physics-Astronomy Building, Box 351580, Seattle, WA 98195-1700, USA}
\email{gkhullar@uw.edu}

\author[0000-0003-1370-5010]{Michael D. Gladders}
\affiliation{Department of Astronomy and Astrophysics, University of Chicago, 5640 South Ellis Avenue, Chicago, IL, 60637, USA}
\affiliation{Kavli Institute for Cosmological Physics, University of Chicago, 5640 South Ellis Avenue, Chicago, IL, 60637, USA}
\email{gladders@uchicago.edu}

\author[0000-0002-1857-2088]{Boris H\"au{\upshape{\ss}}ler}
\affiliation{European Southern Observatory, Alonso de Córdova 3107, Vitacura, Santiago de Chile, Chile}
\email{bhaeussl@eso.org}

\author[0000-0002-3400-6991]{Cian Roche} 
%\altaffiliation{Las Campanas Observatory}
\affiliation{Department of Physics, Massachusetts Institute of Technology, Cambridge, MA 02139, USA}
%\affiliation{MIT Kavli Institute for Astrophysics and Space Research, Massachusetts Institute of Technology, Cambridge, MA 02139, USA}
\email{roche@mit.edu}

%\author[]{Stephan Ducrepin} 
%\affiliation{Department of Physics, Massachusetts Institute of %Technology, Cambridge, MA 02139, USA}
%\email{mcdonald@space.mit.edu}

\author[0000-0001-7665-5079]{Lindsey E. Bleem}
\affiliation{High-Energy Physics Division, Argonne National Laboratory, 9700 South Cass Avenue, Lemont, IL, 60439, USA}
\affiliation{Kavli Institute for Cosmological Physics, University of Chicago, 5640 South Ellis Avenue, Chicago, IL, 60637, USA}
\affiliation{Department of Astronomy and Astrophysics, University of Chicago, 5640 South Ellis Avenue, Chicago, IL, 60637, USA}
\email{lbleem@anl.gov}

\author[0000-0003-4175-571X]{Benjamin Floyd}
\affiliation{Institute of Cosmology and Gravitation, University of Portsmouth, Dennis Sciama Building, Portsmouth, PO1 3FX, UK}
\email{benjamin.floyd@port.ac.uk}

\author[0009-0004-7337-7674]{Raven Gassis}
\affiliation{Department of Physics, University of Cincinnati, Cincinnati, OH 45221, USA}
\email{gassismr@mail.uc.edu}

\author[0000-0002-4577-8217]{Sebastian Grandis}
\affiliation{Universit\"at Innsbruck, Institut f\"ur Astro- und Teilchenphysik, Technikerstr. 25/8, 6020 Innsbruck, Austria}
\email{Sebastian.Grandis@uibk.ac.at}

\author[0000-0002-5222-1337]{Arnab Sarkar} 
%\altaffiliation{Las Campanas Observatory}
\affiliation{Department of Physics, University of Arkansas, 825 W Dickson Street, Fayetteville, AR, 72701 USA}
\affiliation{MIT Kavli Institute for Astrophysics and Space Research, Massachusetts Institute of Technology, Cambridge, MA 02139, USA}
\email{arnabsar@space.mit.edu}

\author[0000-0001-7975-2894]{Veronica Strazzullo}
\affiliation{INAF - Observatory of Trieste, via G. Tiepolo 11, I-34143 Trieste, Italy}
\affiliation{IFPU - Institute for Fundamental Physics of the Universe, via Beirut 2, I-34014 Trieste, Italy}
\email{veronica.strazzullo@inaf.it}

\author[0000-0003-3521-3631]{Taweewat Somboonpanyakul}
\affiliation{Department of Physics, Faculty of Science, Chulalongkorn University, 254 Phayathai Road, Pathumwan, Bangkok 10330, Thailand}
\email{taweewat.s@chula.ac.th}

%\author[]{You!} 
%\altaffiliation{Las Campanas Observatory}
%\affiliation{Elsewhere}
%\email{fakeemail2@google.com}

%\author[gname=Savannah,sname=Africa]{S. Africa}
%\affiliation{South African Astronomical Observatory}
%\affiliation{University of Cape Town, Department of Astronomy}
%\email{fakeemail3@google.com}

%\author{River Europe}
%\affiliation{University of Heidelberg}
%\email{fakeemail4@google.com}

%\author[0000-0000-0000-0003,sname=Asia,gname=Mountain]{Asia Mountain}
%\altaffiliation{Astrosat Post-Doctoral Fellow}
%\affiliation{Tata Institute of Fundamental Research, Department of Astronomy}
%\email{fakeemail5@google.com}

%\author[0000-0000-0000-0004]{Coral Australia}
%\affiliation{James Cook University, Department of Physics}
%\email{fakeemail6@google.com}

%\author[gname=IceSheet]{Penguin Antarctica}
%\affiliation{Amundsen–Scott South Pole Station}
%\email{fakeemail7@google.com}

%\collaboration{all}{The Terra Mater collaboration}

\begin{abstract}

We introduce the Gini index (G) as a metric for studying the emergence of brightest cluster galaxies (BCGs) by quantifying the disparity in the light distribution of galaxy cluster cores. The Gini index collapses the skewness of the bright-end luminosity function to a single statistic, and provides a measure for the light-dominance of the BCG over other cluster members that is more robust than the magnitude gap. We argue that the Gini index is far less sensitive to short-term changes in the cluster dynamical state than the magnitude gap and solves its known degeneracies by taking the entire cluster membership into account, making G a more sensitive metric to long-term BCG assembly. We apply the magnitude gap and Gini index to a progenitor-selected sample of 92 SPT- and Planck-selected cluster cores at $\mathrm{0< z< 1}$  with corresponding HST F110W \& F200LP and SDSS g \& r photometry. Based on both parametric and empirical tests, we find no statistical evidence for evolution in G or $\rm M_{1,2}$ over $\mathrm{0 < z < 1}$, and find that BCGs have not grown significantly (2$\mathrm{\times}$) faster than the cluster population for $\mathrm{z<1}$ at 3$\mathrm{\sigma}$ confidence. This implies that BCGs achieve dominance and establish defining properties early ($\mathrm{z > 1}$) in massive clusters. In general, we find that the Gini index is a valuable and relatively straightforward measurement, with well-controlled systematics, that could be applied to much larger samples to study BCG dominance and emergence.

%talk about moment-moment scatter

%Our broad-band HST photometry produces robust red-sequence and blue-cloud features when combined with stellar population synthesis.          

\end{abstract}


 %\uat{Galaxies}{573} --- \uat{Cosmology}{343}


 \section{Introduction} 
Galaxy clusters contain a rich array of co-evolving components, including hundreds to thousands of (mostly) red galaxies, a hot intracluster medium (ICM), diffuse intracluster light (ICL), and a massive dark matter halo. The processes driving the coevolution of these components are complicated, but many are connected to the evolution and growth of the brightest cluster galaxy (BCG).  BCGs are a special population of galaxies with nearly uniform properties \citep{Tremaine_Richstone_1977,Loh_Strauss_2006,De_Lucia_2007,Von_Der_Linden_2007,Lin_2010,Diaz-Gimenez_2012,Shen_2014, Solanes_2016}. They are the most massive galaxies in the Universe \textbf{($\rm M_{*} \gtrsim 10 ^{12}~ M_{\odot}$)}, with ancient stellar populations of mean ages $\rm > 10~Gyr $ \citep{De_Lucia_2007, Thomas_2010}. They are found in the centers of massive ($\rm M_{500} > 10^{14} M_{\odot}$) galaxy clusters, and often host radio-bright AGN \citep{Von_Der_Linden_2007,Best_2007}. One defining property of BCGs is their luminosity dominance over the other cluster members, with magnitudes typically $\sim$1--2 mags brighter than the next brightest member \citep{Donzelli_2011, Wen_Han_2015}, and little intrinsic scatter in luminosity across the population \citep{Sandage_1972,Schneider_Gun_Hoessel_1983,Hoessel_Schneider_1985,Postman_Lauer_1995}. This is a testament to the extreme hierarchical growth experienced by BCGs throughout their histories.

It is still relatively unknown which channels -- major or minor mergers, gas rich or poor mergers, cooling flows, or in-situ star formation  -- contribute most to the assembly of BCG stellar mass at different redshifts. While a subject of some uncertainty \citep{Collins_Mann_1998,De_Lucia_2007,Whiley_2008,Collins_2009,Stott_2010,Laporte_2013,Zhang_2016,Ragone-Figueroa_2018, Chu_2021}, it is generally accepted that BCGs experience a factor of $\sim$2 increase in stellar mass between $z\sim1$ and $z\sim0$ \citep[e.g.][]{Lidman_2012,Lin_2013}. This question has been harder to address at $\rm z > 1$, where representative samples of clusters have been previously limited -- though recent studies suggest that massive galaxies and potential BCG progenitors may be assembling and quenched as early as $\rm z \sim 1.5$ \citep[e.g.][]{Strazzullo_2019}. One open question is when the BCG designation even becomes relevant: at what redshift do BCGs emerge as dominant, massive galaxies in their host clusters? BCG emergence is an indication that cluster-specific growth channels are operating efficiently to assemble massive galaxies. The redshift and timescale over which BCGs achieve dominance should help constrain the mechanisms of growth, and may coincide with the establishment of other BCG or cluster properties. 

%De_Lucia_2007,Laporte_2013, Ragone-Figueroa_2018

%and may correlate with the establishment of other characteristic properties of BCGs, such as their central position in clust. 

%The redshift and timescale of emergence --should lend insights on the growth channels operating and the establishment of other  

Historically, BCG dominance has been measured using the magnitude gap \citep[e.g.][]{Bautz_Morgan_1970}, $\rm M_{1,n}$ (Equation \ref{M12_equation}), defined as the difference in absolute magnitude between the BCG and nth brightest cluster galaxy, where typically n = 2, 3, or 4 \citep{Jones_2003,Dariush_2010,2012ApJ...752...12H, Golden-Marx_2018, Golden_Marx_2022, Golden_Marx_2025}. For application to stellar masses, we adopt the modified definition of the magnitude gap as the ``mass gap" in Equation \ref{Mass_gap}, where $\rm M_{*_n}$ is the stellar mass of the nth brightest galaxy. In this analysis, we apply the magnitude gap ($\rm M_{1,n}$) to observed clusters and the mass gap ($\rm M_{*_{1,n}}$) to simulated clusters. The behavior of the indicators is effectively identical, and we will refer to $\rm M_{1,n}$ or the ``magnitude gap" for simplicity in general discussion.

\begin{equation}
\rm
M_{1,n} = M_{n}-M_{1}
\label{M12_equation}
\end{equation}

\begin{equation}
\rm
M_{*_{1,n}} = 2.5log_{10}(\frac{M_{*_1}}{M_{*_n}})
\label{Mass_gap}
\end{equation}

In isolation, the magnitude gap would be a measure of cluster age, with larger values of $\rm M_{1,n}$ correlating with older systems and earlier formation times \citep{DOnghia_2005,Dariush_2007,von_Benda_Beckmann_2008,Dariush_2010,2012ApJ...752...12H}. 
%In turn, larger $\rm M_{1,n}$ typically correlates with more dominant BCGs, given the larger timescales allowed for hierarchical assembly in older clusters \citep{DOnghia_2005,Shen_2014,Solanes_2016,Golden-Marx_2018}.  

However, the magnitude gap has several known limitations. Its calculation relies on the correct identification of the BCG and other massive galaxies. $\rm M_{1,n}$ does not take the relative dominance of the most massive galaxies over $\rm M^{*}$ (the satellite population) into account, leading to degeneracy between relatively less evolved systems with $\rm \bar{M}\sim M^{*}$ and those with several massive galaxies magnitudes brighter than $\rm M^{*}$. As a result, $\rm M_{1,n}$ is far more susceptible to changes in cluster dynamical state \citep{Solanes_2016}, driven by transient phenomena like cluster mergers. As found by \citet{Dariush_2010} and \citet{von_Benda_Beckmann_2008}, the magnitude gap criterion for identifying fossil groups diminishes within fractions of the Hubble time. The magnitude gap struggles to capture the true dominant end-state of the BCG. A classic example of this is the Coma cluster, which has effectively no magnitude gap despite being highly evolved.

We introduce the Gini index (G), a widely-used indicator of economic uniformity/disparity, as an alternative indicator of BCG dominance over other cluster members. Our analysis is organized as follows. In Section \ref{Sec: Gini_Index}, we introduce the Gini index, describe its calculation and extrema, and compare it to $\rm M_{*_{1,2}}$ and $\rm M_{*_{1,4}}$ for TNG300-1 clusters. We then apply the Gini index and $\rm M_{1,2}$ to a sample of 92 SZ-selected cluster cores with optical counterparts between $\rm z \in [0,1]$ (Section \ref{sec: sample and observations}) and membership selection described in Section \ref{sec:methods}. Finally, we test the evolution of G and $\rm M_{1,2}$ (Section \ref{sec:gini_evolution}) and discuss implications for BCG emergence (Section \ref{Discussion}). %$\rm M^{*}$

We use $\rm H_0 = 70~km/s/Mpc$ and flat $\Lambda$CDM cosmology with $\rm \Omega_M = 0.3$ and $\rm \Omega_{\Lambda} = 0.7$, except when working with IllustrisTNG TNG300-1 simulation data in Section \ref{subsec: simulations}, where we use Planck 2015 cosmological parameters for consistency. We use AB magnitudes throughout and the \citet{Salpeter_1955} initial mass function (IMF).

\section{The Gini index}
\label{Sec: Gini_Index}

%This dependence on $\rm M_{lim}$ drives the slight richness dependence of G.

Introduced as a sociological indicator by \citet{gini1912variabilità}, the Gini index is a statistic used widely in economics, particularly as a measure of wealth inequality among human populations. In this work, we apply the Gini index to the galaxy population in clusters, where it will serve as a measure of \emph{light} (or stellar mass) inequality among member galaxies. Just as a low Gini index in economics refers to a population with a relatively uniform distribution of wealth, a low Gini index in this work will refer to a galaxy cluster for which the member galaxies are all similarly luminous. Thus, like the magnitude gap, the Gini index can quantify the dominance of the BCG over the other cluster members. As a point of emphasis, this is not equivalent to previous applications of the Gini index to quantify the morphological smoothness of \emph{individual} galaxies \citep[e.g.][]{Abraham_2003,2004AJ....128..163L,Lisker_2008,Florian_2016,Nersesian_2023}. Nor is this equivalent to applications quantifying the smoothness of the ICM in the X-ray \citep[e.g.][]{2022A&A...661A..12G}.

Mathematically, the Gini index is calculated (bar a normalization factor) by taking twice the difference of two integrated distributions: the observed cumulative distribution of the total luminosity contributed by each galaxy (the Lorenz curve), and the theoretical cumulative distribution when all galaxies are equally luminous. As light is concentrated in fewer, more massive galaxies in a sample, the area under the observed cumulative distribution decreases and the Gini index grows. Another, perhaps more intuitive, way to think about G is as a non-parametric measure of the shape or skewness of the luminosity function.  For a sample of $\rm N$ galaxies $\rm i$ with luminosities $\rm L_{i}$, the Gini index G is calculated by Equation \ref{gini_equation_wikipedia}. The definition is equivalent for stellar masses $\rm M_{*_{i}}$.

%This definition is equivalent for galaxy luminosities $\rm L_{i}$. and magnitudes with the substitution $\rm L_i = 10^{-0.4(M_{i})}$

\begin{equation}
\rm
G = \frac{N}{N-1}\frac{\sum_{i = 1}^{N}\sum_{j = 1}^{N} |L_{i} - L_{j}|}{2N\sum_{i = 1}^{N}L_{i}}
\label{gini_equation_wikipedia}
\end{equation}

Typically left out of formalizations for economics, the normalization factor $\rm N/(N-1)$ is used when G is applied to smaller, finite samples of objects to remove dependence on sample size \citep[e.g.][]{Gini_normalization_paper}. We note that the Gini index is insensitive to the choice of magnitude system, as long as it is unbiased. When $\rm L_{i}$ represents a consistent fraction of $\rm L_{tot}$ for each galaxy, this fraction cancels in Equation \ref{gini_equation_wikipedia}.

\begin{figure}[t]
\includegraphics[width=\columnwidth]{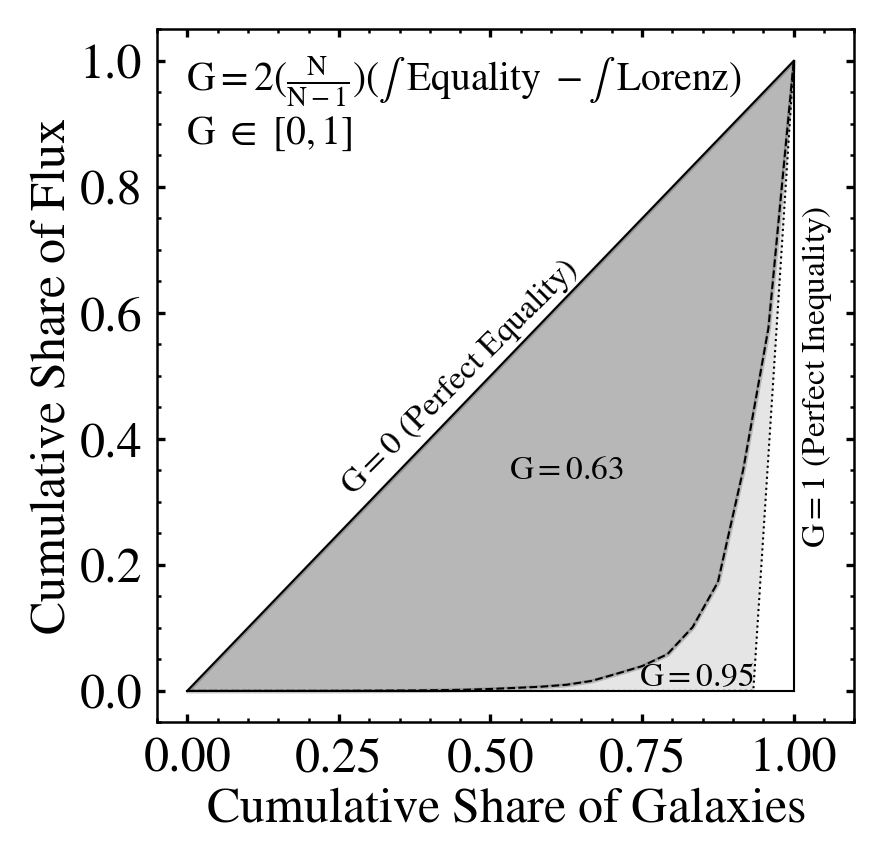}
\caption{An illustration of the calculation of the Gini index. The Lorenz curves for the idealized $\rm G = 0$ and $\rm G = 1$ cases are plotted, as well as those for two galaxy clusters with \textbf{$\rm G = 0.63$} (dashes) and \textbf{$\rm G = 0.95$} (dots). The Gini index is calculated by doubling the area between the 1:1 line of perfect equality and Lorenz curve and then applying the finite sample normalization factor, $\rm N/(N-1)$.
\label{fig:Visual_G_Calc}}
\end{figure}

\begin{figure*}[ht!]
\includegraphics[width=\textwidth]{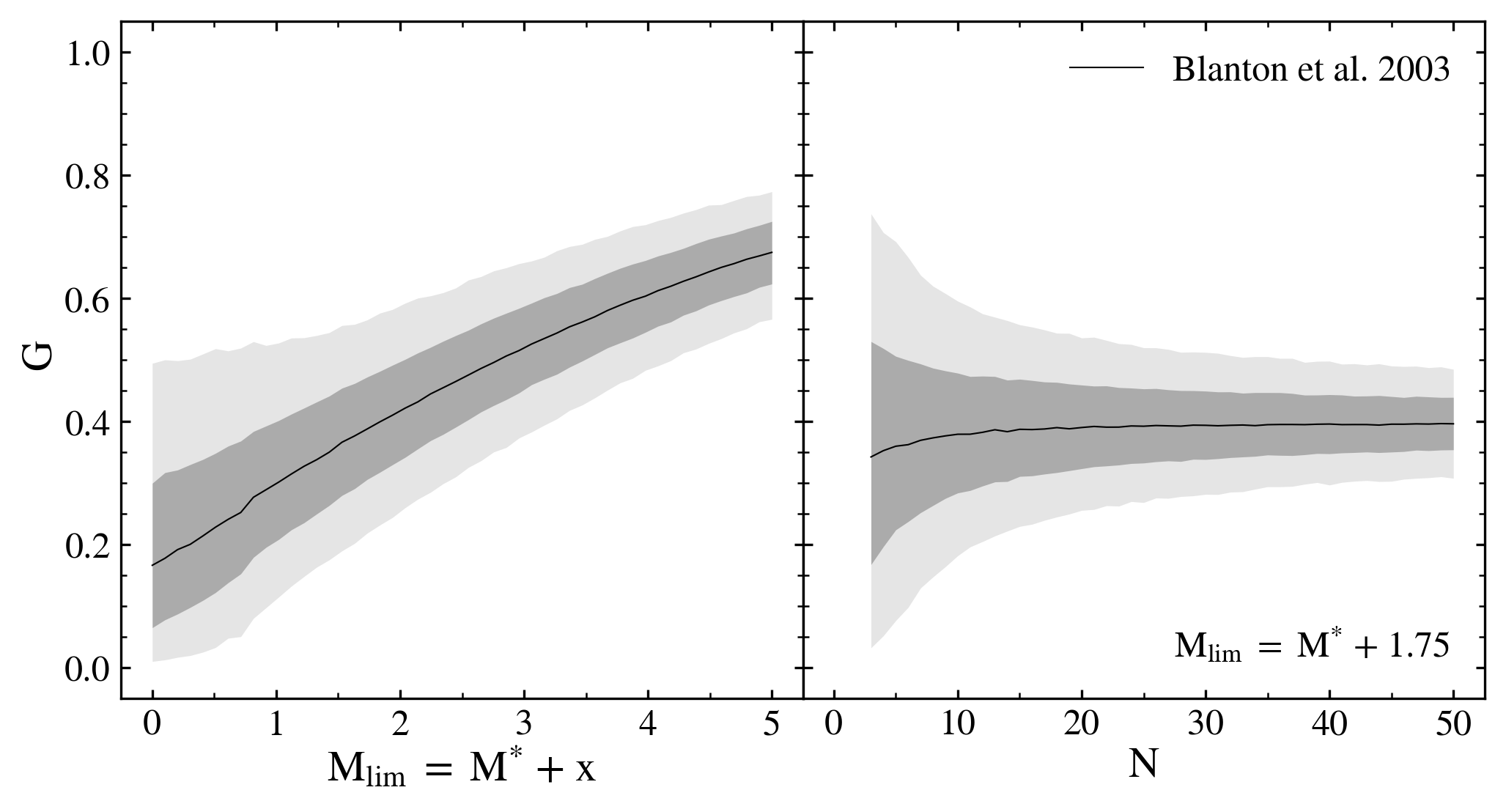}
\caption{The evolution of the Gini index with \textbf{a} varying limiting magnitude ($\rm M_{lim}$) and a varying flat cluster richness (N) cut, applied here to the \citet{Blanton_2003} SDSS r-band Schechter function. Given the application to the luminosity function, this figure displays the distribution of the minimum Gini index, $\rm G_{min}$ (equivalently $\rm G_{field}$) compared to the limiting magnitude and richness, but the trends apply generally. We show the median $\rm G_{min}$ with 68\% and 95\% confidence intervals, calculated by sampling $\rm \Phi(M)$ above $\rm M_{lim}$ and measure G for 10,000 bootstrapping iterations. When limiting richness, we fix $\rm M_{lim} = M^{*} + 1.75$. While the Gini index is largely independent of Schechter model parameters, and thus band and redshift, it is strongly dependent on the limiting magnitude truncating the luminosity function. When $\rm M_{lim}$ is fixed, G(N) is approximately flat at statistically-representative samples of $\rm N \gtrsim 10$. Below $\rm N \sim 10$, G decreases slightly and errors on $\rm G$ become larger. We caution against making a flat richness cut (e.g., always calculating G using the 20 brightest members) on observed clusters, as this indirectly and imprecisely limits $\rm M_{lim}$.
\label{fig:G_v_M_lim_v_N}}
\end{figure*}

%For magnitudes, we advise using the substitution $\rm L_i \Rightarrow 10^{-0.4(M_{i}-\bar{M})}$. A factor of $\rm 10^{0.4\bar{M}}$, where  $\rm \bar{M}$ is the mean sample absolute magnitude, can be completely divided out of equation \ref{gini_equation_wikipedia}, but helps prevent overflow errors upon computation.

%Equation \ref{gini_equation_mags} is an alternate definition better suited for absolute magnitudes $\rm M_{i_k}$ in a given band k. A factor of $\rm 10^{0.4\bar{M}_k}$, where  $\rm \bar{M}_{k}$ is the mean absolute magnitude of the sample, can be completely divided out of equation \ref{gini_equation_mags}, but helps prevent overflow errors upon computation.

%\begin{equation}
%\rm
%G = \frac{\sum_{i = 1}^{n}\sum_{j = 1}^{n} |10^{-0.4(M_{i_k}-\bar{M}_{k})} - 10^{-0.4(M_{j_k}-\bar{M}_{k})}|}{2n\sum_{i = 1}^{n}10^{-0.4(M_{i_k}-\bar{M}_{k})}}
%\label{gini_equation_mags}
%\end{equation}

% Mike: This is too much info. People can figure out luminosity-to-mag conversion.

\subsection{Maximum and Minimum Gini index}

\label{max_min_gini}
The Gini index is a purely positive quantity with a theoretical range between $\rm G = 0$ (all of the galaxies have the same luminosity) and $\rm G = 1$ (one galaxy has all of the luminosity). 
%The maximum Gini index, $\rm G_{max} = 1$, corresponds to a cluster consisting of a single member possessing all observed flux. 
Fossil groups \citep{Ponman_Bertram_1993, Ponman_1994, Vikhlinin_1999, Mulchaey_1999, Jones_2000,Jones_2003}, containing a single, evolved BCG-like galaxy and X-ray emitting ICM in a cluster-sized dark matter halo, are astrophysical systems that approach the $\rm G=1$ idealized limit. On the other hand, the minimum Gini index would be achieved in systems for which all of the member galaxies are of a similar magnitude, such as in a compact group like Stephan's Quintet \citep{1982ApJ...255..382H,1989ApJS...70..687H,2011A&A...533A.142B,Lisenfeld_2017}. 
%However, the range of allowed values of G changes when the index is applied to real, physical systems. <-- I don't think this statement is true...it has to be 0-1.
In reality, the majority of groups and clusters have a mix of bright and faint galaxies, and will lie somewhere in between these extremes.

The Gini index is particularly sensitive to BCG assembly and increases in BCG dominance. This interpretation is most evident in the calculation of the Gini index in Figure \ref{fig:Visual_G_Calc}. When we consider a relatively constant presence of satellites with low probability of merging in massive clusters, the Gini index increases when bright galaxies merge, driving a more sharply-peaked Lorenz curve. Thus, while explicit BCG identification is not required to calculate the Gini index, G is theoretically a metric sensitive to how close the cluster is to concentrating the flux of its galaxy population into a single BCG.

%https://ui.adsabs.harvard.edu/abs/2013A%26A...550A..58V/abstract -- Though when BCG is removed, stellar mass dependence with LF is equal.

%\textbf{A careful reader may wonder how the Gini index is specifically related to BCG growth and dominance, rather than the assembly of massive galaxies in clusters generally.} 

%However, the limit of perfect equality, G = 0, is not possible, given the inherent luminosity function (citations) that governs the distribution of galaxy luminosities. T
The field population, where in-situ galaxy evolution dominates over merger-based growth, is likely the most uniform population of galaxies available and, thus, likely represents the realistic minimum $\rm G$.
%
%Thus, the minimum physical Gini index observed in nature is that which corresponds to the field population at the depth probed by our samples. 
The measurement of $\rm G$ is only weakly dependent on the total number of galaxies included, but strongly dependent on the limiting magnitude applied to the distribution, as we demonstrate in Figure \ref{fig:G_v_M_lim_v_N}. For the limiting magnitudes we are considering in this work ($\rm M_{lim} \in [M^* + 1.5, M^* + 2]$), assuming a luminosity function measured in the field \citep{Blanton_2003}, we find a median \textbf{$\rm G_{field} = 0.35-0.45$.}

As mentioned above, the Gini index scales weakly with the number of galaxies included when $\rm M_{lim}$ is fixed, as shown in Figure \ref{fig:G_v_M_lim_v_N}. %That is, for two galaxy clusters with the same luminosity distribution but different total number of galaxies, we expect to measure the same value of G, as long as the number of galaxies used is $\gtrsim$10.
$\rm G$ is relatively flat with relatively small errors when $\gtrsim$10 galaxies are contributing to the calculation. 
On the other hand, the limiting magnitude (or how many faint galaxies we include) has a strong effect on $\rm G$, as shown in Figure \ref{fig:G_v_M_lim_v_N}, which makes intuitive sense. 
Thus, we will use a fixed limiting magnitude as a general requirement in subsequent sections. We advise against fixing $\rm N$ when calculating $\rm G$ for observed clusters, as this imprecisely varies $\rm M_{lim}$. (see Section \ref{application_benefits_G} and Appendix \ref{Appendix: Richness}). %In general, we note from Figure\, \ref{fig:G_v_M_lim_v_N} that the choice of the limiting magnitude is the most important when making comparisons across samples. 
We note that when we write N or ``richness" here, we are referring to the actual number of galaxies used in the calculation of G, not the inherent cluster richness ($\rm \lambda$).
%So, for two galaxy clusters with the same luminosity distribution but different total number of galaxies, we expect to measure the same value of G, as long as the number of galaxies used is $\gtrsim$10.

%$\rm N>10$ and 

% the luminosity function is probed representatively, at about $\rm N \sim 10$. However, when working with observed clusters, we caution against limiting depth by making cuts on constant richness. This method indirectly limits magnitude inconsistently, as there is natural scatter in the magnitude of the Nth brightest galaxy, limiting the comparability of G across a sample. 

%when limiting depth by making a richness cut. However, we understand this dependence to be driven by  

%you could say something about why this makes sense here-- intuition-building

%and thus needs to be re-calculated for application to different samples. Within limiting magnitude cuts $\rm M_{lim} \in [M^*_{F110W} + 1, M^*_{F110W} + 3]$ (see Sections \ref{lim_mag_richness}, \ref{error estimation}), applied to the \citet{SDSSLuminosityFunction} Schechter function, $\rm G_{min} = 0.35 \pm 0.1$.

\subsection{Behavior of the Gini index}
\label{subsec: simulations}

To further understand the behavior of the Gini index when compared to the magnitude gap, we consider two simple processes that can alter the values of $\rm G$, $\rm M_{1,2}$, and $\rm M_{1,4}$ in galaxy clusters. One is the merger of galaxies, particularly major mergers that lead to the assembly of the BCG or several BCG-like galaxies. In this scenario, $\rm G$ will generally increase as stellar mass is concentrated into fewer galaxies and the relative brightness disparity amongst members increases. While mergers between bright galaxies can cause individual $\rm |L_{i} - L_{j}|$ terms in Equation \ref{gini_equation_wikipedia} between bright galaxies to decrease, the corresponding increase in $\rm |L_{i} - L_{j}|$ terms between bright and $\rm \sim L^{\star}$ and fainter satellite galaxies prevent large decreases in the index.

\begin{figure*}[ht]
\centering
\includegraphics[width=1.7\columnwidth]{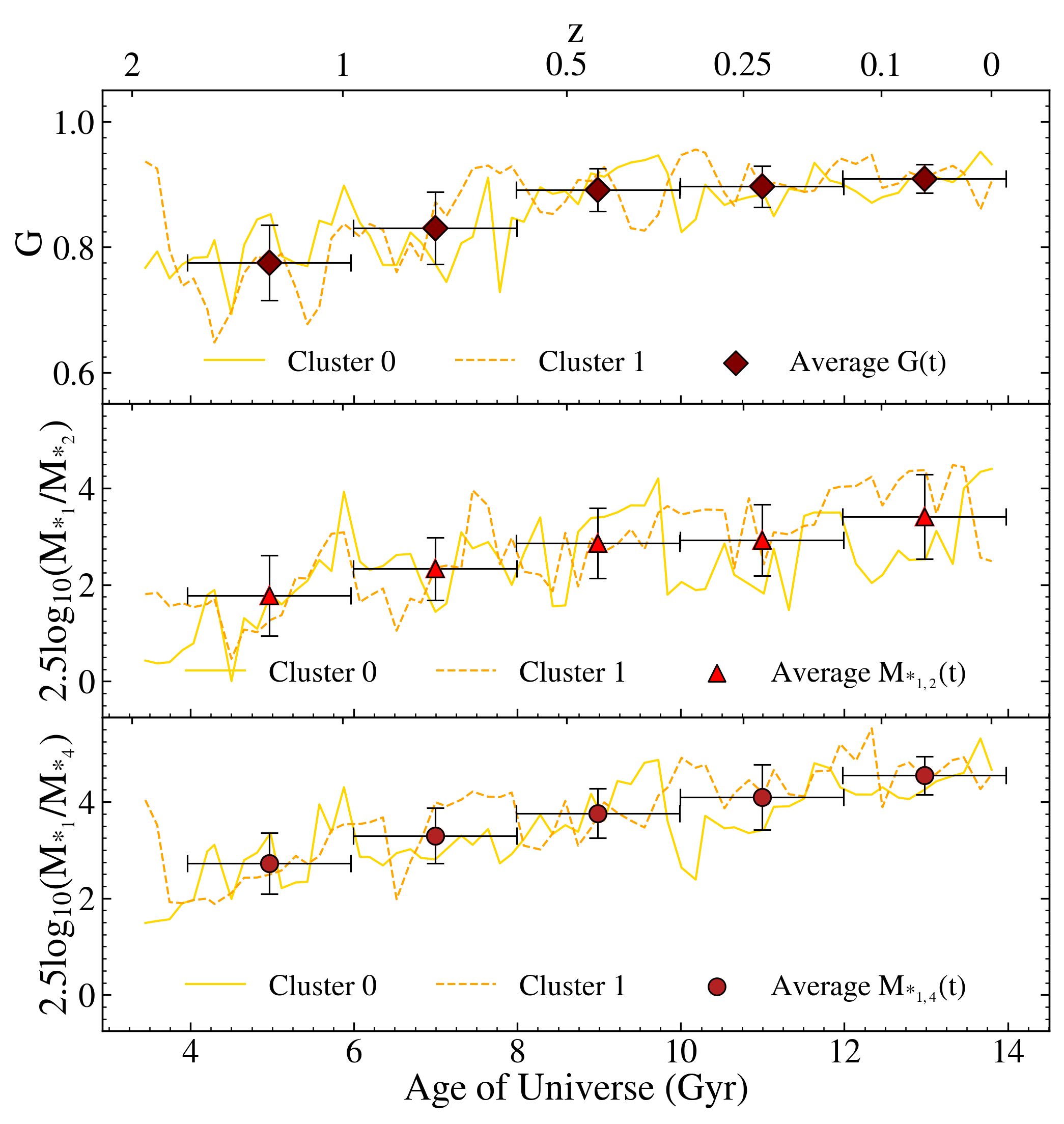}
\caption{The Gini index and mass gaps $\rm M_{*_{1,2}}$ and $\rm M_{*_{1,4}}$ applied to TNG300-1 Clusters 0 and 1 for 10 Gyr of assembly history traced by following the main progenitor branch of the BCG. We select and apply $\rm M_{*_{1,2}}$, $\rm M_{*_{1,4}}$, and G to the 20 brightest galaxies in each cluster at each timestep. The plotted markers are the mean and 1-sigma standard deviations of each indicator in 2 Gyr time bins spanning $\rm t_{age} \in [4~Gyr, 14~Gyr]$, with horizontal error bars denoting the width of each bin. Each indicator shows the increasing trend expected from structural assembly and dynamical relaxation through time as BCGs grow more dominant, although $\rm M_{*_{1,2}}$ is nearly consistent with flat within its mean errors.
\label{fig:simulations}}
\end{figure*}

$\rm M_{1,2}$ is far more sensitive to how mass is concentrated -- the merger of two BCGs can drive $\rm M_{1,2}$ from a value of $\sim$0 to $\sim$4 in a dynamical time. Systems with two massive proto-BCGs and less massive satellites can have identical measurements of $\rm M_{1,2}$ to clusters with many satellite galaxies and no obvious BCG. This known degeneracy has led to wider adoption of $\rm M_{1,n>2}$ by the community, where this problem is less pronounced due to the likelihood that the nth brightest galaxy is more representative of the satellite population \citep{Dariush_2010, Golden-Marx_2018, Golden_Marx_2022, Golden_Marx_2025}. The Gini index is effectively an integral of the generalized magnitude gap over all \textbf{$\rm n = N$}.

A second process that can drive the values of $\rm G$, $\rm M_{1,2}$, and $\rm M_{1,4}$ is the mergers of galaxy clusters \citep{von_Benda_Beckmann_2008}. Consider the following idealized scenario: a rich cluster merges with another rich cluster of equal mass and an identical population of member galaxies which relax together within the new host cluster potential. In this idealized scenario, the identical cluster merger drives $\rm M_{1,2}$ to zero due to the presence of two identical BCGs. $\rm M_{1,4}$ will similarly decrease, though not to zero but rather to the pre-merger value of $\rm M_{1,2}$ due to the re-indexing of the BCG and fourth brightest galaxy in the cluster post-merger. However, due to its definition, the Gini index remains unchanged before and after the cluster merger. The $\rm |L_{i} - L_{j}|$ terms in Equation \ref{gini_equation_wikipedia} for duplicated identical galaxies will cancel. For non-identical galaxies, the cumulative sum of remaining $\rm |L_{i} - L_{j}|$ terms will double the initial value of the sum, but this factor will cancel out upon the doubling of $\rm \sum_{i = 1}^{N}L_{i}$. Though this scenario is fictional, it illustrates why the Gini index is well-suited for measuring the long-term emergence and assembly of the BCG, since it is the most agnostic to short-lived phenomena (like the cluster's dynamical state) and the indexing of individual member galaxies.

%one branch of the BCG merger tree for FoF-identified group0 and group1 in the TNG300-1 z = 0 snapshot
%G_M14_M12_v_time.png
%Gini index and two renditions of the stellar-mass magnitude gap see adapted for stellar mass inputs applied to TNG clusters 0 and 1 throughout the past 10 Gyr. We apply a richness cut of the 20 brightest cluster galaxies, where . The over-plotted markers are the average and 1-sigma standard deviations of each indicator in the 2 Gyr time bins spanning $\rm t_{age} = 4~Gyr$ to $\rm t_{age} = 14~Gyr$. 
%$\rm M{*_{1,2}}$ and $\rm M{*_{1,4}}$ show little evolution in scatter, which is an order of magnitude larger than the scatter in G. G, the indicator most agnostic to the dynamical state of the cluster and labeling of individual galaxies, demonstrates evolution in its scatter. The simultaneous decrease in scatter and sustained value of $\rm \bar{G}$ after $\rm t_{age} = 8~Gyr$ is expected with the emergence and subsequent dominance of the simulated BCGs over the member galaxy stellar mass distribution

To further illustrate their behavior, we apply the Gini index and mass gaps (Equation \ref{Mass_gap}) to the two most massive clusters ($\rm M_{halo} \approx 10^{15}~M_{\odot}$) in the IllustrisTNG TNG300-1 simulation FoF group catalog at $\rm z = 0$  %: group0 ($\rm M_0 = 1.15\times10^{15}~M_{\odot}$) and group1 ($\rm M_1 = 9.64\times10^{14}~M_{\odot}$) 
\citep{nelson2021illustristngsimulationspublicdata,Pillepich_2017,Springel_2017,Nelson_2017,Naiman_2018,Marinacci_2018}. These are the only two clusters available in the group catalog with $\rm M_{500}$ overlapping with our sample selection function in Figure \ref{fig:samples}. We trace the membership of these clusters iteratively using the merger history of the BCG (via \citet{Rodriguez_Gomez_2015} SubLink merger trees), which is identified as the most massive subhalo in groups 0 and 1 at $\rm z = 0$. The BCG is traced to its corresponding main progenitor in previous simulation snapshots ($\rm \bar{\Delta t} = 160 \pm 30~ Myr$), which we use to identify the group (cluster) in which it is embedded and the group's 20 most massive member galaxies at each time-step. Our Clusters 0 and 1 are thus understood as the progenitors of groups 0 and 1 in the redshift zero TNG300-1 group catalog, obtained by following the main progenitor branch of their redshift zero BCGs backwards in time for 10 Gyr.

%\textbf{Cluster 0 and 1 are thus the main progenitor halos of groups 0 and 1, identified by following the main branch of the BCG merger tree for 10 Gyr.}

%thus represent the structures around one branch of the z = 0 BCG's merger tree for 10 Gyr of assembly.

The results of Figure \ref{fig:simulations} support the results of the above thought experiments. G, $\rm M_{*_{1,2}}$, and $\rm M_{*_{1,4}}$ display an overall positively increasing trend with time, generally expected of cluster and BCG co-assembly \citep{Laporte_2013, Zhang_2016}. However, the fractional scatter and its evolution are quite different between G, $\rm M_{*_{1,2}}$, and $\rm M_{*_{1,4}}$, as shown in Figure \ref{fig:sim_frac_err}. 
% although this trend is most evident for $\rm M_{*_{1,4}}$ and G

\begin{figure}[ht!]
\includegraphics[width=\columnwidth]{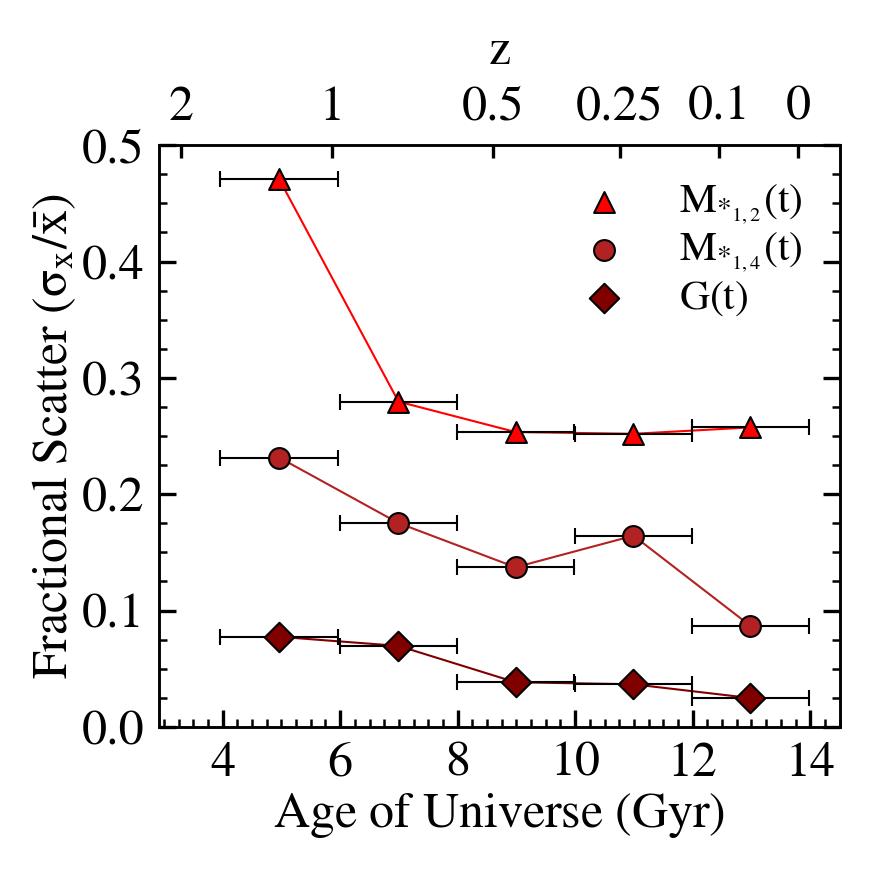}
\caption{Fractional scatter about mean values of G, $\rm M_{{*}_{1,2}}$, and $\rm M_{{*}_{1,4}}$ calculated in Figure \ref{fig:simulations} for TNG300-1 Clusters 0 and 1. $\rm \sigma_x$ is the 1-sigma standard deviation on each indicator, x\textbf{,} within a 2 Gyr time period (denoted by the horizontal error bar on each marker). $\rm \sigma_{G}/\bar{G} < \sigma_{M_{*_{1,4}}}/\bar{M}_{*_{1,4}} < \sigma_{M_{*_{1,2}}}/\bar{M}_{*_{1,2}}$ for all times, which agrees with our understanding that sensitivity to the clusters dynamical state and degeneracies due to galaxy indexing drive more scatter in the magnitude gaps than the Gini index.   
\label{fig:sim_frac_err}}
\end{figure}

The fractional scatter (over a 2 Gyr period) on the Gini index is smaller than that of both $\rm M_{*_{1,2}}$ and $\rm M_{*_{1,4}}$ for all time. The fractional scatter on $\rm M_{*_{1,4}}$ is smaller than that of $\rm M_{*_{1,2}}$ for all time. This agrees with our analytical predictions that $\rm M_{*_{1,2}}$ is the most sensitive to dynamical phenomena, while G is the least sensitive. The reduction in fractional scatter is one of the primary benefits of using G over $\rm M_{*_{1,4}}$. By taking the entire satellite population into account (above some limiting magnitude) , G reduces the fractional scatter on $\rm M_{*_{1,4}}$ by over a factor of 2.5 at each timestep in Figure \ref{fig:sim_frac_err}.

Comparing Figures \ref{fig:simulations} and \ref{fig:sim_frac_err}, the intrinsic scatter on $\rm M_{*_{1,2}}$ is $\rm O(1)$. This means that the predicted evolution of $\rm M_{*_{1,2}}$ between $\rm z=2$ and $\rm z=1$ of $\sim$2 mag is at the same level as the moment-to-moment scatter. Phrased another way, the magnitude gap is too sensitive to the recent dynamical history of the cluster to adequately capture the long-term evolution in this illustrative example. $\rm M_{*_{1,2}}$ has a nearly unchanging average scatter for all times later than 6 Gyr in the example clusters, likely due to its sensitivity to cluster accretion and galaxy indexing.

%likely driven by the Gini Indices lesser degree of sensitivity to  The $\rm 1\sigma$ errors in $\rm M_{*_{1,2}}$ and $\rm M_{*_{1,4}}$ are $\rm O(1)$, which are larger than the $\rm 1\sigma$ in G, which are $\rm O(0.1)$. 

Conversely, the scatter in G and $\rm M_{*_{1,4}}$ decreases nearly monotonically in Figure \ref{fig:sim_frac_err}. The errors on the mean values of G, in particular, decrease by \textbf{$\rm \sim 60$\%} between $\rm t_{age} = 7 - 9$ Gyr as G reaches a nearly fixed mean value \textbf{$\rm \bar{G} = 0.91 \pm 0.02$} after 8 Gyr in Figure \ref{fig:simulations}. This is similar to the understood assembly history of BCGs, which should theoretically involve G increasing to a large, nearly fixed value as the BCG emerges and dominates the luminosity distribution of its host cluster \citep[e.g.][]{2026ApJ..1000..103V}. This behavior indicates that G is capable of capturing increases in BCG dominance, given the evolution of $\rm \bar{G}$ at early times caused by BCG growth at a rate faster than its surrounding cluster. It is possible that evolution in the moment-to-moment scatter in G and $\rm M_{*_{1,4}}$ could also be used to indicate BCG emergence, though a larger dedicated study of simulated clusters would be required to confirm this.
%Maybe put the paragraph below back in?
%  

Overall, we argue that short-term dynamical processes and degeneracies due to mass concentration drive scatter in $\rm M_{*{1,2}}$ that can fully obscure evolution when applied to two massive TNG300-1 clusters. By taking a larger portion of the satellite population into account, G is relatively agnostic to these effects, reducing fractional scatter measured in the more degenerate indicator $\rm M_{*{1,4}}$ by a factor $\rm >2.5$. We argue that of the three metrics, from their performance in this illustrative example, G is the best-suited for measuring BCG assembly and emergence.

\section{Sample Selection and Data Analysis} 
\label{sec: sample and observations}

%% The "ht!" tells LaTeX to put the figure "here" first, at the "top" next
%% and to override the normal way of calculating a float position.
%% The asterisk after "figure" tells the compiler to span multiple columns
%% if a two column style is selected.

\begin{figure*}[ht!]
\centering
\includegraphics[width=1.5\columnwidth]{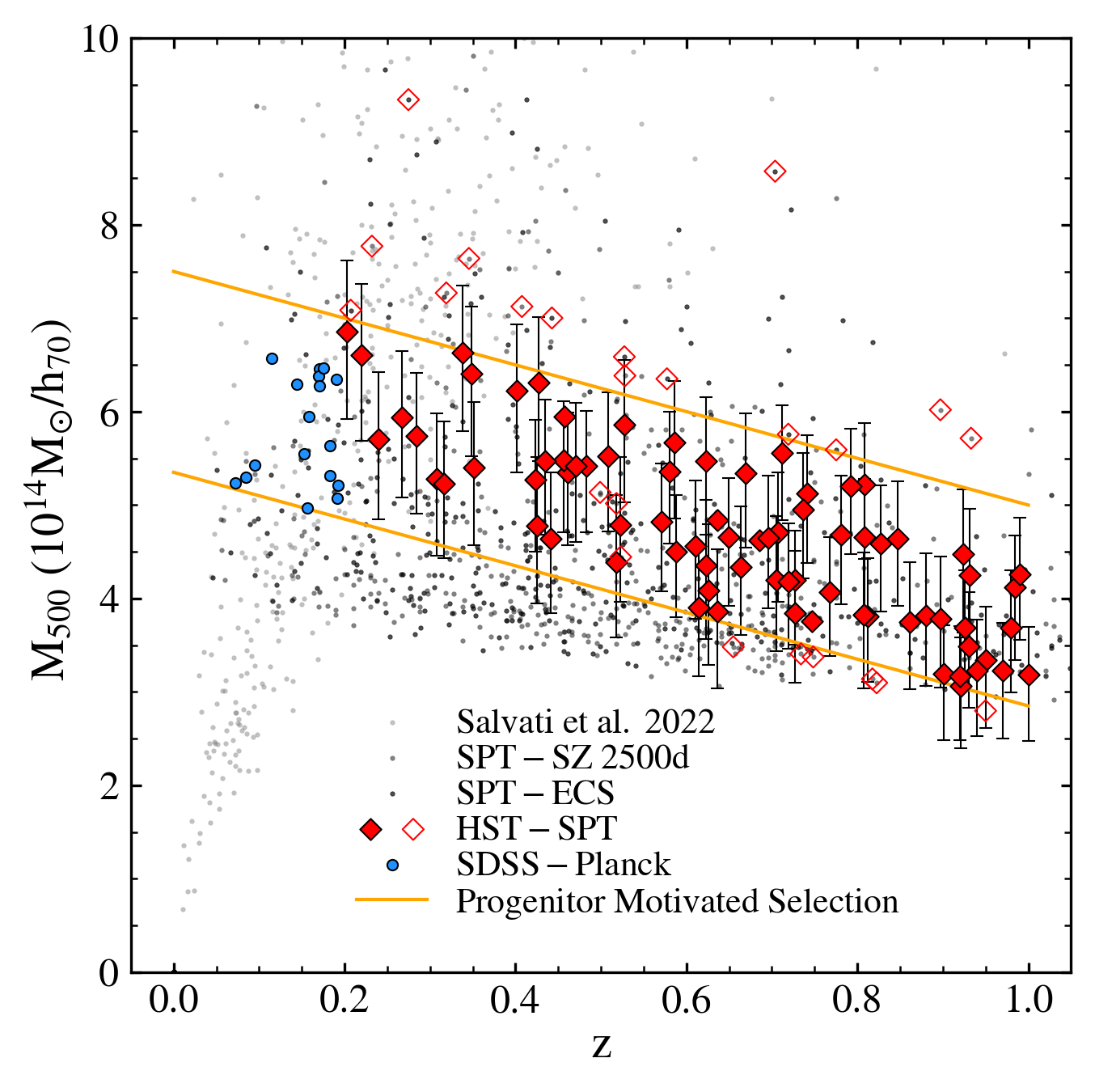}
\caption{Samples and selection function. HST-SPT (with $\rm 1\sigma$ uncertainties on $\rm M_{500})$ and SDSS-Planck are plotted over the SPT-SZ 2500d, SPT-ECS, and \citet{Salvati_2022} Planck PSZ2 samples. Open HST-SPT markers represent clusters that were imaged by HST, but not used in the final sample (see further discussion in Section \ref{subsec: HST observations}). Our selection function is linear and has a slope of $\rm m = -2.5 \times 10^{14} M_{\odot}/h_{70}$ consistent with expected cluster growth rates from simulations \citep{Fakhouri_2010}. Progenitors of approximately Perseus to Coma mass clusters are selected between two intercepts at $\rm z = 0$ : $\rm b_{lower} = 5.35 \times 10^{14} M_{\odot}/h_{70}$ and $\rm b_{upper} = 7.5 \times 10^{14} M_{\odot}/h_{70}$.
\label{fig:samples}}
\end{figure*} 
\subsection{The HST-SPT Cluster Sample}
\label{subsec: HST-SPT}

HST-SPT is a sample of 76 SZ-selected, massive galaxy clusters ($\rm M_{500} > 10^{14} M_{\odot}/h_{70}$) spanning $\rm 0.2 < z < 1.0$ , selected in an unbiased way for follow-up from the SPT-SZ 2500 degree survey (SPT-SZ 2500d) \citep{Bocquet_2019,Bleem_2015} and SPTpol Extended Cluster Survey (SPT-ECS) \citep{Bleem_2020}. All 76 clusters in this sample have shallow HST UVIS+IR imaging in the F110W and F200LP bands (found in MAST at \dataset[10.17909/7x0k-0323]{http://dx.doi.org/10.17909/7x0k-0323}). The combined high resolution photometry and progenitor-motivated selection of HST-SPT is well-suited for evolutionary studies of BCGs. The data can also be leveraged for studies of strong lensing statistics and the concentration-mass relation, or studies of AGN activity and feedback \citep[e.g.][]{Calzadilla_2023}.

%with $\rm \bar{z} \approx 0.72$, 
%This legacy-quality sample of clusters has an associated collection of multi-wavelength data, including Chandra (0.5-10 keV), DES (griz), Spitzer (3.6, 4.5 $\rm \mu m$), and ASKAP (0.9, 1.4 GHz) coverage at the highest redshifts. While we use the sample to study the assembly the of cluster core, it is well-optimized for studies of dark matter content, strong lensing, lensed starburst galaxies, and AGN activity. 

The initial HST-SPT sample (HST 15307; PI Gladders) was selected through a redshift-dependent mass cut on $\rm M_{500}$ ($\rm M_{500c}$ from \citet{Bocquet_2019}), applied to cluster samples from the SPT-SZ 2500d and SPT-ECS surveys. The slope of the mass cut was based on \cite{Fakhouri_2010}, which predicts an approximate doubling of $\rm M_{500}$ from $\rm z \in [0,1.5]$ for massive halos ($\rm M_{500} > 10^{14}M_{\odot}$). At high redshifts, this selection function was flattened to account for the paucity of known low-mass, high-redshift clusters at the time. This yielded a sample of hundreds of candidate clusters, from which 122 clusters were ultimately observed as part of the HST-SNAP program.

%Remove this since we say it below.
For this work, we exclude the highest redshift clusters, opting for a cleaner selection function including clusters at $\rm z<1$ with descendant ($\rm z=0$) masses between $\rm M_{500} = 5.35 \times 10^{14} M_{\odot}/h_{70}$ and $\rm M_{500} = 7.5 \times 10^{14} M_{\odot}/h_{70}$. This reduced sample of 76 clusters allows us to study a coevolving cluster population over $\rm \sim$8\,Gyr, as shown in Figure \ref{fig:samples}. Further rationale for these more stringent cuts is described in Section \ref{subsec: HST observations}.%Thus, HST-SPT is a sample of self-similar progenitor clusters with highly-studied anchor structures in the Local Universe (citations).

%in the Millenium II simulations
%it's $\rm z = 0$ intercepts select for the progenitors of Perseus- to Coma- like clusters. Thus, HST-SPT is a sample of the progenitors of the most well-studied clusters in the local Universe. progenitor-motivated,

For the majority of HST-SPT, we use cluster redshifts published in \citet{Bocquet_2019} (for SPT-SZ 2500d clusters) and \citet{Bleem_2020} (for SPT-ECS clusters). For a few SPT-SZ 2500d clusters, the \citet{Bocquet_2019} redshift was inconsistent with the observed red sequence (RS) upon inspection. In these cases, we opted to use the \citet{Bleem_2015} redshifts instead that preceded the \citet{Bocquet_2019} values. We use the \citet{Bleem_2015} values of $\rm z = 0.87$ for SPT-CLJ0643-4535, $\rm z = 0.88$ for SPT-CLJ2108-4445 and $\rm z = 0.97$ for SPT-CLJ2148-4843.

%is assigned  \citep{Bleem_2015},  is assigned  \citep{Bleem_2015}, and  is assigned  \citep{Bleem_2015}. \citet{Bleem_2015} and or \citet{Bleem_2020}

%, unless they were updated by . For a few clusters, the reported redshift was inconsistent with the observed red sequence upon inspection, and in these cases, we replace the \citet{Bocquet_2019} or \citet{Bleem_2020} redshifts with published alternatives. SPT-CLJ0643-4535 is assigned $\rm z = 0.87$ \citep{Bleem_2015}, SPT-CLJ2108-4445 is assigned $\rm z = 0.88$ \citep{Bleem_2015}, and SPT-CLJ2148-4843 is assigned $\rm z = 0.97$ \citep{Bleem_2015}.

\subsection{HST Observations}
% Check Average time of exposure.
\label{subsec: HST observations}

The Gini index and magnitude gap are indices sensitive to the luminosity distribution of bright galaxies. In observing the HST-SPT sample with the Hubble Space Telescope, we opted to maximize the sensitivity and spatial resolution with which the clusters are observed. To reduce exposure times (as required for SNAP programs), while ensuring sensitivity to a wide range of galaxy SEDs in the optical and near-IR, we chose to use the broadest possible filters on WFC3: F200LP (UVIS) and F110W (IR). F200LP spans from the UV to the NIR, while F110W spans the red optical to the bluer NIR, particularly the I and J bands. F200LP provides greater or equal depth to similar filters, F606W and F814W, for all reasonable galaxy SEDs, and does not require post-flash to reduce CTI when orbit packing, maximizing exposure time and observation depth. We additionally configured the orbits to use three different exposure times-- short, medium, or long-- observing higher redshift clusters for longer times to ensure sufficient depth across the sample. To optimize PSF reconstruction and observational efficiency, we opted for a 3-point line dither in UVIS and a 4-point box dither in the IR. This approach also ensures the effective management of detector defects and cosmic rays. 
%
%With a mean wavelength of $\rm 6043$ \AA~ and bandpass from [1990 \AA, 10778 \AA], ~ , with a bandpass of [8846.79 \AA, 14122.58 \AA] and mean wavelength of 11797.14 \AA
%
%Our chosen broadband filters are well-suited for combining F200LP and F110W observations of each cluster into one SNAP program orbit, which have exposure times with upper limits of 45 min. We optimized our orbit-packing solutions to fit within this limit. 
 %Typical exposure times for these observations were between 30-45 minutes, depending on the redshift. %Our optimization leads to a minimum orbit time of 33.2 min at the lowest observed z ($\rm z \sim 0.2$) up to 45 min at the highest observed z ($\rm z \sim 1.4$).

There is a subset of clusters for which we have HST imaging that are not used in the final HST-SPT sample. As described above, we tightened our selection function from that originally proposed, and thus some observed high- and low-mass clusters have been eliminated from the HST-SPT sample. Additionally, we found that our HST SNAP observations were too shallow to probe deeply into the luminosity function for clusters with $\rm z > 1.0$, which introduced an unphysical decline in richness (N) at high-$z$. This was especially concerning given the dependence of G on limiting magnitude, as our inability to observe fainter galaxies in these high-z clusters imparted an effective magnitude limit different than that of the rest of the sample. Thus, to ensure completeness and statistical comparability, we limited HST-SPT to clusters with $\rm z \le 1.0$, where we can consistently reach depths of $\rm M^{*}+2$. Finally, we removed a selection of clusters from HST-SPT due to foreground contamination in their optical images.  SPT-CLJ0641-5950, SPT-CLJ2040-5342, and SPT-CLJ2050-4213 have bright foreground stars or diffraction spikes co-occurrent or obscuring the identified BCG, causing unreliable photometry in the cluster cores.  

In total, we observed 122 SPT cluster targets with the HST-SNAP program from 2017 to 2019, 76 of which make up the finalized HST-SPT sample used in this work. On average, our clusters are observed for integration times of $\rm 834 \pm 164~s$ in F110W and $\rm 779 \pm 44~s$ in F200LP. Even at high redshifts with relatively short exposures, our broadband observations are sensitive enough to capture the faint end of the red sequence down to $\rm M^{*}+2$ mag ($\rm M_{F110W} \approx -19$) at $\rm z = 1$. %-16 at $\rm z \in [0.2,0.6]

%We use a 3- point line dither in UVIS, and a 4-point box dither in the IR. This approach effectively adresses bad pixel and cosmic ray rejection 
%This is sufficient for bad pixel and CR rejection, and provides a good balance between PSF reconstruction and efficiency.
%Do we expect greater scatter in F110W observation times?

\subsection{HST Data Reduction}
% Check Average time of exposure.
\label{subsec: HST Data Reduction}

Prior to combining the calibrated single images produced by the MAST archive, we apply two corrections. The first addresses a visible mismatch in sky levels in the F200LP data between the two WFC3-UVIS chips. We rectify this issue by subtracting the background sky on each chip, with the sky estimated by fitting an iterated Gaussian model, sigma-clipped on the high end, to the pixel values. This correction has no effect on photometry.
%This correction is purely cosmetic with no effect on photometry, and is done to provide the best resulting data for constructing two-band psuedo-color images for further visual examination.

Secondly, in the IR/F110W data, we correct for the well-known ‘IR blobs’ \citep{2010wfc..rept....6P} with additional flat fielding. To do so we acquired several hundred additional F110W images of similar integration time from the archive (from program GO-15163), which we then combined with most of the F110W data described here using IRAF’s imcombine task, tuned to provide robust rejection of the faint wings of object outliers in this large image stack. We then used \textsc{Galfit} \citep{2002AJ....124..266P} to build a model - using a combination of S\'ersic \citep{1963BAAA....6...41S} profiles - of each identified IR blob across the stacked image. The composite full image of all the fitted blob models is then used to flat-field these features in the F110W data. %At the time that this calibration was computed, neither GO-15163 nor the program described here were entirely complete, however we do not expect significant change in the IR-blobs over the remaining data not directly used in creating this specialized flat-field.
%\textbf{which overlap in many instances with important target objects (galaxy cluster member galaxies and occasionally lensed arcs) observed in these data. Under normal circumstances, the pixels affected by these features are flagged and not used in final data products; in this program, because the data are minimal (three dithered images only, due to the truncated SNAP orbit length and two filter observations within that orbit) we aim to recover the areas affected by IR blobs by careful flat-fielding.  

Data reduction of the WFC3 imaging with the F200LP and F110W filter - tuned as above - then follows the standard procedures established by  STScI, employing the Python3 package Drizzlepac \citep{2012drzp.book.....G,2021AAS...23821602H}. For astrometric matching of images, sub-exposures taken with the same filter were aligned to the same reference grid, using tweakreg operating on tuned catalogs extracted using Source Extractor \citep{1996A&AS..117..393B}; the two filters were also astrometrically matched. These exposures were subsequently combined using astrodrizzle, using a Gaussian kernel, with nominal cosmic ray rejection, a final pixfrac = 1.0 and 0.8 (for the F110W and F200LP, respectively) and pixel scale of $\rm 0.03^{\prime \prime}/pixel$. 

\subsection{GALAPAGOS Modeling}
\renewcommand{\arraystretch}{1.5}
\begin{deluxetable*}{ccccccc}[ht!]
%\digitalasset
\tablewidth{\textwidth}
\tablecaption{Cluster Samples Summary \label{tab:sample_summary}}
\tablehead{
\colhead{Sample Name}& \colhead{\# Clusters}& \colhead{Redshift Range}& \colhead{Description}
}
\startdata 
HST-SNAP & 122 & $\rm 0.2 < z < 1.4$ & \parbox{10cm}{SPT-SZ 2500d and SPT-ECS sample clusters with HST-SNAP program imaging in F200LP and F110W.}\\ [3ex] 
HST-SPT & 76 & $\rm 0.2 < z < 1.0$ & \parbox{10cm}{HST-SNAP clusters within selection function (Figure \ref{fig:samples}) and $\rm z < 1$. Three clusters within selection function removed from final sample due to foreground obscuration of the cluster core.} \\ [4ex] 
SDSS-Planck & 16 & $\rm 0.05 < z < 0.2$ &  \parbox{10cm}{Planck PSZ2 clusters from \citet{Salvati_2022} within selection function (Figure \ref{fig:samples}) in SDSS footprint with $\rm b > 20^\circ$. Six clusters removed from final sample due to unreliable BCG photometry or issues with optical cluster identification.}\\[4ex]
\enddata
\tablecomments{Summary of cluster samples described in Sections \ref{subsec: HST-SPT} - \ref{Subsec:_Planck_Sample}}
\end{deluxetable*}

The calculation of the Gini index requires luminosity estimates for each member galaxy, which requires (i) source identification, (ii) source deblending, and (iii) estimating the total luminosity at large radius. This analysis benefits from the high angular resolution and depth of the HST data, which allows individual galaxies to be identified and separated more cleanly. 

To convert our HST imaging to photometric catalogs, we utilize the \textsc{Galapagos-2} software \citep{2012MNRAS.422..449B,2013MNRAS.430..330H,2022A&A...664A..92H}. The finer details of how this software works can be found in the above references, but we will summarize again here. This software package uses Source Extractor \citep{1996A&AS..117..393B} to identify and deblend sources, which provides rough estimates of the total luminosity, size, and shape of each source. \textsc{Galapagos} then identifies which sources are in overlapping areas, based on their apparent sizes, and groups all overlapping sources into cutouts to be jointly fit. All sources of similar brightness in a cutout are simultaneously fit using \textsc{Galfit} \citep{2002AJ....124..266P}, beginning with the brightest sources. These models are then held fixed while fainter sources in the cutout are fit, until all sources in a cutout have been modeled. For this analysis, we assume a single S\'ersic \citep{1963BAAA....6...41S} profile for each galaxy, with initial guesses for each parameter inferred from the values from Source Extractor. For HST-SPT BCGs, the \textsc{Galfit} F110W S\'ersic magnitudes are, on average, $\rm 0.98 \pm 0.4$ mags brighter than those returned by Source Extractor.

We use a single S\'ersic profile throughout, rather than a more flexible double S\'ersic model. The faintest galaxies at high redshift are well-fit by a single component, but will often result in spurious fits when model flexibility is increased, as the additional degrees of freedom are used to fit nearby galaxies or artifacts.  Allowing for additional model flexibility only at low redshift would likely lead to bias (low-$z$ galaxies would have a more robust extrapolation of their light profile to infinity) that mimics evolution in G and $\rm M_{1,2}$. For example, the luminosities of double and single component BCGs vary by about 0.2 mag systematically \citep{Donzelli_2011}, which can drive moderate changes in $\rm \Delta G$ and $\rm \Delta M_{1,2}$. 

An additional feature of \textsc{Galapagos-2} is that multi-band photometry is fit simultaneously using \textsc{Galfitm} \citep{2022A&A...664A..92H}. Utilizing this feature, we fit data in the F110W and F200LP bandpasses at the same time, requiring that the position, ellipticity, orientation, scale radius, and S\'ersic index, n, are constant between the two fits, with only the total luminosity varying. This choice forces a flat color gradient for each galaxy, but ensures that the significantly-shallower F200LP fits do not have the freedom to settle on anomalous values. We utilize the library of PSFs available from STScI\footnote{https://www.stsci.edu/hst/instrumentation/wfc3/data-analysis/psf} for both WFC3-UVIS and WFC3-IR, rotating the model PSF to match our observation before using it as a convolution kernel for the model fitting.

For each field, we visually inspect the best-fit model image after the fitting procedure is complete. In most cases, we make a second pass through \textsc{Galapagos} after manually masking spurious sources (e.g., diffraction spikes), which considerably improves the fitting process. We repeat this procedure until all galaxies are well-modeled, and there are no obvious artifacts being included in the modeling process. 

Ultimately, this process yielded structural parameters for tens of thousands of galaxies. We use the two-band integrated brightnesses from this process as inputs for our color-magnitude plots and for the calculation of G and $\rm M_{1,2}$.

\subsection{SDSS-Planck Low-Redshift Sample}
\label{Subsec:_Planck_Sample}

We use the same selection function applied to HST-SPT to select a low-redshift comparison sample ($\rm z = 0.05 - 0.20$) from the Planck PSZ2 cluster catalog \citep{Planck2016}. While the SPT beam radius of $\sim$ 1$^{\prime}$ is optimized for SZ cluster detection at high redshift, high pass filtering used in the CMB maps to reduce atmospheric contamination limits the telescope's sensitivity to large angular scales, and thus low-z clusters. Planck's $\sim$ 10$^{\prime}$ beam is better suited for cluster detection at low z, given the larger angular size of low-z clusters on the sky. To fairly compare $\rm M_{500}$ between SPT and Planck clusters, we adopt the de-biased marginalized masses for Planck PSZ2 from \citet{Salvati_2022}, given their agreement with \citet{Bocquet_2019} masses for clusters observed by both SPT and Planck. For the corresponding optical data, we select Planck PSZ2 clusters in the Sloan Digital Sky Survey (SDSS) footprint, bar PSZ2 G127.71-69.55, which does not have an obvious optical overdensity of galaxies within 10' of its SZ-center. 

We note that the SDSS-Planck and HST-SPT samples are well-matched in terms of their relative angular resolution. The SDSS pixel scale is 0.4$^{\prime\prime}$, which corresponds to 0.4 - 1.32 kpc over the range of the SDSS-Planck sample ($\rm z\in [0.05,0.2]$). Over the range of HST-SPT ($\rm z \in [0.2,1.0]$), the WFC3-IR pixel scale of 0.13$^{\prime\prime}$ corresponds to 0.43 - 1.15 kpc.

%We adopt the measurement of $\rm (1-b)_{Eddington} = 0.77^{+0.02}_{-0.025}$ from \citet{Bleem_2020}, which is consistent with the literature ().

%which is consistent with several other studies ()

We query the SDSS Galaxy table \citep{2025arXiv250707093S} for all objects within a 10$^{\prime}$ radius around the published PSZ2 SZ center in the g and r bands. Previous studies of SDSS galaxies have used combinations of g-i \citep[e.g.][]{Gavazzi_2010} or u-r \citep[e.g.][]{Baldry_2004} to capture the 4000 \AA ~break. We find, for our population of quiescent galaxies between $\rm z = 0.05 - 0.20$, that g-r produced the tightest observed red sequence and distinguishable blue cloud feature when combined with our red sequence modeling (see Section \ref{subsec: RS modeling}). This limits contamination when making our membership selection via a color-cut, which we use for SDSS-Planck despite the availability of spectroscopic redshifts to best complement HST-SPT. We use extinction-corrected modelMags when making our color-cut, due to their optimization for measuring galaxy colors. Extinction-corrected Petrosian magnitudes, which are optimized for accurate total flux measurements, are used for all luminosity-sensitive methods, including cluster center identification (Section \ref{subsec: core_ID}), making limiting magnitude cuts (Section \ref{subsec: lim_mag_richness}), and calculating G and $\rm M_{1,2}$. We opt for Petrosian magnitudes over cmodelMags to avoid photometric issues related to poor surface brightness fitting in crowded and blended cluster cores. 

We note that Petrosian magnitudes, though relatively unbiased \citep{2002AJ....124.1810S}, display monotonic magnitude deficits for more concentrated galaxies when compared with magnitudes from well-fit S\'ersic profiles ($\sim$ 0.2 mags dimmer for n = 4, for example) \citep{2005AJ....130.1535G}. This leads to a moderate possible decrease in $\rm \Delta G$, but is a relatively random effect that only systemically affects the SDSS-Planck sample. To understand the impact of the different magnitude systems on our analysis, we artificially brighten the SDSS-Planck Petrosian magnitudes by 0.2 mag to mimic S\'ersic magnitudes. The HST-SPT sample is left unchanged and the analysis is repeated. We show in Tables \ref{tab:evolutionary_parameters} and \ref{tab:K-S_test} that the results of this injection test are consistent within $\rm 1\sigma$ to those from our main sample. We also note that our result is robust when only the unaffected HST-SPT sample is considered in Sections \ref{parametric_evolution} and \ref{empirical_evolution}. %\textbf{Additionally, the S\'ersic index, n, does not evolve strongly in elliptical galaxies over $\rm z = 0 - 0.2$} \citep{2013MNRAS.428.1460B}, \textbf{making the effect relatively redshift independent.}

%This leads to a moderate possible decrease of $\rm \Delta G \sim -0.02$ if the flux of the BCG alone is underestimated by 0.2 mag, which is an upper estimate. 

We make a minimal selection of quality cuts based on photometric flags when querying SDSS. We select objects flagged with BINNED1, BINNED2, or BINNED4 detections, but otherwise do not reject objects flagged as SATURATED or with mismatched BLENDED, DEBLENDED, and CHILD flags, as these cuts were shown to preferentially eliminate bright galaxies in some clusters in early phases of the project. We choose to optimize for completeness of member galaxies detection over pristine member photometry given the sensitivity of G and $\rm M_{1,2}$ to the collective luminosity distribution of cluster galaxies. To eliminate foreground stars, we limit SDSS-Planck to clusters at least 20 degrees above the galactic plane. Additionally, we visually inspect each cluster and cut sources corresponding to bright foreground stars and diffraction spikes, along with any galaxies obscured by these features, as we did for the HST-SPT sample.

We removed five clusters from the SDSS-Planck sample at this stage due to photometric issues with their identified BCGs. PSZ2 G033.81+77.18 has a BCG with no photometry in SDSS and PSZ2 G067.52+34.75 has a BCG obscured by a bright foreground star. Additionally, PSZ2 G077.90-26.63, PSZ2 G115.25-72.07, and PSZ2 G217.09+40.15 have BCGs that were treated as point sources when deblended, and are flagged as DEBLENDED\_AS\_PSF. In total, we have sufficient data for 16 galaxy clusters in the SDSS-Planck sample (see Figure \ref{fig:samples}).

%to eliminate galaxies with unreliable photometry or other data quality issues.
%This step is also crucial for eliminating foreground stars, though we additionally limit SDSS-Planck to clusters at least 20 degrees above the galactic plane to limit stellar crowding. We select objects flagged with detections in BINNED1, BINNED2, or BINNED4 photometric flags, while rejecting objects that are de-blended incorrectly by ensuring the BLENDED, DEBLENDED, and CHILD flags match. Finally, we reject objects that are indicated as saturated. This combination provides a good balance between rejecting foreground stars while preserving most luminous core galaxies. We cut three clusters from the sample at this stage due to photometric issues with their BCGs-- the most critical galaxies in our analysis. PSZ2 G033.81+77.18 has a BCG that is flagged as SATURATED by SDSS. Additionally, PSZ2 G115.25-72.07 and PSZ2 G217.09+40.15 have BCGs that were treated as point sources when deblended, and are flagged as DEBLENDED\_AS\_PSF.  

%https://ui.adsabs.harvard.edu/abs/2001AJ....122.2267E/abstract
%These cuts are typically used when selecting for luminous and red galaxies in SDSS. While these quality cuts are used for the 

\section{Measuring the Gini index} 
\label{sec:methods}
%% The "ht!" tells LaTeX to put the figure "here" first, at the "top" next
%% and to override the normal way of calculating a float position.
%% The asterisk after "figure" tells the compiler to span multiple columns
%% if a two column style is selected.

%Should i just call M_F110W here M?
\begin{figure*}[ht!]
\centering
\includegraphics[width=\textwidth]{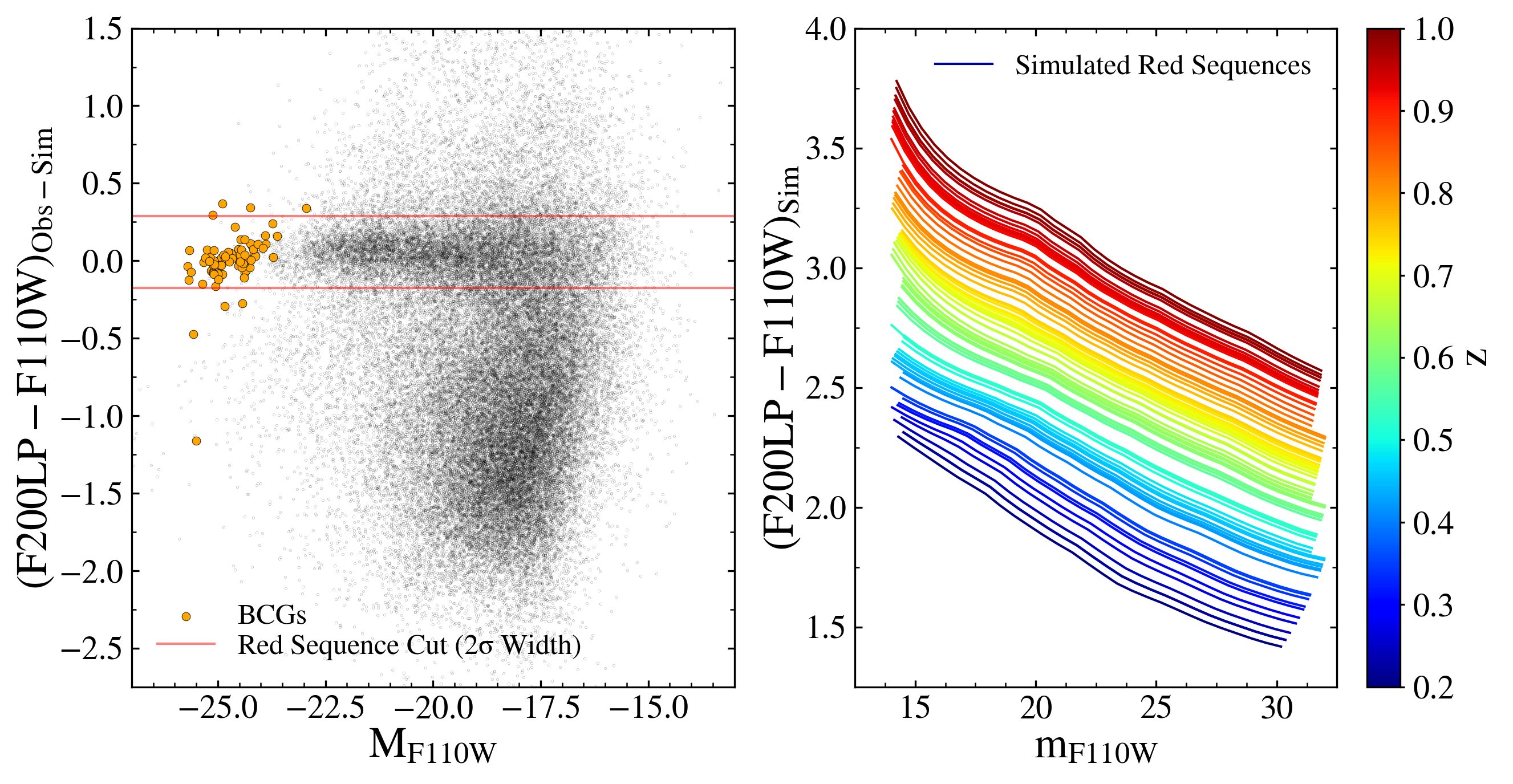}
\caption{The left panel shows the HST-SPT stacked sample plotted in color-offset (observed - simulated red sequence colors) and absolute magnitude space. The right panel shows the FSPS simulated red sequences for the HST-SPT sample. The y-intercepts of the simulated red sequences scales strongly with redshift, while the slope remains relatively constant, fixed by our mass-metallicity relationship (Equation \ref{eq: adjusted MZR}). An example color cut is shown, which uses a red sequence width $\rm width_{RS} = \pm 2\sigma_{RS}$ around the measured sample mean $\rm \mu_{RS}$. Despite our use of broad filters, our red-sequence modeling produces a well-defined red sequence and blue cloud for HST-SPT. 
\label{fig:RS_CMD}}
\end{figure*}

%Summarize cuts up here?
To measure the Gini index of HST-SPT and Planck-SDSS clusters, we need to accomplish the following goals. First, we identify the member galaxies for each cluster. Lacking access to spectroscopic redshift information for individual galaxies, we rely on a red sequence (RS) selection for membership. Next, we identify the positions of the cluster cores about which our measurement will be centered. Finally, we apply cuts on cluster-centric radius and absolute magnitude to produce statistically comparable samples of galaxies across all redshifts. Below, we describe our methods in reference to the HST-SPT sample in detail, but note that we repeat the identical processes for the SDSS-Planck anchor sample, unless explicitly stated otherwise. 

\subsection{Red Sequence Modeling, K-Corrections, and Membership Selection}
\label{subsec: RS modeling}

We use a color cut to constrain the membership of each sample cluster. Cluster galaxies are known to preferentially reside on the red sequence (RS) in color-magnitude space \citep{2000AJ....120.2148G}. When modeled linearly, the intercept of the red sequence scales strongly with redshift, while its slope is fixed by the stellar mass -- stellar metallicity relationship (stellar MZR) \citep{Kodama_Arimoto_1997}. For our purposes, this method of membership selection is reliable without redshift information for individual galaxies. We work with color offsets, which are the difference between the observed F200LP-F110W color and simulated red sequence colors: $\rm (F200LP-F110W)_{Obs-Sim} = (F200LP-F110W)_{Obs}-(F200LP-F110W)_{Sim}$. Positive color offsets correspond to galaxies with $\rm (F200LP-F110W)_{Obs}$ colors redder than the simulated red sequence, while negative color offsets correspond to observed colors bluer than than the simulated red sequence. Color offsets remove redshift dependence from the red sequence feature and allow for red sequence stacking across the entire HST-SPT sample. We find that the stacked red sequence feature is clearly visible even in the broad F110W and F200LP bands (Figure \ref{fig:RS_CMD}). In color offset space, the stacked red sequence feature should flatten if our simulated red sequences sufficiently model the slope of observed red sequences. 

%Small differences between the observed and simulated red sequence offset and slope (likely due to differences in star-formation histories (SFHs)) will drive scatter for individual galaxies, contributing to the width of the red sequence in color-offset space.

We simulate individual cluster red sequences by modeling the colors of typical elliptical galaxies at each cluster redshift. To achieve this aim, we produce simple stellar populations (SSPs) using the Flexible Stellar Population Synthesis library (FSPS) \citep{Conroy_2009_FSPS, Conroy_2010_FSPS,ben_johnson_2024_12447779}. We assume that the old population is formed in a single burst at $\rm z = 5$ with a Salpeter IMF, characteristic of typical SFHs for quiescent cluster galaxies \citep{De_Lucia_2007,Thomas_2010}. The SSPs are then ``observed'' at the cluster redshift in the same bandpasses as our observations. This process is repeated for a broad range of possible galaxy stellar masses and thus stellar metallicity inputs, allowing for the measurement of the red sequence over the dynamic range of the observed sample magnitudes.  

We adopt the linear form of the \citet{Leethochawalit_2018} quiescent-galaxy $\rm z = 0.4$ stellar MZR for metallicity inputs to FSPS to generate a realistic slope for our simulated red sequences. This relationship is consistent with other relationships in the literature over our sample redshift range. When we applied the MZR as published to HST-SPT, our stacked red sequence feature was not flat, indicating that our simulated red sequences were not fully capturing the slope of the observed red sequences.  
To correct this, we empirically adjusted the slope and intercept of the MZR to flatten the stacked red sequence and better fit our observed data. We note that even with these adjustments, our empirical stellar MZR (Equation \ref{eq: adjusted MZR}) is still consistent with other published studies of quiescent galaxies \citep{Gallazzi_2014,Choi_2014,Leethochawalit_2018, Khullar_2022, Vaughan_2022, Bevacqua_2024}. For consistency, we use the adjusted MZR when modeling SDSS-Planck red sequences as well. 
%Here, $\rm M_*$ is a Chabrier mass consistent with \citet{Leethochawalit_2018}.

\begin{equation}
\rm
\langle [Z/H]\rangle = -0.34 + 0.32~log_{10}(\frac{M_*}{10^{10} M_\odot})
\label{eq: adjusted MZR}
\end{equation}

%\noindent{}As an aside, we note that the fact that the slope of the red sequence can be ``flattened'' by varying the MZR implies that we could constrain the slope of the MZR and its evolution using these data as first demontrated by \cite{1998ApJ...501..571G}. We will address this in an upcoming follow-up paper.

Finally, we use our FSPS-modeled SSPs to calculate appropriate K-corrections for the samples, both to account for the reddening of galaxy SEDs within the F110W- and r-bands and their differing bandpasses. We simulate F110W- and r-band magnitudes for our $\rm z_{born} = 5$ Salpeter-IMF SSP across $\rm z \in [0,1]$, calculating the change in apparent magnitude measured as a function of redshift relative to an arbitrary normalization point. These corrections are added in addition to the distance modulus when converting between apparent and absolute magnitudes. We use the resulting rest-frame corrected absolute magnitudes for the remainder of the work, which are now band-agnostic and allow $\rm M_{F110W}$ and $\rm M_{r}$ to be compared directly. 

Stacking the color-magnitude diagram for galaxies after these corrections are applied reveals well-defined red sequence and blue cloud features (Figure \ref{fig:RS_CMD}). We fit a sum of three Gaussians to the distribution of HST-SPT color offsets to capture the red sequence, blue cloud, and background components. 

The galaxy population shown in Figure \ref{fig:RS_CMD} contains cluster galaxies, field galaxies,  and  fore-background contamination. Both the blue cloud and red sequence features are contaminated, to different extent, by interlopers. In Section \ref{G_Contamination} we explore the impact of contamination of the red sequence on our measurements of G. The blue cloud cluster galaxy population is not relevant to the main goals of this work and thus we do not particularly focus on estimating its contamination by interlopers.

The HST-SPT color offset red sequence has a mean color offset of $\rm \mu_{RS} = 0.055$ mag and a width of $\rm \sigma_{RS} = 0.12$ mag. The width of the HST-SPT RS is primarily driven by our use of broad filters and small differences between the model and true SFH of individual galaxies. When making the color-cut for individual HST-SPT clusters, we allow $\rm \mu_{RS}$ to re-center within $\pm 0.25$ mag of its predicted color if a greater overdensity of galaxies exists within $\rm \pm 2\sigma_{RS}$ of the shifted mean. This accounts primarily for uncertainties in cluster redshift, which may be a secondary driver of the HST-SPT RS width, and other secondary systematics likely related to our assumed SFHs breaking down at higher sample redshifts. We did not find it necessary to repeat this re-centering of $\rm \mu_{RS}$ for the SDSS-Planck sample of nearby clusters in the narrow SDSS  g- and r-bands, where modeling the red sequence is more straightforward and redshifts are generally more secure. For the SDSS-Planck sample, we measured  $\rm \mu_{RS} = -0.074$ mag and a width of $\rm \sigma_{RS} = 0.046$ mag.

We apply a color cut to the HST-SPT sample in color offset space, where we treat all galaxies in each HST field as if they are located at the cluster redshift. Galaxies with color errors $\rm < \pm 0.25$ mag and with color offsets that fall within $\rm \pm [2,3] \sigma_{RS}$ of the re-centered $\rm \mu_{RS}$ are taken to be cluster members (see Section \ref{error estimation} for more details about cut ranges). Given our redshift-dependent red sequence modeling, our flat color cut on color-offset is equivalent to making a sloped cut in color-magnitude space, with the slope matching the predicted red sequence slope. The probability of extreme foreground and background contamination of a cluster's membership is low, given the nearly monotonic variance of the RS with z and the narrow fields of view considered ($\rm \sim 2.3^\prime \times 2.1^\prime$ for WFC3). 
%absolute magnitude --
%Only galaxies with the correct F200LP-F110W and $\rm M_{F110W}$ pairs will be located on the red sequence, due to it's nearly monotonic variation with magnitude.

%Color offsets are defined as the difference between the Galapagos-modeled single-sersic profile color and the color of the simulated red sequence at the galaxy's absolute magnitude for the given cluster.

\subsection{BCG Identification}
\label{sub: BCG ID}
%Discuss the case of BCG systems in a little more detail, add snapshots of BCGs
We visually inspect every HST-SPT cluster and identify one BCG. In our high-resolution HST images, we look for the co-occurrence of strong lensing features and a large and bright galaxy or systems of several bright galaxies at the peak of the optical galaxy density. Most of our identified BCGs have colors  that are consistent with their cluster's $\rm 2\sigma_{RS}$ color cut ($\rm 72/76$ for HST-SPT and $\rm 14/16$ for SDSS-Planck) and are selected as members regardless. However, we force the small number of BCGs outside the red sequence to be included as members of their respective clusters in some of our calculations of G and $\rm M_{1,2}$, as described in subsequent sections.

Four of the added BCGs are those with color-offsets bluer than $\rm \mu_{RS}-2\sigma_{RS}$. BCGs can naturally fall outside the red sequence due to ICM-cooling processes or galaxy-galaxy interactions that incite star formation, or contamination from lensing \citep{Cerulo_2019, Chu_2021}. The BCG of SPT-CLJ2250-4808 is the bluest outlier with $\rm (F200LP-F110W)_{Obs-Sim} = -1.16$, but features a clumpy blue morphology characteristic of ongoing star formation. The remaining outlier BCGs with reddened color offsets may suggest small uncertainties in the cluster redshifts. %\textbf{analysis is BCG independent...}

%We find that three of the BCGs with color offsets redder than 0.25 are captured by their individual cluster's color cut with $\rm \mu_{RS}$ re-centering, indicating potential uncertainty with the cluster redshift. However, the other red BCGs, particularly those with color offsets $\rm > 0.5$, seem to be reddened by other physical processes or systematic effects. %We explore the impact of adding Non-RS BCGs to our color-cut selected members in Section \ref{Discussion}.

%The reddest BCG, with a color offset of ~~ from SPT-CLJ ~~ was indicated by galapagos to  
%***Look at what happened with redder-than typical BCGs -- I can expand more upon this section if needed***

% Mike Here -- Hi Mike
\subsection{Cluster Core Identification}
\label{subsec: core_ID}
% Add equation?
We calculate a light-weighted average position to locate each cluster core on the sky, where the density of red-sequence galaxies is the highest. These ``centers-of light'' are analogous to calculating the center-of-mass for galaxy luminosities, and are robust against BCG misidentification. Starting at the peak of the SZ detection, we calculate the F110W luminosity-weighted average position of red-sequence member galaxies within a 2$^{\prime}$ radius circle, assuming that all galaxies that survive the red-sequence cut are located at the cluster redshift. 

The 2$^{\prime}$ radius encompasses most of the WFC3 field of view, and provides a coarse estimate of the center-of-light. We repeat this process iteratively five times, to allow the light-weighted position to be made more accurate through gradient ascent as it algorithmically approaches the peak of the F110W luminosity density. Typically, this coarse-averaging is needed if the center of the SZ detection is significantly offset from the HST pointing. 

We iteratively repeat the centering process within 1$^{\prime}$ and 0.5$^{\prime}$ radii to fine-tune the center-of-light. This process typically places the finalized algorithm center within 0.1$^{\prime}$ of the identified BCG upon inspection, though this can vary in cases of mergers or when clusters have several bright galaxies. The centers-of-light can be calculated without explicit identification of a BCG, and are one position about which the Gini index is calculated. For comparison, we additionally calculate G and $\rm M_{1,2}$ about the BCG coordinates -- another common identifier of the cluster center \citep{2004ApJ...617..879L,Lauer_2014,2012MNRAS.420.2120M,10.1093/mnras/staa3286}. %Two additional steps are needed to center the cluster core more precisely. We calculate a finer center of light within a 1$^{\prime}$ radius around the coarse center of light. This intermediate result is then the starting position for an iterative averaging within a radius of 0.5$^{\prime}$, which is repeated five times. 

We choose a range of physical apertures from $\rm r_{core} = 0.15-0.2R_{500}$ around the identified center-of-light to define the cluster core (see Section \ref{error estimation} for more details about cut ranges). This range is based on the physical field of view of the lowest redshift clusters in our sample, and ensures statistically similar galaxy populations are being selected across the full sample, given the relative universality of the luminosity function with $\rm R_{500}$ \citep{Popesso_2005_conf,Popesso_2005,Moretti_2015,Wen_Han_2015} and the dependence of the quiescent fraction and spatial concentration of galaxies on $\rm R_{500}$ \citep{Goto_2004, Hennig_2017}. %The fraction of $\rm R_{500}$ probed is limited by the field of view (FOV) of the lowest redshift clusters in HST-SPT.

%This radius corresponds to the approximate physical size of the field of view (FOV) of the lowest redshift clusters in HST-SPT.Only member galaxies within the defined cluster core are used in calculating the Gini index.

This process is repeated in full for the SDSS-Planck low-$z$ sample. However, to account for the lower mean redshift and thus larger mean angular size of SDSS-Planck clusters on the sky, we scale the angular radii over which we perform our coarse, intermediate, and fine-tuned averaging by a factor of three. This factor allows averaging to be performed over approximately the same physical areas for HST-SPT and SDSS-Planck clusters.

\subsection{Limiting Magnitude Cuts} \label{subsec: lim_mag_richness}
%The HST F110W- and F200LP-bands are broader than the SDSS g- and r-bands. Thus, to compare the F110W- and r-band magnitudes between the HST-SNAP and Planck-SDSS samples over their full redshift range, we calculate k-corrections. The rest-frame $\rm t_{age} = 5~Gyr$ elliptical galaxy SED from the SWIRE template library \citep{Polletta_2007ApJ...663...81P} is shifted to the observed redshift. A measurement of galaxy brightness is made by summing the template's transmitted normalized spectral flux over the F110W- or r-band throughput, which is then normalized by subtracting the brightness measured in the F110W-band at $\rm z = 0.4$.

We control for the number of galaxies included in the calculation of the Gini index by either imposing a limiting absolute magnitude or a fixed number of galaxies (via a flat, fixed richness cut discussed in Section \ref{max_min_gini}, not to be confused with intrinsic physical richness, $\rm \lambda$) across all redshifts. To determine an appropriate limiting magnitude, we fit a single-component Schechter function to the HST-SPT red sequence. We measure the characteristic magnitude $\rm M^{*} = -20.94 \pm 0.43$ for the $\rm 2\sigma_{RS}$ HST-SPT red sequence. We apply a limiting magnitude cut of $\rm M_{lim} \in \rm M_\star + [1.5,2]$, the lower limit of which is determined by the depth of the $\rm z \sim 1$ HST-SPT cluster observations (see Section \ref{error estimation} for more details about cut ranges). While the observations of lower redshift clusters reach deeper into the luminosity function, this choice of limiting magnitude ensures a statistically fair comparison across all redshifts. %We similarly make a flat richness cut of the $\rm N_{lim} \in [15,25]$ brightest galaxies, fixed by the available richness at $\rm z~\sim 1$. We count up to $\rm N_{lim}$ in cases where $\rm N < N_{lim}$ due to the restrictiveness of previous cuts.

% talk about how Mlim allows for evolution but Nlim does not

%If the richness cut exceeds the number of galaxies in the core after the proceeding cuts, we  Flat richness cut doesn't work as z changes -- different galaxies.

\subsection{Error Estimation}

\label{error estimation}
Typically, when using a bootstrapping procedure to estimate uncertainties on the Gini index or magnitude gap, we would resample within the uncertainties on the absolute magnitudes used to calculate the indicators. However, the errors on our absolute magnitudes, which are $< \rm 5\%$, are not the largest sources of uncertainty on estimates of $\rm G$ and $\rm M_{1,2}$ for each cluster. Rather, the median values of these indicators are far more influenced by our choices of cuts used to define the population of member galaxies that they are applied to. 

We define reasonable ranges through which the membership, size, and limiting magnitude cuts could plausibly vary. Our color cut allows for $\rm width_{RS} \in [\pm 2\sigma_{RS}, \pm 3\sigma_{RS}]$ around each re-centered $\rm \mu_{RS}$, our size cut allows for $\rm r_{core} \in [0.15, 0.20]~R_{500}$, and our magnitude limit allows for $\rm M_{lim} \in M^{*} + [1.5,2]$. We calculate $\rm G$ or $\rm M_{1,2}$ 1000 times for each cluster, allowing each cut quantity to vary randomly within the specified ranges. We then plot the median values and $\rm 68\%$ confidence intervals from the distributions of G and $\rm M_{1,2}$ formed by this modified bootstrapping. 

%This randomization can cause a set of individual cuts applied to a cluster to become restrictive enough so that $\rm G$ or $\rm M_{12}$ cannot be calculated. In these cases, we assume the cluster is such that it contains a BCG and 2nd brightest member just dimmer our limiting magnitude. Thus, we set G = 1, as appropriate for a fossil group, and $\rm M_{1,2} = M_{lim} - M_1$.

%where  Additionally, particularly when non-RS BCGs are not added manually, our centering algorithms can fail. In total, the algorithm failed for ~~~ clusters (~~~) on individual runs, but never beyond a threshold of 1\% of the bootstrapping runs.
%SPT-CLJ2311-4203, SPT-CLJ2146-4846, and SPT-CLJ2148-6116

In addition to allowing cut values to vary, we alternate whether all (non-RS+RS) BCGs or just BCGs within our color cut (RS BCGs) are used in the calculation of G and $\rm M_{1,2}$ (Section \ref{sub: BCG ID})\textbf{,} and whether BCG or algorithm-derived cluster centers are used (Section \ref{subsec: core_ID}). The permutations of these choices make up the columns 1-3 in Figures \ref{fig:G_M12_v_z_systematics}, \ref{fig:N_v_z_systematics}, and \ref{fig:G_M12_N_richness_cut}, and are designed to illustrate the effect of bias from our choice of cuts in our results.

%, and whether we limit depth through richness or magnitude cuts (Section \ref{lim_mag_richness}).

%(see Table \ref{tab:cut_ranges}).

%\begin{deluxetable*}{ccc}[h!]
%\digitalasset
%\tablewidth{\columnwidth}
%\tablecaption{\label{tab:cut_ranges}}
%\tablehead{
%\colhead{Cut Quantity} & \colhead{Minimum} & \colhead{Maximum}
%}
%\startdata
%$\rm width_{RS}$ around $\rm \mu_{RS}$ & $\rm 1\sigma_{RS}$ & $\rm 3\sigma_{RS}$ \\
%$\rm r_{core}$ & $\rm 150~kpc $& $\rm 250~kpc $ \\
%$\rm M_{lim}$ & $\rm M^*_{F110W} + 2$ & $\rm M^*_{F110W} + 4$ \\
%$\rm N$ & $\rm km~s^{-1}$ & FWHM \\
%\enddata
%\tablecomments{}
%\end{deluxetable*}

%\subsection{Maximum and Minimum Gini index}

\section{The Redshift Evolution of the Gini index and Magnitude Gap} \label{sec:gini_evolution}

%\begin{figure*}[h!]
%\centering
%\includegraphics[width=\textwidth]%{G_z_skinny_plot.png}
%\caption{ The Gini index vs. redshift distribution of the HST-SPT and SDSS-Planck samples for three core center and BCG color-cut permutations:  BCG centers with non-RS BCGs (panel 1), light-weighted algorithm centers with non-RS BCGs (panel 2), and light-weighted algorithm centers without non-RS BCGs (panel 3). We plot the best fit line (solid) with 16th and 84th percentile error regions (shaded) and fit intrinsic scatter (dashed) for each panel. The redshift distribution of the Gini index is relatively similar across the cut permutations, apart from a decrease in G for a selection of individual clusters as non-RS BCGs are removed from the calculation in panel 3 and the remaining RS galaxies have a more equal luminosity distribution. The intrinsic scatter is an order of magnitude smaller than that produced by the established indicator, $\rm M_{1,2}$, in Figure \ref{fig:M12_v_z_systematics}. Each fit suggests a weakly decreasing trend in G with z that is consistent with non-evolution within $\rm 2\sigma$ over $\rm z\in [0,1]$. 
%\label{fig:G_v_z_systematics}}
%\end{figure*}

\begin{figure*}[ht!]
\centering
\includegraphics[width=\textwidth]{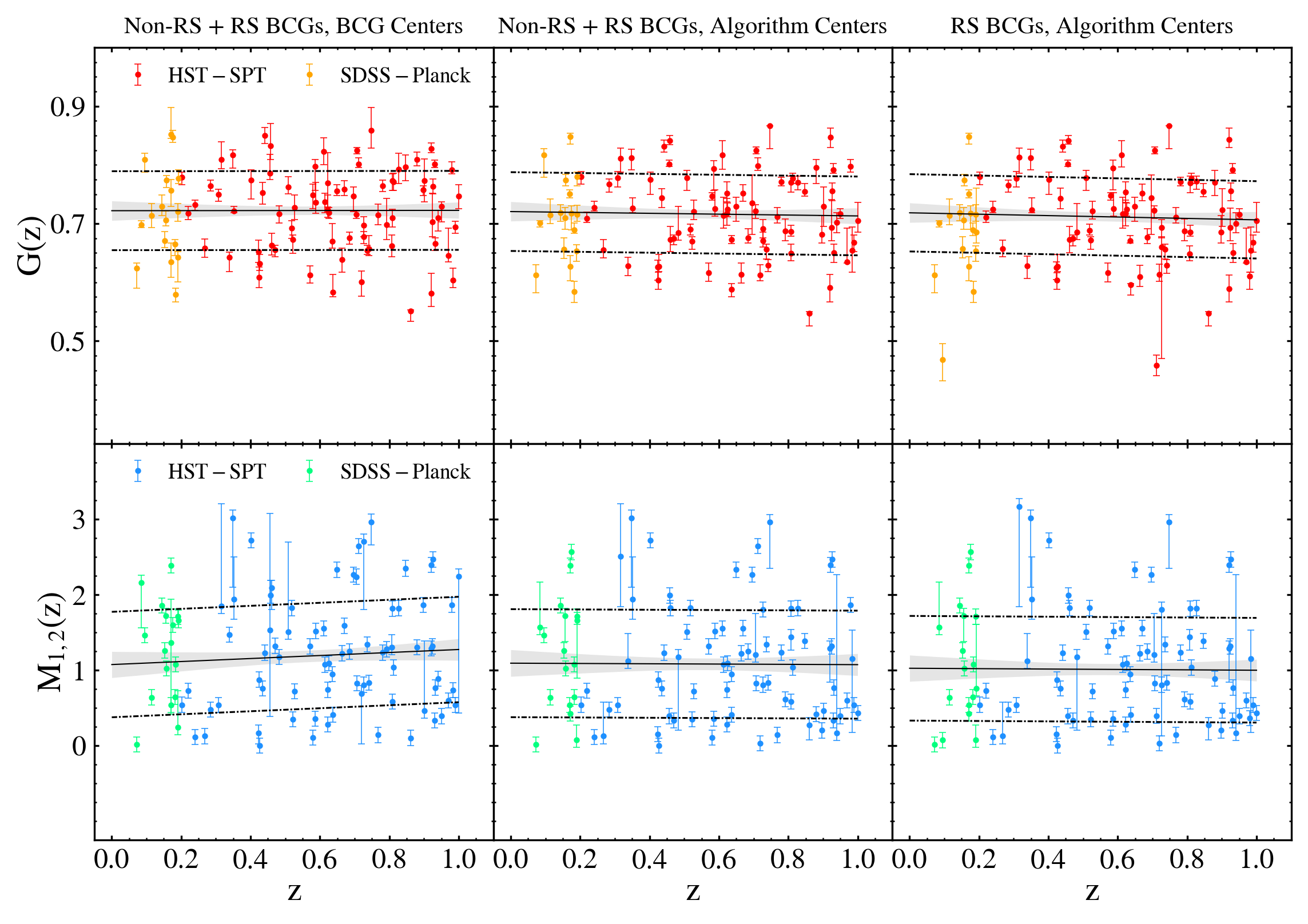}
\caption{ The Gini index (upper) and Magnitude Gap (lower) vs. redshift of the HST-SPT and SDSS-Planck samples for three cluster center types and BCG color-cut permutations:  BCG centers with All BCGs (non-RS+RS) (panel 1), light-weighted algorithm centers with all BCGs (non-RS+RS) (panel 2), and light-weighted algorithm centers with only RS BCGs (panel 3). We plot the best fit line (solid) with 16th and 84th percentile error regions (shaded) and fit intrinsic scatter (dashed) for each distribution. The redshift distribution of the Gini index is relatively similar across the cut permutations, with a weakly-decreasing trend consistent with non-evolution within \textbf{$\rm 1\sigma$}. $\rm M_{1,2}(z)$ features a fit fractional scatter that is \textbf{$\rm \sim 7$} times greater than that of \rm G(z), and is fit with a wider variety of linear profiles that are consistent with non-evolution within $\rm 1\sigma$. Our application of linear regression and the KS-test conclude that G and $\rm M_{1,2}$ are unevolving for $\rm z < 1$, though the $\rm M_{1,2}$ result is noisier.
\label{fig:G_M12_v_z_systematics}}
\end{figure*}
%There is a decrease in G for a selection of individual clusters as non-RS BCGs are removed from the calculation in panel 3 and the remaining RS galaxies have a more equal luminosity distribution.
%This result is similar to that obtained for simulated TNG300 clusters in Figure \ref{fig:simulations} over this z regime.

%My phrasing is pretty awkward throughout this section, so some help is appreciated.

After selecting our populations of core member galaxies, we calculate the Gini index and magnitude gap for each cluster in  the HST-SPT and SDSS-Planck samples. We retrieve three iterations of G and $\rm M_{1,2}$ vs. redshift, based on the variation of center types and included BCGs made in Section \ref{error estimation}. We present two tests of the redshift evolution of G and $\rm M_{1,2}$: one using linear regression and the other using the KS statistic.

\subsection{Parametric Evidence of non-Evolution}
\label{parametric_evolution}

\begin{deluxetable*}{cccccc}[ht!]
%\digitalasset
\tablewidth{\textwidth}
\tablecaption{Parametric Evolution Results \label{tab:evolutionary_parameters}}
\tablehead{\colhead{} &\colhead{}&\colhead{}&\multicolumn{3}{c}{SDSS-Planck + HST-SPT} \\
\colhead{Indicator}& \colhead{All BCGs?}& \colhead{Center Type}& \colhead{m} & \colhead{b} & \colhead{$\rm \sigma$} }
\startdata
G & Yes & BCG & $\rm 0.00\pm 0.03$ & $\rm 0.72\pm 0.01$ & $\rm 0.07\pm 0.01$  \\
G & Yes & Algorithm & $\rm -0.01\pm0.03$ & $\rm 0.72\pm 0.01$ & $\rm 0.07\pm 0.01$  \\
G & No & Algorithm &  $\rm -0.01\pm 0.03$ & $\rm 0.71\pm 0.01$ & $\rm 0.07\pm 0.01$  \\
$\rm M_{1,2}$ & Yes & BCG &  $\rm 0.2\pm 0.3$ & $\rm 1.19\pm0.08$ & $\rm 0.70\pm 0.06$ \\
$\rm M_{1,2}$ & Yes & Algorithm & $\rm 0.02\pm 0.3$ & $\rm 1.08\pm 0.08$ & $\rm 0.72\pm 0.07$
\\
$\rm M_{1,2}$ & No & Algorithm & $\rm -0.03\pm 0.3$ & $\rm 1.01\pm0.08$ & $\rm 0.69\pm 0.06$  \\ \tableline
\colhead{}&\colhead{}&\colhead{}&\multicolumn{3}{c}{HST-SPT} \\
\colhead{Indicator}& \colhead{All BCGs?}& \colhead{Center Type}& \colhead{m} & \colhead{b} & \colhead{$\rm \sigma$} \\\tableline
G & Yes & BCG & $\rm -0.01\pm 0.04$ & $\rm 0.72\pm 0.01$ & $\rm 0.07\pm 0.01$  \\
G & Yes & Algorithm & $\rm -0.03\pm0.04$ & $\rm 0.72\pm 0.01$ & $\rm 0.07\pm 0.01$  \\
G & No & Algorithm &  $\rm -0.05\pm 0.04$ & $\rm 0.71\pm 0.01$ & $\rm 0.07\pm 0.01$  \\
$\rm M_{1,2}$ & Yes & BCG &  $\rm 0.53\pm 0.4$ & $\rm 1.17\pm0.09$ & $\rm 0.71\pm 0.07$ \\
$\rm M_{1,2}$ & Yes & Algorithm & $\rm 0.2\pm 0.4$ & $\rm 1.06\pm 0.09$ & $\rm 0.71\pm 0.07$
\\
$\rm M_{1,2}$ & No & Algorithm & $\rm 0.06\pm 0.4$ & $\rm 0.99\pm0.08$ & $\rm 0.68\pm 0.07$  \\
\tableline
\colhead{}&\colhead{}&\colhead{}&\multicolumn{3}{c}{SDSS-Planck Injection Test} \\
\colhead{Indicator}& \colhead{All BCGs?}& \colhead{Center Type}& \colhead{m} & \colhead{b} & \colhead{$\rm \sigma$} \\\tableline
G & Yes & BCG & $\rm 0.00\pm 0.03$ & $\rm 0.72\pm 0.01$ & $\rm 0.07\pm 0.01$  \\
G & Yes & Algorithm & $\rm -0.01\pm0.03$ & $\rm 0.72\pm 0.01$ & $\rm 0.07\pm 0.01$  \\
G & No & Algorithm &  $\rm -0.02\pm 0.03$ & $\rm 0.71\pm 0.01$ & $\rm 0.07\pm 0.01$  \\
$\rm M_{1,2}$ & Yes & BCG &  $\rm 0.2\pm 0.3$ & $\rm 1.18\pm0.08$ & $\rm 0.69\pm 0.06$ \\
$\rm M_{1,2}$ & Yes & Algorithm & $\rm -0.02\pm 0.3$ & $\rm 1.08\pm 0.08$ & $\rm 0.72\pm 0.07$
\\
$\rm M_{1,2}$ & No & Algorithm & $\rm -0.03\pm 0.3$ & $\rm 1.01\pm0.08$ & $\rm 0.69\pm 0.06$  \\ 
\enddata
\tablecomments{Summary of parameters fit to the G and $\rm M_{1,2}$ (``Indicator") data in Section \ref{parametric_evolution} for the full SDSS-Planck + HST-SPT sample, the HST-SPT sample, and the SDSS-Planck (with magnitude offset) + HST-SPT sample (``SDSS-Planck Injection Test") discussed in Section \ref{Subsec:_Planck_Sample}. Columns 2-3 specify the choice of BCG or light-weighted (``Algorithm") cluster centers (``Center Type") and if non-RS BCGs are included (``Yes"/``No")  in the calculation of G and $\rm M_{1,2}$ (``All BCGs?"). The fit slope (m), intercept (b) and intrinsic scatter ($\sigma$) are given in columns 4-6 with $\rm 1\sigma$ normally-distributed errors.\\}
\end{deluxetable*}
%Summary of parameters fit  plotted in Figure \ref{fig:G_M12_v_z_systematics} and referenced throughout Section \ref{parametric_evolution}. Columns 2-4 describe the permuted choices of including non-RS BCGs (``Non-RS BCGs"), centering the core on the algorithm derived centers or BCG centers (``Center Type"), and limiting depth cut type (``Depth Cut") used in each indicator calculation in Column 1 (``Indicator").     

 %, where Column 1 ("Result") matches each set of parameters to their corresponding panel. Columns 2-4 describe the permuted choices of including non-RS BCGs ("Non-RS BCGs"), centering the core on the algorithm derived centers or BCG centers ("Center Type"), and the type of limiting depth cut applied to the data ("Depth Cut").

% Mike's note: the selection isn't quite linear in M,z space...or shouldn't be at least

In Figure \ref{fig:G_M12_v_z_systematics} we show the primary results from this work. At a glance, a few things are obvious from this figure. First, there is no clear evolution in either $\rm G$ or $\rm M_{1,2}$ over the redshift range spanned by this sample. Second, the Gini index produces considerably less fractional scatter than the magnitude gap, as we predicted above. And, finally, our choices of cluster center and whether or not to include non-RS BCGs do not significantly impact the finding of non-evolution, though they can drive significant scatter in $\rm G$ or $\rm M_{1,2}$ for individual clusters (as discussed below). To make each of these assertions more quantitative, we turn to parametric modeling of the evolution.

%We apply a model-dependent test of evolution to the redshift distributions of G and $\rm M_{1,2}$. We model G(z) and $\rm M_{1,2}$(z) as having a linear evolution with gaussian scatter. This choice is not motivated by theory, but rather as a means of quantifying any obvious evolution. The null evolution case is possible if a flat line is consistent with our data. 

% Figures \ref{fig:G_v_z_systematics} and \ref{fig:M12_v_z_systematics}  displays the redhsGini index and Magnitude gap  vs. redshift distributions for the HST-SPT and SDSS-Planck samples.We show individual values of G for each sample cluster, as well as median values of G in individual redshift bins. These median values of G suggest a weak to moderate decreasing trend with redshift. 

We fit a linear model of the form $\rm f(z) = m(z-\bar{z})+b$ with intrinsic scatter ($\rm \sigma$) to G(z) and $\rm M_{1,2}(z)$ using the LtsFit package \citep{Cappellari_2013}. %\begin{equation}
%    \rm f(z) = m(z-\bar{z})+b
%    \label{eq:linear_model_w_pivot}
%\end{equation}
This method allows for more robust linear regression in cases where scatter, such as that caused by variation in cluster assembly histories, is intrinsic to the distribution. We fit about a pivot of the full sample mean redshift $\rm \bar{z} = 0.56$ and reject $\rm 3\sigma$ outliers with the FAST Least Trimmed Squares technique \citep{FAST_LTS}. When fitting only the HST-SPT sample, we use $\rm \bar{z}_{HST-SPT} = 0.65$. The resulting fit slope, intercept, and intrinsic scatter parameters with $\rm 1\sigma$ normally-distributed errors are reported in Table \ref{tab:evolutionary_parameters}. These fits, and their uncertainties, are shown in Figure \ref{fig:G_M12_v_z_systematics}. %The shaded error regions are formed by randomly sampling within the $\rm 1\sigma$ fit parameter errors for 10,000 iterations and plotting the 16th and 84th percentile confidence intervals around the median fit at each z. 
As expected when fitting about $\rm \bar{z}$, the error regions are largest at the highest and lowest sample redshifts and minimized at the pivot redshift.%, where fit variation propagates to be the largest.  

As an indicator, the Gini index produces far less intrinsic scatter across the full sample than $\rm M_{1,2}$. When applied to HST-SPT and SDSS-Planck, we find median fit values of $\rm \sigma_G = 0.07$ ($\rm \sigma_G/\bar{G} = 10$\%) and $\rm \sigma_{M_{1,2}} = 0.7$ mag ($\rm \sigma_{M_{1,2}}/\bar{M_{1,2}} = 65$\%). This is consistent with our understanding of the Gini index as an indicator that is less sensitive to cluster dynamical state and individual galaxy indexing, and more sensitive to the dominance of massive galaxies over $\rm M^{*}$.

The sensitivity of $\rm M_{1,2}$ to the definition of the BCG and 2nd-brightest member galaxy increases its sensitivity to choices that can redefine them for individual clusters, such as the inclusion of non-RS BCGs, and location and size of the cluster core. $\rm M_{1,2}$ thus produces larger and more variant error regions on individual outlier points than G across the panels of Figure \ref{fig:G_M12_v_z_systematics}. 

%, and brightness of galaxies included in the calculation

The response of G between panels is more predictable, such as the decrease in G with the removal of non-RS BCGs for a few individual clusters between the 2nd and 3rd panel of Figure \ref{fig:G_M12_v_z_systematics}, due to the more equal resulting luminosity distribution of the remaining RS galaxies when non-RS BCGs are not included in the calculation of G. G appears to be less sensitive to the choice of cluster center than $\rm M_{1,2}$, likely due to its use of the entire galaxy population in the core.  Regardless, the parameters fit to the distributions of G and $\rm M_{1,2}$ are relatively insensitive to choices made in the analysis, and are consistent across the panels of Figure \ref{fig:G_M12_v_z_systematics} within $\rm 1\sigma$.

%The sensitivity of $\rm M_{1,2}$ to the definition of the BCG and 2nd-brightest member galaxy increases its sensitivity to choices that can redefine them for individual clusters, such as the inclusion of non-RS BCGs, location and size of the cluster core, and brightness of galaxies included in the calculation. $\rm M_{1,2}$ thus produces larger and more variant error regions on individual points than G across the panels of Figure \ref{fig:G_M12_v_z_systematics}. 

%The response of G between panels is more predictable, such as the decrease in G from the removal of non-RS BCGs for a few individual clusters between the 2nd and 3rd panel of Figure \ref{fig:G_M12_v_z_systematics}, due to the more equal resulting luminosity distribution of the remaining RS galaxies. Thus, we understand the larger intrinsic scatter and variation in individual points to be a driver of the slightly larger variation in the parameters fit to the $\rm M_{1,2}$ distributions across cut permutations. In contrast, the parameters fit to the distributions of G are relatively insensitive to choices made in the analysis.

The results of this fitting procedure are consistent with the conclusion that G and $ \rm M_{1,2}$ do not evolve between $\rm z \in [0,1]$. The slopes fit to G(z) and $\rm M_{1,2}(z)$ in Table \ref{tab:evolutionary_parameters} are consistent with non-evolution within $\rm 1 \sigma$. This is further highlighted by the fact that the $\rm 1 \sigma$ error region (grey shading) on the mean fits is consistent with a horizontal line in Figure \ref{fig:G_M12_v_z_systematics}, further supporting the null evolution hypothesis for G and $\rm M_{1,2}$.

\subsection{Empirical Evidence of non-Evolution}
\label{empirical_evolution}
\begin{deluxetable*}{ccccccccccccc}[ht!]
%\digitalasset
\tablewidth{\columnwidth}
\tablecaption{Empirical Evolution Results \label{tab:K-S_test}}
\tablehead{\colhead{}&\colhead{}&\twocolhead{Full Sample} & \twocolhead{HST-SPT} & \twocolhead{High-Low z} & \twocolhead{Injection Test}\\
\colhead{All BCGs?}& \colhead{Center Type} & \colhead{$\rm p_{G}$} & \colhead{$\rm p_{M_{1,2}}$} & \colhead{$\rm p_{G}$} & \colhead{$\rm p_{M_{1,2}}$}&\colhead{$\rm p_{G}$} & \colhead{$\rm p_{M_{1,2}}$} &\colhead{$\rm p_{G}$} & \colhead{$\rm p_{M_{1,2}}$}
}
\startdata
Yes & BCG &  0.99 & 0.81 & 0.99 & 0.81 & 0.99 & 0.90 & 1.0 & 0.81 \\
Yes & Algorithm & 0.89 & 0.71 & 0.69 & 0.80 & 0.69 & 0.90 & 0.89 & 0.71 \\
No & Algorithm & 0.81 & 0.59 & 0.32 & 0.57 & 0.90 & 0.67 & 0.81 & 0.59\\
\enddata
\tablecomments{Summary of p-values returned from the application of the KS-test to G ($\rm p_{G}$) and $\rm M_{1,2}$ ($\rm p_{M_{1,2}}$) data in Section \ref{empirical_evolution}. Columns 1-2 specify the choice of BCG or light-weighted (``Algorithm") cluster centers (``Center Type") and if non-RS BCGs are included (``Yes"/``No")  in the calculation of G and $\rm M_{1,2}$ (``All BCGs?"). The KS-test is applied to the distributions of four sub-samples: ``Full Sample" (Column 3) compares the combined HST-SPT and SDSS-Planck sample split across $\rm \bar{z} = 0.56$, ``HST-SPT" (Column 4) compares the HST-SPT sample split across $\rm \bar{z} = 0.56$,``High-Low z" (Column 5) compares the lowest ($\rm z < 0.32 $) and highest ($\rm z > 0.79 $)  z-quartiles of the combined HST-SPT and SDSS-Planck sample, and ``Injection Test" (Column 6) compares the combined HST-SPT and SDSS-Planck sample (with magnitude offset, discussed in Section \ref{Subsec:_Planck_Sample}) split across $\rm \bar{z} = 0.56$. 
}
\end{deluxetable*}

As an independent test of the redshift evolution of G and $\rm M_{1,2}$, we apply the two sample Kolmogorov–Smirnov (KS) test \citep{Smirnoff_1939,Hodges_1958} to redshift-binned G and $\rm M_{1,2}$ data. In our evolution-focused context, the rejection of the null hypothesis ($\rm p < 0.05$) implies the chosen indicator evolves with redshift, as the underlying distribution driving the data changes with redshift. Unlike the parametric test of evolution, where we model G(z) and $\rm M_{1,2}(z)$ with Gaussian scatter about a linear functional form, the KS test is model independent. 

We partition HST-SPT and Planck-SDSS into three sets of sub-samples listed in Table \ref{tab:K-S_test} to test for evolution. The ``Full Sample'' comparison involves splitting the sample in half about $\rm \bar{z} = 0.56$, while ``HST-SPT" divides and compares only the HST-SPT sample (excluding SDSS-Planck data) about $\rm \bar{z}$. ``High-Low z" compares the lowest z quartile ($\rm z < 0.32$) of the combined HST-SPT and Planck-SDSS sample to the highest z quartile ($\rm z > 0.79$). Finally, we apply the KS test to the combined HST-SPT and SDSS-Planck sample with an added magnitude offset (``Injection Test", see Section \ref{Subsec:_Planck_Sample}), split across $\rm \bar{z} = 0.56$. When applying the KS test, we find that no significant evolution in G or $\rm M_{1,2}$ is detected across all sub-samples. %This implies that there is no significant evolution in the distribution of the brightest galaxies in the cores of clusters over $\rm z\in [0,1]$.

%When applied to $\rm M_{1,2}$ data, for both low-z subsamples, the null hypothesis is never rejected. However, this is not the case for the Gini index. We recover $\rm p < 0.05$ for runs A, B, C, and F when testing the HST-SPT sample and for runs B, C, and F when testing the HST-SPT and SDSS-Planck samples (see Table \ref{tab:K-S_test}). We can not reject the null hypothesis for runs D and E for either low-z subsample, where non-RS BCGs are added to the sample and a richness cut is made to limit depth. We suspect that our richness cut may introduce unintended biases that may obscure evidence of redshift evolution (discussed further in Section \ref{systematics}). Regardless, the KS test further supports the conclusion that the Gini index is an indicator sensitive to structural growth. %indicator sensitive to evolution, evolutionary snesitive indicator

\subsection{Richness Evolution \& Systematics}
\label{systematics}

 \begin{figure*}[ht!]
\centering
\includegraphics[width=\textwidth]{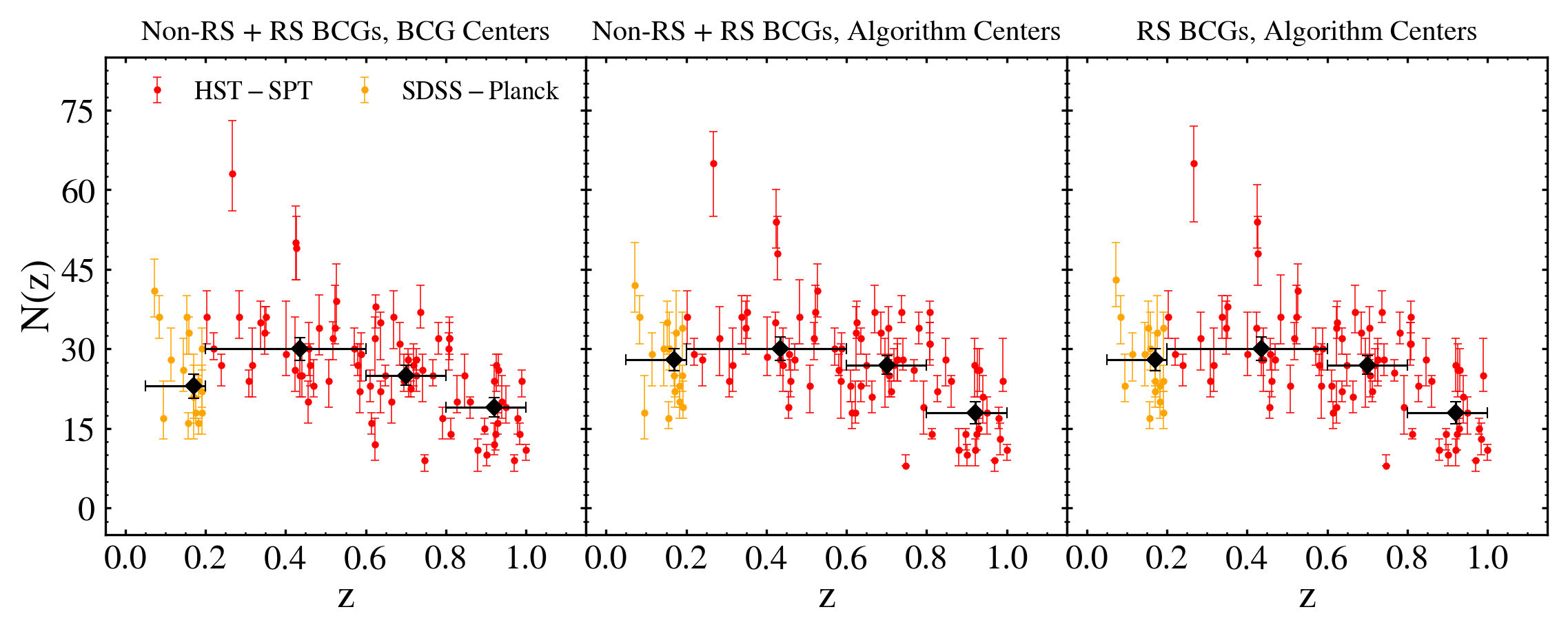}
\caption{Cluster richness (N) vs. redshift for HST-SPT and SDSS-Planck. We plot the median richness in the redshift bins $\rm z \in [0.05,0.2],[0.2,0.6],[0.6,0.8],[0.8,1.0]$ in black, with standard errors on the median approximated by $\rm SE = \frac{\sigma}{\sqrt{N}}$. Horizontal error bars are used to represent the width of the redshift bin where the median is calculated. The decrease in N appears to be driven by proportionality with our decreasing $\rm M_{500}$ progenitor selection and the decreasing red galaxy fraction with cluster redshift (Butcher-Oemler effect), which is most pronounced for $\rm z \in [0.8, 1.0]$. The difference in mean N in the lowest-z bins is driven by data availability and suggests the HST-SPT and SDSS-Planck samples are well-matched. While the declining RS-galaxy fraction causes N to approach values in the $\rm z \in [0.8, 1.0]$ bin that drive G to decrease slightly (see Figure \ref{fig:G_v_M_lim_v_N}), this does not drive significant evolution of G in Figure 
\ref{fig:G_M12_v_z_systematics}.
\label{fig:N_v_z_systematics}}
\end{figure*}

Several possible sources of systematic errors exist within the process of obtaining our results. One possible source is our combination of different cluster samples with different observations. Additionally, the membership, size, and luminosity cuts made on the cluster cores can impact the galaxy populations and number of member galaxies probed. We thus examine the distribution of number of member galaxies, N, used in our calculations of G and $\rm M_{1,2}$ for evidence of non-physical evolution driven by underlying systematics. 

Generally, N is expected to scale with $\rm M_{500}$ \citep{10.1111/j.1365-2966.2010.16406.x, Andreon_Berg_2012,Chiu_2022}.
In the BCG-centered cores, the median N in the redshift bins 
$\rm z \in [0.2,0.6]$ and $\rm z \in [0.8,1.0]$ decreases by a factor of 1.6  from $\rm N = 30 \pm 2.1$ to $\rm N = 19 \pm 1.8$, similar to HST-SPT's selected factor of 1.4 decrease in $\rm M_{500}$ between $\rm z \in [0.4, 0.9]$. Given that we select for membership via a color-cut on the red sequence, this effect is coupled with the Butcher–Oemler effect \citep{Butcher_Oemler_1984}: a decreasing fraction of red sequence galaxies in rich cluster cores with $\rm z \gtrsim 0.2$ \citep[e.g.][]{2019ApJ...876...40P}. We hypothesize that the Butcher–Oemler effect leads to the slightly enhanced decline in richness over that expected from the selection function at high-z. This effect is most prominent between $\rm z \in [0.8, 1.0]$, where the red sequence is the sparsest.

This effect could also be caused, in part, by increased sensitivity in our membership selection to our assumed SFHs in Section \ref{subsec: RS modeling}. At higher redshifts, where $\rm z_{cluster}$ is closer to $\rm z_{born}$, the younger ages of the simulated SSPs used in our red sequence modeling may cause more variation between the observed and simulated red sequences, producing a sparser stacked red sequence feature as these differences drive scatter. We note that the decline in richness for the majority of HST-SPT at $\rm z \sim 1$ does not appear to be driven by our HST observation depth, which caused a discontinuous decrease to $\rm N<10$ for all clusters $\rm z > 1$. 

%Generally, N is expected to scale proportionally with the decrease in $\rm M_{500}$ chosen by our selection function \citep{10.1111/j.1365-2966.2010.16406.x} (citations), most evident in Figure \ref{fig:N_v_z_systematics} at $\rm z > 0.6$ $\rm z < 1$ does not appear to be predominantly driven by incompleteness caused by our HST observation depth, which caused a discontinous decrease in N for $\rm z > 1$ unlike the continous, sloped trend present in Figure \ref{fig:N_v_z_systematics}.  

The redshift distribution of cluster richness, N, in Figure \ref{fig:N_v_z_systematics} suggests the SDSS-Planck and HST-SPT samples are well-matched. The median N falls slightly for the SDSS-Planck sample from the lowest-z HST-SPT value, and scatter towards lower N increases when comparing SDSS-Planck to low-z HST-SPT clusters. However, this appears to be an effect of the $\rm M_{500}$ distribution within our selection function, rather than an effect related to our K-corrections or $\rm M_{500}$ de-biasing between the SZ surveys. In Figure \ref{fig:samples}, SDSS-Planck clusters are more concentrated near the lower $\rm M_{500}$ threshold than HST-SPT clusters within $\rm z \in [0.2,0.4]$, accounting for the lower N reached by SDSS-Planck in Figure \ref{fig:N_v_z_systematics}. We attempted to incorporate more massive Planck clusters in the sample near the upper $\rm M_{500}$ threshold, but found no overlap within SDSS or DES for this mass range. 

The evolution of richness across our sample appears to be driven by understood phenomena. On average, over the full sample, N is above the threshold at which G decreases slightly with richness (Figure \ref{fig:G_v_M_lim_v_N}), limiting the ability of richness to drive evolution in G or its scatter. In the $\rm z \in [0.8, 1.0]$ bin, where N approaches the sensitivity threshold for G due to the lower fraction of RS-galaxies in cluster cores, the expected decrease in G ($\rm \Delta G < 0.05$ from a drop in $\rm N = 10 \rightarrow 3$ in Figure \ref{fig:G_v_M_lim_v_N}) from this effect does not produce significant non-physical evolution in our result. That is, a lower number of galaxies in the highest-redshift bin ought to lead to stronger evolution in G, which we do not observe.
Thus, we expect these results to be robust to any evolution in the red sequence richness over our sample. 

\subsection{Robustness of Results Against Membership Contamination}
\label{G_Contamination}

\begin{figure*}[ht!]
\centering
\includegraphics[width=1.5\columnwidth]{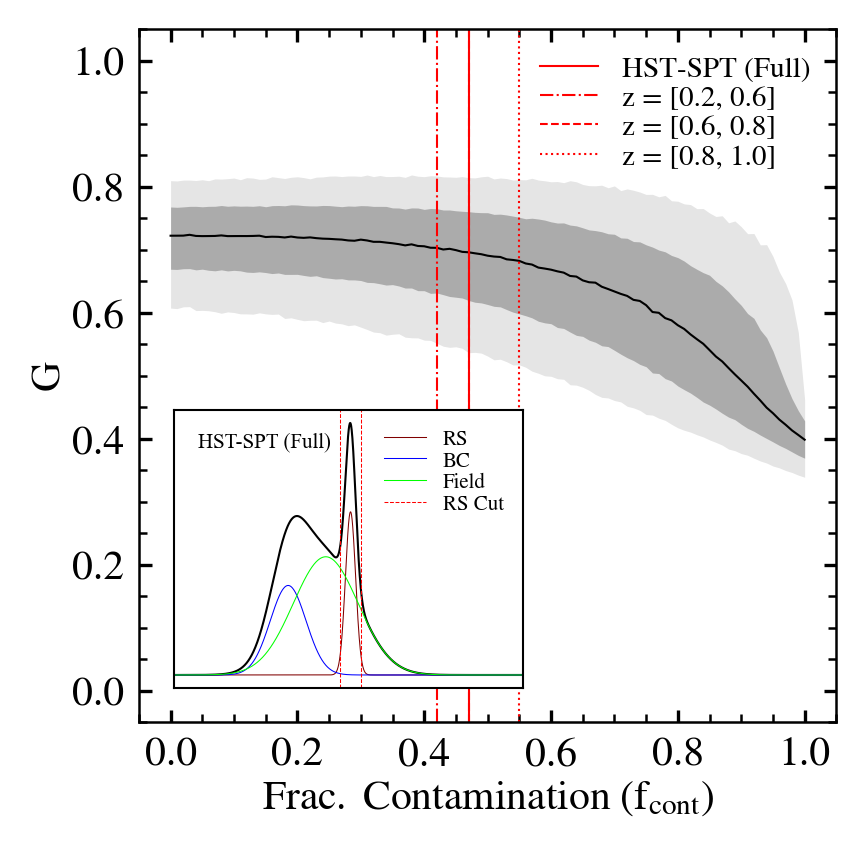}
\caption{Gini index vs. Fraction of Contamination ($\rm f_{cont}$), where G (Median -- black line, shading -- 68\% and 95\% C.I.s after 10,000 bootstrapping iterations) is calculated with its $\rm 1\sigma$ and $\rm 2\sigma$ errors for samples of $\rm N = 100$ galaxies drawn from a ``cluster" distribution of bright, red sequence core galaxies from HST-SPT and Planck-SDSS and a ``field" distribution given by the \citet{Blanton_2003} luminosity function. Here, $\rm f_{cont} = 0$ corresponds to the value of G calculated for a purely  ``cluster"-drawn sample, and $\rm f_{cont} = 1$ corresponds to the value of G calculated for a purely ``field"-drawn sample. The inset demonstrates the three-component Gaussian fitting of the color offset distribution used to estimate $\rm f_{cont}$ for the full HST-SPT sample, as well as HST-SPT split across three z-bins: $\rm z \in [0.2,0.6], [0.6,0.8], [0.8,1.0]$. While our measured $\rm f_{cont}$ is relatively high and increases with redshift, G is robust at the levels of contamination estimated in our cluster membership selection.\label{fig:G_contamination}}
\end{figure*}

We discuss in Section \ref{subsec: RS modeling} that our membership selection is not optimized to eliminate contamination from interlopers. For simplicity, we treat foreground and background interloper galaxies, as well as galaxies outside of clusters at $\rm z_{Cluster}$, as field galaxies. As we show in Figure \ref{fig:G_v_field_discussion}, the populations of bright galaxies in the field and cluster cores produce significantly different values of G. Additionally, as discussed in Section \ref{systematics}, there is evidence that our stacked red sequence feature becomes sparser at higher redshifts, particularly due to the inherent decreased fraction of red sequence galaxies at high-z, and the possible breakdown of assumptions folded into our membership selection. We explore the impact of field contamination on our membership selection further, given the ability of field galaxies to significantly decrease G, particularly at high-z where the red sequence signal to noise is weaker. 

To quantify the impact of field contamination on G, we draw samples of galaxies from two distinct field and cluster distributions. We form the cluster distribution by combining K-corrected absolute magnitudes from SDSS-Planck and HST-SPT galaxies that survive the most stringent color ($\rm width_{RS} = 2\sigma_{RS}$), median size ($\rm r_{core} =  0.175~R_{500}$) and magnitude ($\rm M_{lim} = M^{*} + 1.75$) cuts implemented in Section \ref{error estimation}. 
%\textbf{We estimate the probability density function of this data with kernel density estimation with a guassian kernel of bandwidth 0.5 mag, and draw ``cluster" samples from this distribution. 
``Field" samples are drawn from the \citet{Blanton_2003} luminosity function. We draw samples of varying percentages of N = 100 field and cluster galaxies and calculate G for each, forming the black line in Figure \ref{fig:G_contamination}. 

We then estimate the fraction of field and blue cloud galaxies selected by our red sequence cut. We fit a three-component \textbf{Gaussian} profile to the distribution of color offsets for the full HST-SPT sample, as well as the $\rm z \in [0.2,0.6], [0.6,0.8],~and~[0.8,1.0]$ redshift bins separately. We show how the distribution of color offsets for the full HST-SPT sample was fit for the red sequence, blue cloud, and field contamination in the inset panel of Figure \ref{fig:G_contamination}. The contamination fraction is given by the ratio of the areas under the field and blue cloud distributions to the area under the total color-offset distribution within the $\rm \mu_{RS} \pm2\sigma_{RS}$ red sequence cut. We assume the contamination fraction will be highest for the HST-SPT sample, given our use of broad bands and the much higher cluster redshifts spanned. Thus, we do not repeat this process for our low-z SDSS-Planck anchor sample.

We find that the contamination fraction is relatively high, with a value of $\rm f_{cont} = 0.47$ for the full sample. The contamination fraction increases with redshift ($\rm z \in [0.2,0.6]: f_{cont} = 0.42, $$\rm ~z \in [0.6,0.8]: f_{cont} = 0.47,$ $\rm ~z\in [0.8,1.0]: f_{cont} = 0.55$), which we expect as the red sequence feature becomes more sparse at high-z. However, the high contamination fraction for HST-SPT does not significantly impact G, with only a $\rm 6\%$ decrease in G when compared to the non-contaminated case in the most affected bin ($\rm z\in [0.8,1.0]$), which falls within the $\rm \sigma_G/\bar{G} = 10\%$ fractional scatter fit in Section \ref{parametric_evolution}. While our method of membership selection allows for high levels of possible \xout{field} contamination, we show that our calculation of G is relatively robust against this effect. This is likely caused by the relatively lower $\rm M^{*}$ of the field -- most interloper galaxies sampled are likely similar in brightness to low-mass satellites, which do not significantly drive G. 

\section{Discussion} 
\label{Discussion}

\subsection{Implications of a Non-evolving Gini index}

\begin{figure*}[ht!]
\centering
\includegraphics[width=1.5\columnwidth]{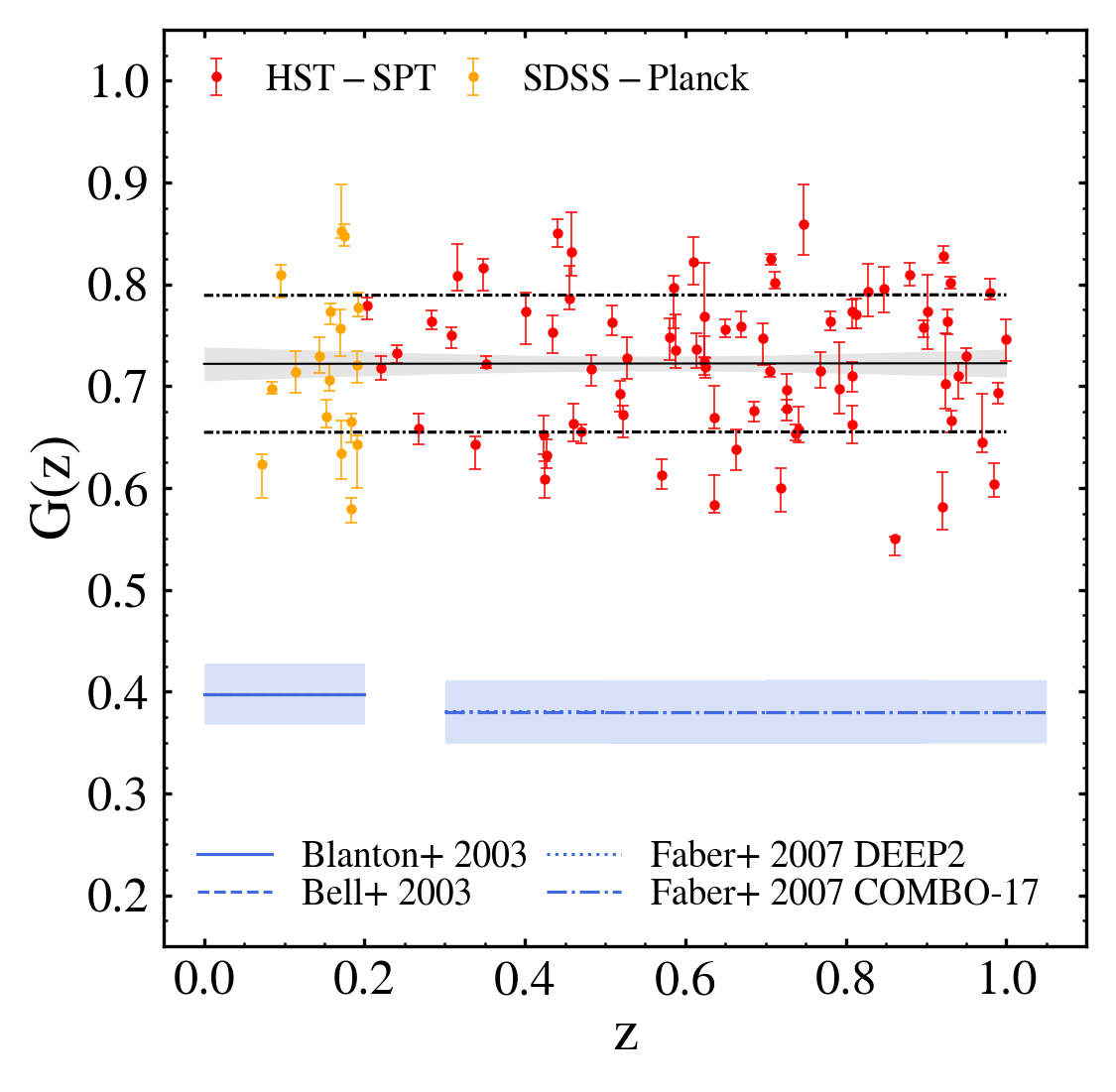}
\caption{ The Gini index of cluster cores vs. the galaxy field. We reproduce panel 1 of Figure \ref{fig:G_M12_v_z_systematics}, where G is calculated for red sequence galaxies in BCG-centered cores of the HST-SPT and SDSS-Planck samples with all BCGs (non-RS+RS) included. We additionally calculate the median Gini index (blue lines) with 16th and 84th percentile C.I.s for several published luminosity functions (\citet{Blanton_2003} r-band, \citet{Bell_2003} r-band, and \citet{Faber_2007} for all galaxies), representing the full galaxy field.
We sample each luminosity function above the median cut applied to our observed data ($\rm M_{lim} = M^{*}+1.75$), and find that $\rm G_{field}$ and thus the shape of the full luminosity function does not evolve strongly over $\rm z \in [0,1]$. $\rm G > G_{field}$ and is non-evolving for the cluster cores across $\rm z \in [0,1]$, implying that BCGs are dominant and growing with the red sequence for $\rm z \le 1$. 
\label{fig:G_v_field_discussion}}
\end{figure*}

%BCG has established early

Our parametric and empirical tests of evolution in Section \ref{sec:gini_evolution} suggest that the Gini index and magnitude gap are non-evolving in the $\rm z \in [0,1]$ regime. While transient, dynamical phenomena can drive significant scatter in $\rm M_{1,2}$, the Gini index is more sensitive to massive galaxy assembly and the dominance of the BCG over the general cluster population.

Our primary results have two major implications. One is that the luminosity distribution of red sequence galaxies in cluster cores is more skewed towards higher luminosities than the full luminosity function of the field. In Figure \ref{fig:G_v_field_discussion}, we compare the Gini index for our BCG-centered cores (panel 1 of Figure \ref{fig:G_M12_v_z_systematics}) with the Gini index calculated from large field surveys based on the published luminosity functions \citep{Blanton_2003,Bell_2003,Faber_2007}, assuming a consistent limiting magnitude. The Gini index of our sample clusters is a factor of two larger than the field Gini index for $\rm z \in [0,1]$. At $\rm M < -19$ in the cluster core, this is driven by the presence of the BCG and other massive galaxies -- which are not present in the field -- that skew the luminosity distribution of red sequence galaxies in cluster cores brighter. This result is consistent with studies of the cluster luminosity function that show either an increase in $\rm M^*$ or require a second component at the bright end, due to the excess of bright, red galaxies in clusters over the field \citep{De_Propris_2003,2013A&A...550A..58V,McNaught_Roberts_2014, Lan_2016}. 

The increased fraction of faint galaxies in clusters over the field \citep{Lan_2016} should also drive G to increase, though this population ($\rm M > -18$) is not targeted in our analysis due to our conservative choice of limiting magnitude. The enhanced Gini index in clusters is a testament to the unique growth channels of massive galaxies and the environmental quenching of faint galaxies in the cluster environment that produce a higher ratio of faint to bright galaxies than the field. $\rm G_{field}$ is relatively unevolving over $\rm z \in [0,1]$, which additionally implies that the shape of the full field luminosity function at the bright end does not evolve strongly during this epoch. 

The second major result is that our $\rm 3\sigma$ measurement of non-evolution in $\rm G$ over $\rm z \in [0,1]$ is consistent with BCGs emerging early in their host clusters, before $\rm z = 1$, as BCGs transitioning from non-dominant to dominant, massive galaxies should result in the general increase of G as mergers of bright ($\rm M < -19$) galaxies in the core concentrate light into fewer members. The non-evolution of G implies that the concentration of light is not drastically changing over $\rm z\in [0,1]$. Clusters are not drastically closer to concentrating their light into fewer galaxies (per our interpretation of G in Section \ref{max_min_gini}) through mechanisms related to the assembly of BCGs. The growth channels that concentrate light into the (eventual) BCG and skew the bright end of the cluster luminosity function must be operating before $\rm z = 1$, consistent with the interpretation that BCGs grow slowly and steadily with their host clusters over the past $\sim$8\,Gyr.

%Given the relative dominace of BCGs at $\rm z\sim 0$ \textbf{The non-evolution of G }\textbf{Phrased another way, the relative concentration of light} more massive galaxies (including the BCG)

%\textbf{Said a different way, the dominance of BCGs relative to their host clusters are statistically comparable at $\rm z \sim 0$ and $\rm z \sim 1$}.

The narrative that the BCG is established early is consistent with several other studies of its defining properties, \citep{Aragon_Salamanca_1998,De_Lucia_2007,Stott_2011,Santos_2012,Ragone-Figueroa_2018,Chu_2021,2024ApJ...976..169C,Ding_2024,2026ApJ..1000..103V}, and other studies of massive galaxy assembly in the cluster environment \citep{Khullar_2022}. %This \textbf{supports the choices of} studies that rely on the BCG as a proxy for the cluster center, cluster redshift, or simply to identify high-$\rm z$ clusters. 
%\textbf{We note that this result confirms the model particularily for massive ($\rm M_{500} > 10^{14} M_{\odot}$) clusters}.

%\textbf{We note that our results pair well with results from simulations, interpretations of BCG centers from simulations. Though BCgs in simulations dpon't always agree..} %Established BCGs imply their use as cluster property proxies, such as centroid-locators, is robust up to $\rm z = 1$. 
%This conclusion is further supported by our application of G to TNG300 clusters (Figure \ref{fig:simulations}), which is consistent with non-evolution within the errors of our z-binned medians.

%\citet{Ragone_Figueroa_2020} found that the BCG is aligned with the cluster's major axis by $\rm z = 2$, ... talk about centering, star formation and inside-out formation times, AGN activity, etc.

%There is evidence of early co-evolution of BCGs with their host clusters, such as the centralization and alignment of BCGs with the cluster major axis by z = 2 \citep{Ragone_Figueroa_2020}. 

One interesting caveat of this result is that we know that BCGs are actively growing during this period, nearly doubling in stellar mass between $\rm z \in [0,1]$ \citep{Lidman_2012}. Thus, across our sample, BCGs are actively accreting stellar mass, and it might be reasonable to expect that this growth alone should produce an evolutionary trend in G between the highest and lowest redshift clusters. We explore the effect of BCG growth on G during this epoch with a thought experiment conducted with our TNG300-1 clusters. 

Doubling the stellar mass of the BCG at the 6 Gyr timestep to estimate the \citet{Lidman_2012} predicted growth leads to a modest increase of $\rm \Delta G = +0.06 $ at z = 0 when the satellite population is left unchanged. However, $\rm z \in [0,1]$ coincides with enrichment of the red sequence \citep{Martinet_2015}, with studies predicting a doubling of the RS stellar mass during this period, driven by satellite galaxies \citep{Bell_2004,De_Lucia_2004,De_Lucia_2007_RS,Brown_2007,Faber_2007,Bell_2008,Rudnick_2009,Stott_2009}. Simply doubling the number of satellite galaxies in TNG300-1 clusters to mimic a richer RS above $\rm M_{lim}$ with the predicted BCG growth rates leads to $\rm \Delta G = +0.02 $ at z = 0, which does not account for the mass growth of individual RS galaxies that would further decrease $\rm \Delta G$. Thus, the non-evolution of G does not imply that the BCG and red sequence are stagnant over $\rm z \in [0,1]$. Rather, our result is consistent with the interpretation that the growth of the BCG and enrichment and growth of the red sequence occur together at similar rates. The early assembly of satellite galaxies may further support shared histories of growth between the red sequence and BCGs \citep{Khullar_2022}. Both a stagnant and co-growing BCG+red sequence population can lead to non-evolution in G over $\rm 0 < z<1$ at the level detected by our analysis, supporting the interpretation that relative  BCG dominance is not changing over $\rm 0 < z<1$, and was established before this epoch.     % This is a really nice bit of text, great work-- thank you <3

%with some studies predicting a doubling of the RS number density during this period

Do we ever expect to observe evolution in BCG dominance through G? At some point, the cluster population may resemble the field population more closely, which would require the two tracks in Figure \ref{fig:G_v_field_discussion} converging. However, it has been shown that an excess of bright galaxies exists in protoclusters when compared to the field \citep{Hatch_2009,Koyama_2012,2018MNRAS.473.1977S}, that likely would increase G above $\rm G_{field}$. The answer likely lies at high redshifts, possibly up to the era of protoclusters. Quantifying BCG emergence may require determining a more realistic $\rm G_{min}$ for clusters, given evidence of massive galaxy assembly before virialization. Future studies at high-z will  require a more careful treatment of cluster membership, as the red sequence continues to become less populated at $\rm z>1$, to avoid biasing the Gini index low and artificially recovering evolution. Regardless, our analysis, particularly our parametric test of evolution, should be sensitive to a slope from evolution of \textbf{$\rm m = -0.06$} within $\rm 3\sigma$ (via the slope uncertainties in Table \ref{tab:evolutionary_parameters}).

\subsection{Application and Benefits of the Gini index}
\label{application_benefits_G}

The Gini index provides a non-parametric measure of the light dominance of the BCG over the satellite population in galaxy clusters. More broadly, the Gini index provides a model-independent probe of the shape of the luminosity function, by essentially quantifying the bright-end skewness of the underlying luminosity function in a single parameter. Comparing measurements of G allows for a study of the deviation of the luminosity function from a chosen standard, without any consideration of functional form. %For our BCG-focused analysis, our measurement of G captures how skewed the bright-end of luminosity function is above the field, without fitting for a double Schecter function. 
By applying different magnitude limits and membership selection, one could imagine repurposing the Gini index to quantify the evolution in the faint upturn of the luminosity function, without assuming an underlying functional form.

%BCG dominance has traditionally been measured by the magnitude gap, but $\rm M_{1,2}$ is more sensitive to transient phenomenon, and is a better measure for cluster dynamical state. G can be calculated without identifying BCGs, and 

When applying the Gini index to our sample, we found it necessary to apply K-corrections to account for SED-aging across $\rm z \in [1,0]$ and band differences between HST-SPT and SDSS-Planck when calculating absolute magnitudes. It was most straightforward to work with fully K-, band-, and distance-corrected absolute magnitudes when designing and applying cuts to our samples. When comparing across different works, such corrections, as well as matching magnitude and radial cuts, are critical to facilitate a statistically fair comparison.

We strongly recommend making cuts on absolute magnitude as opposed to fixing the number of included sources via some ``brightest N galaxies'' approach, given the insensitivity of G to $\rm M_{lim}$ (see Figure \ref{fig:G_v_M_lim_v_N}). Forcing a fixed number of galaxies to be included in the calculation across a sample effectively applies a varying magnitude cut to each cluster given by the magnitude of the Nth brightest galaxy, driving non-physical scatter in G (for a visualization of this effect, please see Appendix \ref{Appendix: Richness}, where we reproduce Figures \ref{fig:G_M12_v_z_systematics} and \ref{fig:N_v_z_systematics} for a flat richness cut). %We also caution against comparing G between studies that use different bands and limiting depths. 

%This analysis would traditionally reuqire  

%. One could 

%deviations

%. For our study, 

%Centrals are good up to z = 1

%It limits scatter introduced by the 

%may be smeared out by dwarf gaalxies. 
%double vs single schecter function-- massive galaxies dominating light

%Implications of result. for cluster growth 

%How to use it --K-correction, Collapses luminosity function into one number -- non analytic/ parametric 

%The way the Gini index evolves is consistent with our expectations based on how clusters grow.

%What does the emergence of the BCG mean?

%Minimum Gini

%Inplications of observed evolution

%Mass evolution

\section{Conclusion}

We present the Gini index as an alternate indicator of brightest cluster galaxy dominance and emergence in galaxy clusters. The Gini index is less sensitive to the cluster dynamical state and member galaxy indexing than the magnitude gap, and accounts for the relative brightness of massive galaxies over $\rm M^{*}$, providing a more complete picture of structural assembly with fewer degeneracies. When applied to observed cluster cores, G produces $\rm \sim 7$ times less fractional scatter than the more commonly used ``magnitude gap'', $\rm M_{1,2}$.

We calculate the Gini index, G, for each cluster in a sample of 92 Planck- and SPT-selected galaxy clusters, lying on a common evolutionary track spanning $\rm z=0$ to $\rm z=1$.  These estimates are based on optical data from SDSS (low-$\rm z$) and HST (high-$\rm z$). Over the redshift range $\rm 0 < z < 1$, we find no statistically significant evidence for evolution in G or M$_{1,2}$. In addition, we find that $\rm G_{cluster}$ (red sequence, cluster core) $ \rm \sim 2G_{field}$ (all galaxies) for all $\rm z < 1$. This has two major implications: (i) We find no evidence for evolution in G, and rule out that BCGs have grown significantly (2$\rm \times$) faster than the cluster population between $\rm z=1$ and $\rm z=0$ at 3$\rm \sigma$ confidence. Our result is consistent with BCGs in massive clusters emerging from the galaxy luminosity distribution of their host clusters early, establishing their relative dominance over red sequence satellite galaxies before $\rm z = 1$. This means that the use of BCGs as proxies for massive host cluster properties up to $\rm z = 1$ should be robust; and (ii) We find no evidence for evolution in G during a known period of BCG stellar mass growth. The enrichment of the red sequence and mass growth of red sequence galaxies can suppress evolution in G if they occur at a comparable rate to the growth of the BCG. In other words, given the measured constraints on the evolution of G, we can conclude that BCGs established their dominance at $\rm z>1$. We find no evidence that they have been growing faster than other member galaxies since $\rm z=1$.
%\end{itemize}

The Gini index generally probes deviations in the shape of the luminosity function, allowing for versatile application to different scientific contexts. G is highly sensitive to the choice of limiting magnitude when considering a galaxy population, and is not easily comparable between studies with different choices unless the underlying shape of the luminosity function is fully known. This non-parametric approach to quantifying BCG evolution can be easily and rapidly applied to large samples. Additionally, the calculation of G does not require the identification of a BCG, and we show that our results are consistent across permutations of BCG-dependent or -independent cut choices. G is less sensitive than the often-used magnitude gap to short-term evolution (e.g., cluster-cluster mergers) and is more sensitive to the long term assembly history of the galaxy population.

\begin{acknowledgments}

This research is based on observations made with the NASA/ESA Hubble Space Telescope obtained from the Space Telescope Science Institute, which is operated by the Association of Universities for Research in Astronomy, Inc., under NASA contract NAS 5–26555. 
Support for US investigators in program HST-GO-15307 was provided by NASA through a grant from the Space Telescope Science Institute, which is operated by the Association of Universities for Research in Astronomy, Inc., under NASA contract NAS5-26555. K.\ Fosdick and M.\ McDonald also acknowledge support from the National Science Foundation via AAG grant number AST-2510234. Argonne National Laboratory’s work was supported by the U.S. Department of Energy, Office of High Energy Physics, under contract DE-AC02-06CH11357.
For the purpose of open access, the authors have applied a Creative Commons Attribution (CC BY) license to any Author Accepted Manuscript version arising.

\end{acknowledgments}

\bibliography{sample7}{}

@article{Hennig_2017,
   title={Galaxy Populations in Massive Galaxy Clusters to $z$ = 1.1: Color Distribution, Concentration, Halo Occupation Number and Red Sequence Fraction},
   ISSN={1365-2966},
   url={http://dx.doi.org/10.1093/mnras/stx175},
   DOI={10.1093/mnras/stx175},
   journal={Monthly Notices of the Royal Astronomical Society},
   publisher={Oxford University Press (OUP)},
   author={Hennig, C. and Mohr, J. J. and Zenteno, A. and Desai, S. and Dietrich, J. P. and Bocquet, S. and Strazzullo, V. and Saro, A. and Abbott, T. M. C. and Abdalla, F. B. and Bayliss, M. and Benoit-Lévy, A. and Bernstein, R. A. and Bertin, E. and Brooks, D. and Capasso, R. and Capozzi, D. and Carnero, A. and Kind, M. Carrasco and Carretero, J. and Chiu, I. and D’Andrea, C. B. and daCosta, L. N. and Diehl, H. T. and Doel, P. and Eifler, T. F. and Evrard, A. E. and Fausti-Neto, A. and Fosalba, P. and Frieman, J. and Gangkofner, C. and Gonzalez, A. and Gruen, D. and Gruendl, R. A. and Gupta, N. and Gutierrez, G. and Honscheid, K. and Hlavacek-Larrondo, J. and James, D. J. and Kuehn, K. and Kuropatkin, N. and Lahav, O. and March, M. and Marshall, J. L. and Martini, P. and McDonald, M. and Melchior, P. and Miller, C. J. and Miquel, R. and Neilsen, E. and Nord, B. and Ogando, R. and Plazas, A. A. and Reichardt, C. and Romer, A. K. and Rozo, E. and Rykoff, E. S. and Sanchez, E. and Santiago, B. and Schubnell, M. and Sevilla-Noarbe, I. and Smith, R. C. and Soares-Santos, M. and Sobreira, F. and Stalder, B. and Stanford, S.A. and Suchyta, E. and Swanson, M. E. C. and Tarle, G. and Thomas, D. and Vikram, V. and Walker, A. R. and Zhang, Y.},
   year={2017},
   month=jan, pages={stx175} }

@book{gini1912variabilità,
  title={Variabilit{\`a} e mutabilit{\`a}: contributo allo studio delle distribuzioni e delle relazioni statistiche. [Fasc. I.]},
  author={Gini, C.},
  series={Studi economico-giuridici pubblicati per cura della facolt{\`a} di Giurisprudenza della R. Universit{\`a} di Cagliari},
  url={https://books.google.com/books?id=fqjaBPMxB9kC},
  year={1912},
  publisher={Tipogr. di P. Cuppini}
}

@article{Fakhouri_2010,
   title={The merger rates and mass assembly histories of dark matter haloes in the two Millennium simulations: Merger rates},
   volume={406},
   ISSN={0035-8711},
   url={http://dx.doi.org/10.1111/j.1365-2966.2010.16859.x},
   DOI={10.1111/j.1365-2966.2010.16859.x},
   number={4},
   journal={Monthly Notices of the Royal Astronomical Society},
   publisher={Oxford University Press (OUP)},
   author={Fakhouri, Onsi and Ma, Chung-Pei and Boylan-Kolchin, Michael},
   year={2010},
   month=jun, pages={2267–2278} }

@article{Bleem_2020,
   title={The SPTpol Extended Cluster Survey},
   volume={247},
   ISSN={1538-4365},
   url={http://dx.doi.org/10.3847/1538-4365/ab6993},
   DOI={10.3847/1538-4365/ab6993},
   number={1},
   journal={The Astrophysical Journal Supplement Series},
   publisher={American Astronomical Society},
   author={Bleem, L. E. and Bocquet, S. and Stalder, B. and Gladders, M. D. and Ade, P. A. R. and Allen, S. W. and Anderson, A. J. and Annis, J. and Ashby, M. L. N. and Austermann, J. E. and Avila, S. and Avva, J. S. and Bayliss, M. and Beall, J. A. and Bechtol, K. and Bender, A. N. and Benson, B. A. and Bertin, E. and Bianchini, F. and Blake, C. and Brodwin, M. and Brooks, D. and Buckley-Geer, E. and Burke, D. L. and Carlstrom, J. E. and Rosell, A. Carnero and Carrasco Kind, M. and Carretero, J. and Chang, C. L. and Chiang, H. C. and Citron, R. and Moran, C. Corbett and Costanzi, M. and Crawford, T. M. and Crites, A. T. and da Costa, L. N. and de Haan, T. and De Vicente, J. and Desai, S. and Diehl, H. T. and Dietrich, J. P. and Dobbs, M. A. and Eifler, T. F. and Everett, W. and Flaugher, B. and Floyd, B. and Frieman, J. and Gallicchio, J. and García-Bellido, J. and George, E. M. and Gerdes, D. W. and Gilbert, A. and Gruen, D. and Gruendl, R. A. and Gschwend, J. and Gupta, N. and Gutierrez, G. and Halverson, N. W. and Harrington, N. and Henning, J. W. and Heymans, C. and Holder, G. P. and Hollowood, D. L. and Holzapfel, W. L. and Honscheid, K. and Hrubes, J. D. and Huang, N. and Hubmayr, J. and Irwin, K. D. and James, D. J. and Jeltema, T. and Joudaki, S. and Khullar, G. and Klein, M. and Knox, L. and Kuropatkin, N. and Lee, A. T. and Li, D. and Lidman, C. and Lowitz, A. and MacCrann, N. and Mahler, G. and Maia, M. A. G. and Marshall, J. L. and McDonald, M. and McMahon, J. J. and Melchior, P. and Menanteau, F. and Meyer, S. S. and Miquel, R. and Mocanu, L. M. and Mohr, J. J. and Montgomery, J. and Nadolski, A. and Natoli, T. and Nibarger, J. P. and Noble, G. and Novosad, V. and Padin, S. and Palmese, A. and Parkinson, D. and Patil, S. and Paz-Chinchón, F. and Plazas, A. A. and Pryke, C. and Ramachandra, N. S. and Reichardt, C. L. and Remolina González, J. D. and Romer, A. K. and Roodman, A. and Ruhl, J. E. and Rykoff, E. S. and Saliwanchik, B. R. and Sanchez, E. and Saro, A. and Sayre, J. T. and Schaffer, K. K. and Schrabback, T. and Serrano, S. and Sharon, K. and Sievers, C. and Smecher, G. and Smith, M. and Soares-Santos, M. and Stark, A. A. and Story, K. T. and Suchyta, E. and Tarle, G. and Tucker, C. and Vanderlinde, K. and Veach, T. and Vieira, J. D. and Wang, G. and Weller, J. and Whitehorn, N. and Wu, W. L. K. and Yefremenko, V. and Zhang, Y.},
   year={2020},
   month=mar, pages={25} }

@article{Bleem_2015,
   title={GALAXY CLUSTERS DISCOVERED VIA THE SUNYAEV-ZEL’DOVICH EFFECT IN THE 2500-SQUARE-DEGREE SPT-SZ SURVEY},
   volume={216},
   ISSN={1538-4365},
   url={http://dx.doi.org/10.1088/0067-0049/216/2/27},
   DOI={10.1088/0067-0049/216/2/27},
   number={2},
   journal={The Astrophysical Journal Supplement Series},
   publisher={American Astronomical Society},
   author={Bleem, L. E. and Stalder, B. and de Haan, T. and Aird, K. A. and Allen, S. W. and Applegate, D. E. and Ashby, M. L. N. and Bautz, M. and Bayliss, M. and Benson, B. A. and Bocquet, S. and Brodwin, M. and Carlstrom, J. E. and Chang, C. L. and Chiu, I. and Cho, H. M. and Clocchiatti, A. and Crawford, T. M. and Crites, A. T. and Desai, S. and Dietrich, J. P. and Dobbs, M. A. and Foley, R. J. and Forman, W. R. and George, E. M. and Gladders, M. D. and Gonzalez, A. H. and Halverson, N. W. and Hennig, C. and Hoekstra, H. and Holder, G. P. and Holzapfel, W. L. and Hrubes, J. D. and Jones, C. and Keisler, R. and Knox, L. and Lee, A. T. and Leitch, E. M. and Liu, J. and Lueker, M. and Luong-Van, D. and Mantz, A. and Marrone, D. P. and McDonald, M. and McMahon, J. J. and Meyer, S. S. and Mocanu, L. and Mohr, J. J. and Murray, S. S. and Padin, S. and Pryke, C. and Reichardt, C. L. and Rest, A. and Ruel, J. and Ruhl, J. E. and Saliwanchik, B. R. and Saro, A. and Sayre, J. T. and Schaffer, K. K. and Schrabback, T. and Shirokoff, E. and Song, J. and Spieler, H. G. and Stanford, S. A. and Staniszewski, Z. and Stark, A. A. and Story, K. T. and Stubbs, C. W. and Vanderlinde, K. and Vieira, J. D. and Vikhlinin, A. and Williamson, R. and Zahn, O. and Zenteno, A.},
   year={2015},
   month=jan, pages={27} }

@article{Bocquet_2019,
   title={Cluster Cosmology Constraints from the 2500 deg2 SPT-SZ Survey: Inclusion of Weak Gravitational Lensing Data from Magellan and the Hubble Space Telescope},
   volume={878},
   ISSN={1538-4357},
   url={http://dx.doi.org/10.3847/1538-4357/ab1f10},
   DOI={10.3847/1538-4357/ab1f10},
   number={1},
   journal={The Astrophysical Journal},
   publisher={American Astronomical Society},
   author={Bocquet, S. and Dietrich, J. P. and Schrabback, T. and Bleem, L. E. and Klein, M. and Allen, S. W. and Applegate, D. E. and Ashby, M. L. N. and Bautz, M. and Bayliss, M. and Benson, B. A. and Brodwin, M. and Bulbul, E. and Canning, R. E. A. and Capasso, R. and Carlstrom, J. E. and Chang, C. L. and Chiu, I. and Cho, H-M. and Clocchiatti, A. and Crawford, T. M. and Crites, A. T. and Haan, T. de and Desai, S. and Dobbs, M. A. and Foley, R. J. and Forman, W. R. and Garmire, G. P. and George, E. M. and Gladders, M. D. and Gonzalez, A. H. and Grandis, S. and Gupta, N. and Halverson, N. W. and Hlavacek-Larrondo, J. and Hoekstra, H. and Holder, G. P. and Holzapfel, W. L. and Hou, Z. and Hrubes, J. D. and Huang, N. and Jones, C. and Khullar, G. and Knox, L. and Kraft, R. and Lee, A. T. and Linden, A. von der and Luong-Van, D. and Mantz, A. and Marrone, D. P. and McDonald, M. and McMahon, J. J. and Meyer, S. S. and Mocanu, L. M. and Mohr, J. J. and Morris, R. G. and Padin, S. and Patil, S. and Pryke, C. and Rapetti, D. and Reichardt, C. L. and Rest, A. and Ruhl, J. E. and Saliwanchik, B. R. and Saro, A. and Sayre, J. T. and Schaffer, K. K. and Shirokoff, E. and Stalder, B. and Stanford, S. A. and Staniszewski, Z. and Stark, A. A. and Story, K. T. and Strazzullo, V. and Stubbs, C. W. and Vanderlinde, K. and Vieira, J. D. and Vikhlinin, A. and Williamson, R. and Zenteno, A.},
   year={2019},
   month=jun, pages={55} }

@ARTICLE{1996A&AS..117..393B,
       author = {{Bertin}, E. and {Arnouts}, S.},
        title = "{SExtractor: Software for source extraction.}",
      journal = {\aaps},
         year = "1996",
        month = "Jun",
       volume = {117},
        pages = {393-404},
          doi = {10.1051/aas:1996164},
       adsurl = {https://ui.adsabs.harvard.edu/abs/1996A&AS..117..393B}
}

@article{Planck2016,
   title={Planck2015 results: XXVII. The secondPlanckcatalogue of Sunyaev-Zeldovich sources},
   volume={594},
   ISSN={1432-0746},
   url={http://dx.doi.org/10.1051/0004-6361/201525823},
   DOI={10.1051/0004-6361/201525823},
   journal={Astronomy \& Astrophysics},
   publisher={EDP Sciences},
   author={Ade, P. A. R. and Aghanim, N. and Arnaud, M. and Ashdown, M. and Aumont, J. and Baccigalupi, C. and Banday, A. J. and Barreiro, R. B. and Barrena, R. and Bartlett, J. G. and Bartolo, N. and Battaner, E. and Battye, R. and Benabed, K. and Benoît, A. and Benoit-Lévy, A. and Bernard, J.-P. and Bersanelli, M. and Bielewicz, P. and Bikmaev, I. and Böhringer, H. and Bonaldi, A. and Bonavera, L. and Bond, J. R. and Borrill, J. and Bouchet, F. R. and Bucher, M. and Burenin, R. and Burigana, C. and Butler, R. C. and Calabrese, E. and Cardoso, J.-F. and Carvalho, P. and Catalano, A. and Challinor, A. and Chamballu, A. and Chary, R.-R. and Chiang, H. C. and Chon, G. and Christensen, P. R. and Clements, D. L. and Colombi, S. and Colombo, L. P. L. and Combet, C. and Comis, B. and Couchot, F. and Coulais, A. and Crill, B. P. and Curto, A. and Cuttaia, F. and Dahle, H. and Danese, L. and Davies, R. D. and Davis, R. J. and de Bernardis, P. and de Rosa, A. and de Zotti, G. and Delabrouille, J. and Désert, F.-X. and Dickinson, C. and Diego, J. M. and Dolag, K. and Dole, H. and Donzelli, S. and Doré, O. and Douspis, M. and Ducout, A. and Dupac, X. and Efstathiou, G. and Eisenhardt, P. R. M. and Elsner, F. and Enßlin, T. A. and Eriksen, H. K. and Falgarone, E. and Fergusson, J. and Feroz, F. and Ferragamo, A. and Finelli, F. and Forni, O. and Frailis, M. and Fraisse, A. A. and Franceschi, E. and Frejsel, A. and Galeotta, S. and Galli, S. and Ganga, K. and Génova-Santos, R. T. and Giard, M. and Giraud-Héraud, Y. and Gjerløw, E. and González-Nuevo, J. and Górski, K. M. and Grainge, K. J. B. and Gratton, S. and Gregorio, A. and Gruppuso, A. and Gudmundsson, J. E. and Hansen, F. K. and Hanson, D. and Harrison, D. L. and Hempel, A. and Henrot-Versillé, S. and Hernández-Monteagudo, C. and Herranz, D. and Hildebrandt, S. R. and Hivon, E. and Hobson, M. and Holmes, W. A. and Hornstrup, A. and Hovest, W. and Huffenberger, K. M. and Hurier, G. and Jaffe, A. H. and Jaffe, T. R. and Jin, T. and Jones, W. C. and Juvela, M. and Keihänen, E. and Keskitalo, R. and Khamitov, I. and Kisner, T. S. and Kneissl, R. and Knoche, J. and Kunz, M. and Kurki-Suonio, H. and Lagache, G. and Lamarre, J.-M. and Lasenby, A. and Lattanzi, M. and Lawrence, C. R. and Leonardi, R. and Lesgourgues, J. and Levrier, F. and Liguori, M. and Lilje, P. B. and Linden-Vørnle, M. and López-Caniego, M. and Lubin, P. M. and Macías-Pérez, J. F. and Maggio, G. and Maino, D. and Mak, D. S. Y. and Mandolesi, N. and Mangilli, A. and Martin, P. G. and Martínez-González, E. and Masi, S. and Matarrese, S. and Mazzotta, P. and McGehee, P. and Mei, S. and Melchiorri, A. and Melin, J.-B. and Mendes, L. and Mennella, A. and Migliaccio, M. and Mitra, S. and Miville-Deschênes, M.-A. and Moneti, A. and Montier, L. and Morgante, G. and Mortlock, D. and Moss, A. and Munshi, D. and Murphy, J. A. and Naselsky, P. and Nastasi, A. and Nati, F. and Natoli, P. and Netterfield, C. B. and Nørgaard-Nielsen, H. U. and Noviello, F. and Novikov, D. and Novikov, I. and Olamaie, M. and Oxborrow, C. A. and Paci, F. and Pagano, L. and Pajot, F. and Paoletti, D. and Pasian, F. and Patanchon, G. and Pearson, T. J. and Perdereau, O. and Perotto, L. and Perrott, Y. C. and Perrotta, F. and Pettorino, V. and Piacentini, F. and Piat, M. and Pierpaoli, E. and Pietrobon, D. and Plaszczynski, S. and Pointecouteau, E. and Polenta, G. and Pratt, G. W. and Prézeau, G. and Prunet, S. and Puget, J.-L. and Rachen, J. P. and Reach, W. T. and Rebolo, R. and Reinecke, M. and Remazeilles, M. and Renault, C. and Renzi, A. and Ristorcelli, I. and Rocha, G. and Rosset, C. and Rossetti, M. and Roudier, G. and Rozo, E. and Rubiño-Martín, J. A. and Rumsey, C. and Rusholme, B. and Rykoff, E. S. and Sandri, M. and Santos, D. and Saunders, R. D. E. and Savelainen, M. and Savini, G. and Schammel, M. P. and Scott, D. and Seiffert, M. D. and Shellard, E. P. S. and Shimwell, T. W. and Spencer, L. D. and Stanford, S. A. and Stern, D. and Stolyarov, V. and Stompor, R. and Streblyanska, A. and Sudiwala, R. and Sunyaev, R. and Sutton, D. and Suur-Uski, A.-S. and Sygnet, J.-F. and Tauber, J. A. and Terenzi, L. and Toffolatti, L. and Tomasi, M. and Tramonte, D. and Tristram, M. and Tucci, M. and Tuovinen, J. and Umana, G. and Valenziano, L. and Valiviita, J. and Van Tent, B. and Vielva, P. and Villa, F. and Wade, L. A. and Wandelt, B. D. and Wehus, I. K. and White, S. D. M. and Wright, E. L. and Yvon, D. and Zacchei, A. and Zonca, A.},
   year={2016},
   month=sep, pages={A27} }

@article{Leethochawalit_2018,
   title={Evolution of the Stellar Mass–Metallicity Relation. I. Galaxies in the z ∼ 0.4 Cluster Cl0024},
   volume={856},
   ISSN={1538-4357},
   url={http://dx.doi.org/10.3847/1538-4357/aab26a},
   DOI={10.3847/1538-4357/aab26a},
   number={1},
   journal={The Astrophysical Journal},
   publisher={American Astronomical Society},
   author={Leethochawalit, Nicha and Kirby, Evan N. and Moran, Sean M. and Ellis, Richard S. and Treu, Tommaso},
   year={2018},
   month=mar, pages={15} }

@article{Salvati_2022,
   title={Combining Planck and SPT Cluster Catalogs: Cosmological Analysis and Impact on the Planck Scaling Relation Calibration},
   volume={934},
   ISSN={1538-4357},
   url={http://dx.doi.org/10.3847/1538-4357/ac7ab4},
   DOI={10.3847/1538-4357/ac7ab4},
   number={2},
   journal={The Astrophysical Journal},
   publisher={American Astronomical Society},
   author={Salvati, L. and Saro, A. and Bocquet, S. and Costanzi, M. and Ansarinejad, B. and Benson, B. A. and Bleem, L. E. and Calzadilla, M. S. and Carlstrom, J. E. and Chang, C. L. and Chown, R. and Crites, A. T. and Haan, T. de and Dobbs, M. A. and Everett, W. B. and Floyd, B. and Grandis, S. and George, E. M. and Halverson, N. W. and Holder, G. P. and Holzapfel, W. L. and Hrubes, J. D. and Lee, A. T. and Luong-Van, D. and McDonald, M. and McMahon, J. J. and Meyer, S. S. and Millea, M. and Mocanu, L. M. and Mohr, J. J. and Natoli, T. and Omori, Y. and Padin, S. and Pryke, C. and Reichardt, C. L. and Ruhl, J. E. and Ruppin, F. and Schaffer, K. K. and Schrabback, T. and Shirokoff, E. and Staniszewski, Z. and Stark, A. A. and Vieira, J. D. and Williamson, R.},
   year={2022},
   month=aug, pages={129} }

@article{Blanton_2003,
   title={The Galaxy Luminosity Function and Luminosity Density at Redshiftz= 0.1},
   volume={592},
   ISSN={1538-4357},
   url={http://dx.doi.org/10.1086/375776},
   DOI={10.1086/375776},
   number={2},
   journal={The Astrophysical Journal},
   publisher={American Astronomical Society},
   author={Blanton, Michael R. and Hogg, David W. and Bahcall, Neta A. and Brinkmann, J. and Britton, Malcolm and Connolly, Andrew J. and Csabai, Istvan and Fukugita, Masataka and Loveday, Jon and Meiksin, Avery and Munn, Jeffrey A. and Nichol, R. C. and Okamura, Sadanori and Quinn, Thomas and Schneider, Donald P. and Shimasaku, Kazuhiro and Strauss, Michael A. and Tegmark, Max and Vogeley, Michael S. and Weinberg, David H.},
   year={2003},
   month=aug, pages={819–838} }

@misc{nelson2021illustristngsimulationspublicdata,
      title={The IllustrisTNG Simulations: Public Data Release}, 
      author={Dylan Nelson and Volker Springel and Annalisa Pillepich and Vicente Rodriguez-Gomez and Paul Torrey and Shy Genel and Mark Vogelsberger and Ruediger Pakmor and Federico Marinacci and Rainer Weinberger and Luke Kelley and Mark Lovell and Benedikt Diemer and Lars Hernquist},
      year={2021},
      eprint={1812.05609},
      archivePrefix={arXiv},
      primaryClass={astro-ph.GA},
      url={https://arxiv.org/abs/1812.05609}, 
}

@article{Pillepich_2017,
   title={First results from the IllustrisTNG simulations: the stellar mass content of groups and clusters of galaxies},
   volume={475},
   ISSN={1365-2966},
   url={http://dx.doi.org/10.1093/mnras/stx3112},
   DOI={10.1093/mnras/stx3112},
   number={1},
   journal={Monthly Notices of the Royal Astronomical Society},
   publisher={Oxford University Press (OUP)},
   author={Pillepich, Annalisa and Nelson, Dylan and Hernquist, Lars and Springel, Volker and Pakmor, Rüdiger and Torrey, Paul and Weinberger, Rainer and Genel, Shy and Naiman, Jill P and Marinacci, Federico and Vogelsberger, Mark},
   year={2017},
   month=dec, pages={648–675} }

@article{Springel_2017,
   title={First results from the IllustrisTNG simulations: matter and galaxy clustering},
   volume={475},
   ISSN={1365-2966},
   url={http://dx.doi.org/10.1093/mnras/stx3304},
   DOI={10.1093/mnras/stx3304},
   number={1},
   journal={Monthly Notices of the Royal Astronomical Society},
   publisher={Oxford University Press (OUP)},
   author={Springel, Volker and Pakmor, Rüdiger and Pillepich, Annalisa and Weinberger, Rainer and Nelson, Dylan and Hernquist, Lars and Vogelsberger, Mark and Genel, Shy and Torrey, Paul and Marinacci, Federico and Naiman, Jill},
   year={2017},
   month=dec, pages={676–698} }

@article{Nelson_2017,
   title={First results from the IllustrisTNG simulations: the galaxy colour bimodality},
   volume={475},
   ISSN={1365-2966},
   url={http://dx.doi.org/10.1093/mnras/stx3040},
   DOI={10.1093/mnras/stx3040},
   number={1},
   journal={Monthly Notices of the Royal Astronomical Society},
   publisher={Oxford University Press (OUP)},
   author={Nelson, Dylan and Pillepich, Annalisa and Springel, Volker and Weinberger, Rainer and Hernquist, Lars and Pakmor, Rüdiger and Genel, Shy and Torrey, Paul and Vogelsberger, Mark and Kauffmann, Guinevere and Marinacci, Federico and Naiman, Jill},
   year={2017},
   month=nov, pages={624–647} }

@article{Naiman_2018,
   title={First results from the IllustrisTNG simulations: a tale of two elements – chemical evolution of magnesium and europium},
   volume={477},
   ISSN={1365-2966},
   url={http://dx.doi.org/10.1093/mnras/sty618},
   DOI={10.1093/mnras/sty618},
   number={1},
   journal={Monthly Notices of the Royal Astronomical Society},
   publisher={Oxford University Press (OUP)},
   author={Naiman, Jill P and Pillepich, Annalisa and Springel, Volker and Ramirez-Ruiz, Enrico and Torrey, Paul and Vogelsberger, Mark and Pakmor, Rüdiger and Nelson, Dylan and Marinacci, Federico and Hernquist, Lars and Weinberger, Rainer and Genel, Shy},
   year={2018},
   month=mar, pages={1206–1224} }

@article{Marinacci_2018,
   title={First results from the IllustrisTNG simulations: radio haloes and magnetic fields},
   ISSN={1365-2966},
   url={http://dx.doi.org/10.1093/mnras/sty2206},
   DOI={10.1093/mnras/sty2206},
   journal={Monthly Notices of the Royal Astronomical Society},
   publisher={Oxford University Press (OUP)},
   author={Marinacci, Federico and Vogelsberger, Mark and Pakmor, Rüdiger and Torrey, Paul and Springel, Volker and Hernquist, Lars and Nelson, Dylan and Weinberger, Rainer and Pillepich, Annalisa and Naiman, Jill and Genel, Shy},
   year={2018},
   month=aug }

@article{Rodriguez_Gomez_2015,
   title={The merger rate of galaxies in the Illustris simulation: a comparison with observations and semi-empirical models},
   volume={449},
   ISSN={0035-8711},
   url={http://dx.doi.org/10.1093/mnras/stv264},
   DOI={10.1093/mnras/stv264},
   number={1},
   journal={Monthly Notices of the Royal Astronomical Society},
   publisher={Oxford University Press (OUP)},
   author={Rodriguez-Gomez, Vicente and Genel, Shy and Vogelsberger, Mark and Sijacki, Debora and Pillepich, Annalisa and Sales, Laura V. and Torrey, Paul and Snyder, Greg and Nelson, Dylan and Springel, Volker and Ma, Chung-Pei and Hernquist, Lars},
   year={2015},
   month=mar, pages={49–64} }

@article{Cappellari_2013,
   title={The ATLAS3D project – XV. Benchmark for early-type galaxies scaling relations from 260 dynamical models: mass-to-light ratio, dark matter, Fundamental Plane and Mass Plane},
   volume={432},
   ISSN={0035-8711},
   url={http://dx.doi.org/10.1093/mnras/stt562},
   DOI={10.1093/mnras/stt562},
   number={3},
   journal={Monthly Notices of the Royal Astronomical Society},
   publisher={Oxford University Press (OUP)},
   author={Cappellari, Michele and Scott, Nicholas and Alatalo, Katherine and Blitz, Leo and Bois, Maxime and Bournaud, Frédéric and Bureau, M. and Crocker, Alison F. and Davies, Roger L. and Davis, Timothy A. and de Zeeuw, P. T. and Duc, Pierre-Alain and Emsellem, Eric and Khochfar, Sadegh and Krajnović, Davor and Kuntschner, Harald and McDermid, Richard M. and Morganti, Raffaella and Naab, Thorsten and Oosterloo, Tom and Sarzi, Marc and Serra, Paolo and Weijmans, Anne-Marie and Young, Lisa M.},
   year={2013},
   month=may, pages={1709–1741} }

@article{FAST_LTS,
	author = {Rousseeuw, Peter J. and Van Driessen, Katrien},
	date = {2006/01/01},
	doi = {10.1007/s10618-005-0024-4},
	id = {ROUSSEEUW2006},
	isbn = {1573-756X},
	journal = {Data Mining and Knowledge Discovery},
	number = {1},
	pages = {29--45},
	title = {Computing LTS Regression for Large Data Sets},
	url = {https://doi.org/10.1007/s10618-005-0024-4},
	volume = {12},
	year = {2006}}

@article{10.1111/j.1365-2966.2010.16406.x,
    author = {Andreon, S. and Hurn, M. A.},
    title = {The scaling relation between richness and mass of galaxy clusters: a Bayesian approach},
    journal = {Monthly Notices of the Royal Astronomical Society},
    volume = {404},
    number = {4},
    pages = {1922-1937},
    year = {2010},
    month = {05},
    issn = {0035-8711},
    doi = {10.1111/j.1365-2966.2010.16406.x},
    url = {https://doi.org/10.1111/j.1365-2966.2010.16406.x},
    eprint = {https://academic.oup.com/mnras/article-pdf/404/4/1922/3789311/mnras0404-1922.pdf},
}

@article{Lan_2016,
   title={The galaxy luminosity function in groups and clusters: the faint-end upturn and the connection to the field luminosity function},
   volume={459},
   ISSN={1365-2966},
   url={http://dx.doi.org/10.1093/mnras/stw898},
   DOI={10.1093/mnras/stw898},
   number={4},
   journal={Monthly Notices of the Royal Astronomical Society},
   publisher={Oxford University Press (OUP)},
   author={Lan, Ting-Wen and Ménard, Brice and Mo, Houjun},
   year={2016},
   month=apr, pages={3998–4019} }

@article{Lidman_2012,
   title={Evidence for significant growth in the stellar mass of brightest cluster galaxies over the past 10 billion years: Brightest cluster galaxies},
   volume={427},
   ISSN={1365-2966},
   url={http://dx.doi.org/10.1111/j.1365-2966.2012.21984.x},
   DOI={10.1111/j.1365-2966.2012.21984.x},
   number={1},
   journal={Monthly Notices of the Royal Astronomical Society},
   publisher={Oxford University Press (OUP)},
   author={Lidman, C. and Suherli, J. and Muzzin, A. and Wilson, G. and Demarco, R. and Brough, S. and Rettura, A. and Cox, J. and DeGroot, A. and Yee, H. K. C. and Gilbank, D. and Hoekstra, H. and Balogh, M. and Ellingson, E. and Hicks, A. and Nantais, J. and Noble, A. and Lacy, M. and Surace, J. and Webb, T.}, 
    year={2012}}

@article{Martinet_2015,
   title={The evolution of the cluster optical galaxy luminosity function betweenz= 0.4 and 0.9 in the DAFT/FADA survey},
   volume={575},
   ISSN={1432-0746},
   url={http://dx.doi.org/10.1051/0004-6361/201423796},
   DOI={10.1051/0004-6361/201423796},
   journal={Astronomy \& Astrophysics},
   publisher={EDP Sciences},
   author={Martinet, Nicolas and Durret, Florence and Guennou, Loïc and Adami, Christophe and Biviano, Andrea and Ulmer, Melville P. and Clowe, Douglas and Halliday, Claire and Ilbert, Olivier and Márquez, Isabel and Schirmer, Mischa},
   year={2015},
   month=mar, pages={A116} }

@article{Dariush_2010,
   title={The mass assembly of galaxy groups and the evolution of the magnitude gap},
   ISSN={1365-2966},
   url={http://dx.doi.org/10.1111/j.1365-2966.2010.16569.x},
   DOI={10.1111/j.1365-2966.2010.16569.x},
   journal={Monthly Notices of the Royal Astronomical Society},
   publisher={Oxford University Press (OUP)},
   author={Dariush, Ali A. and Raychaudhury, Somak and Ponman, Trevor J. and Khosroshahi, Habib G. and Benson, Andrew J. and Bower, Richard G. and Pearce, Frazer},
   year={2010},
   month=apr }

@article{Bell_2003,
   title={The Optical and Near‐Infrared Properties of Galaxies. I. Luminosity and Stellar Mass Functions},
   volume={149},
   ISSN={1538-4365},
   url={http://dx.doi.org/10.1086/378847},
   DOI={10.1086/378847},
   number={2},
   journal={The Astrophysical Journal Supplement Series},
   publisher={American Astronomical Society},
   author={Bell, Eric F. and McIntosh, Daniel H. and Katz, Neal and Weinberg, Martin D.},
   year={2003},
   month=dec, pages={289–312} }

@article{Faber_2007,
   title={Galaxy Luminosity Functions toz∼1 from DEEP2 and COMBO‐17: Implications for Red Galaxy Formation},
   volume={665},
   ISSN={1538-4357},
   url={http://dx.doi.org/10.1086/519294},
   DOI={10.1086/519294},
   number={1},
   journal={The Astrophysical Journal},
   publisher={American Astronomical Society},
   author={Faber, S. M. and Willmer, C. N. A. and Wolf, C. and Koo, D. C. and Weiner, B. J. and Newman, J. A. and Im, M. and Coil, A. L. and Conroy, C. and Cooper, M. C. and Davis, M. and Finkbeiner, D. P. and Gerke, B. F. and Gebhardt, K. and Groth, E. J. and Guhathakurta, P. and Harker, J. and Kaiser, N. and Kassin, S. and Kleinheinrich, M. and Konidaris, N. P. and Kron, R. G. and Lin, L. and Luppino, G. and Madgwick, D. S. and Meisenheimer, K. and Noeske, K. G. and Phillips, A. C. and Sarajedini, V. L. and Schiavon, R. P. and Simard, L. and Szalay, A. S. and Vogt, N. P. and Yan, R.},
   year={2007},
   month=aug, pages={265–294} }

@article{von_Benda_Beckmann_2008,
   title={The fossil phase in the life of a galaxy group},
   volume={386},
   ISSN={1365-2966},
   url={http://dx.doi.org/10.1111/j.1365-2966.2008.13221.x},
   DOI={10.1111/j.1365-2966.2008.13221.x},
   number={4},
   journal={Monthly Notices of the Royal Astronomical Society},
   publisher={Oxford University Press (OUP)},
   author={von Benda-Beckmann, Alexander M. and D’Onghia, Elena and Gottlöber, Stefan and Hoeft, Matthias and Khalatyan, Arman and Klypin, Anatoly and Müller, Volker},
   year={2008},
   month=jun, pages={2345–2352} }

@article{Ponman_Bertram_1993,
	author = {Ponman, T. J. and Bertram, D.},
	date = {1993/05/01},
	doi = {10.1038/363051a0},
	id = {Ponman1993},
	isbn = {1476-4687},
	journal = {Nature},
	number = {6424},
	pages = {51--54},
	title = {Hot gas and dark matter in a compact galaxy group},
	url = {https://doi.org/10.1038/363051a0},
	volume = {363},
	year = {1993}}

@article{Ponman_1994,
	author = {Ponman, T. J. and Allan, D. J. and Jones, L. R. and Merrifield, M. and McHardy, I. M. and Lehto, H. J. and Luppino, G. A.},
	date = {1994/06/01},
	doi = {10.1038/369462a0},
	id = {Ponman1994},
	isbn = {1476-4687},
	journal = {Nature},
	number = {6480},
	pages = {462--464},
	title = {A possible fossil galaxy group},
	url = {https://doi.org/10.1038/369462a0},
	volume = {369},
	year = {1994}}

@article{Vikhlinin_1999,
   title={X-Ray Overluminous Elliptical Galaxies: A New Class of Mass Concentrations in the Universe?},
   volume={520},
   ISSN={0004-637X},
   url={http://dx.doi.org/10.1086/312134},
   DOI={10.1086/312134},
   number={1},
   journal={The Astrophysical Journal},
   publisher={American Astronomical Society},
   author={Vikhlinin, A. and McNamara, B. R. and Hornstrup, A. and Quintana, H. and Forman, W. and Jones, C. and Way, M.},
   year={1999},
   month=jul, pages={L1–L4} }

@ARTICLE{Jones_2000,
       author = {{Jones}, L.~R. and {Ponman}, T.~J. and {Forbes}, Duncan A.},
        title = "{Multiwavelength observations of an evolved galaxy group: an end-point of galaxy merging?}",
      journal = {\mnras},
         year = 2000,
        month = feb,
       volume = {312},
       number = {1},
        pages = {139-150},
          doi = {10.1046/j.1365-8711.2000.03118.x},
       adsurl = {https://ui.adsabs.harvard.edu/abs/2000MNRAS.312..139J}
}

@ARTICLE{Mulchaey_1999,
       author = {{Mulchaey}, John S. and {Zabludoff}, Ann I.},
        title = "{The Isolated Elliptical NGC 1132: Evidence for a Merged Group of Galaxies?}",
      journal = {\apj},
         year = 1999,
        month = mar,
       volume = {514},
       number = {1},
        pages = {133-137},
          doi = {10.1086/306952},
archivePrefix = {arXiv},
       eprint = {astro-ph/9810458},
 primaryClass = {astro-ph},
       adsurl = {https://ui.adsabs.harvard.edu/abs/1999ApJ...514..133M}
}

@ARTICLE{Schneider_Gun_Hoessel_1983,
       author = {{Schneider}, D.~P. and {Gunn}, J.~E. and {Hoessel}, J.~G.},
        title = "{CCD photometry of Abell clusters. I. Magnitudes and redshifts for 84 brightest cluster galaxies.}",
      journal = {\apj},
         year = 1983,
        month = jan,
       volume = {264},
        pages = {337-355},
          doi = {10.1086/160602},
       adsurl = {https://ui.adsabs.harvard.edu/abs/1983ApJ...264..337S}
}

@ARTICLE{Postman_Lauer_1995,
       author = {{Postman}, Marc and {Lauer}, Tod R.},
        title = "{Brightest Cluster Galaxies as Standard Candles}",
      journal = {\apj},
         year = 1995,
        month = feb,
       volume = {440},
        pages = {28},
          doi = {10.1086/175245},
       adsurl = {https://ui.adsabs.harvard.edu/abs/1995ApJ...440...28P}
}

@ARTICLE{Sandage_1972,
       author = {{Sandage}, A.},
        title = "{The redshift-distance relation. II. The Hubble diagram and its scatter for first-ranked cluster galaxies: A formal value for q$_{0}$.}",
      journal = {\apj},
         year = 1972,
        month = nov,
       volume = {178},
        pages = {1-24},
          doi = {10.1086/151763},
       adsurl = {https://ui.adsabs.harvard.edu/abs/1972ApJ...178....1S}
}

@ARTICLE{Hoessel_Schneider_1985,
       author = {{Hoessel}, J.~G. and {Schneider}, D.~P.},
        title = "{CCD Observations of Abell clusters. IV. Surface photometry of 175 brightest cluster galaxies.}",
      journal = {\aj},
         year = 1985,
        month = sep,
       volume = {90},
        pages = {1648-1664},
          doi = {10.1086/113871},
       adsurl = {https://ui.adsabs.harvard.edu/abs/1985AJ.....90.1648H}
}

@article{Golden-Marx_2018,
	author = {Golden-Marx, Jesse B. and Miller, Christopher J.},
	doi = {10.3847/1538-4357/aac2bd},
	journal = {The Astrophysical Journal},
	month = {jun},
	number = {1},
	pages = {2},
	publisher = {The American Astronomical Society},
	title = {The Impact of Environment on the Stellar Mass--Halo Mass Relation},
	url = {https://doi.org/10.3847/1538-4357/aac2bd},
	volume = {860},
	year = {2018}}

@ARTICLE{Solanes_2016,
       author = {{Solanes}, Jos{\'e} M. and {Perea}, Jaime D. and {Darriba}, Laura and {Garc{\'\i}a-G{\'o}mez}, Carlos and {Bosma}, Albert and {Athanassoula}, Evangelia},
        title = "{Forming first-ranked early-type galaxies through hierarchical dissipationless merging}",
      journal = {\mnras},
         year = 2016,
        month = sep,
       volume = {461},
       number = {1},
        pages = {321-343},
          doi = {10.1093/mnras/stw1278},
       adsurl = {https://ui.adsabs.harvard.edu/abs/2016MNRAS.461..321S}
}

@ARTICLE{Jones_2003,
       author = {{Jones}, L.~R. and {Ponman}, T.~J. and {Horton}, A. and {Babul}, A. and {Ebeling}, H. and {Burke}, D.~J.},
        title = "{The nature and space density of fossil groups of galaxies}",
      journal = {\mnras},
         year = 2003,
        month = aug,
       volume = {343},
       number = {2},
        pages = {627-638},
          doi = {10.1046/j.1365-8711.2003.06702.x},
archivePrefix = {arXiv},
       eprint = {astro-ph/0304257},
 primaryClass = {astro-ph},
       adsurl = {https://ui.adsabs.harvard.edu/abs/2003MNRAS.343..627J}
}

@ARTICLE{DOnghia_2005,
       author = {{D'Onghia}, E. and {Sommer-Larsen}, J. and {Romeo}, A.~D. and {Burkert}, A. and {Pedersen}, K. and {Portinari}, L. and {Rasmussen}, J.},
        title = "{The Formation of Fossil Galaxy Groups in the Hierarchical Universe}",
      journal = {\apjl},
         year = 2005,
        month = sep,
       volume = {630},
       number = {2},
        pages = {L109-L112},
          doi = {10.1086/491651},
archivePrefix = {arXiv},
       eprint = {astro-ph/0505544},
 primaryClass = {astro-ph},
       adsurl = {https://ui.adsabs.harvard.edu/abs/2005ApJ...630L.109D}
}

@article{Abraham_2003,
   title={A New Approach to Galaxy Morphology. I. Analysis of the Sloan Digital Sky Survey Early Data Release},
   volume={588},
   ISSN={1538-4357},
   url={http://dx.doi.org/10.1086/373919},
   DOI={10.1086/373919},
   number={1},
   journal={The Astrophysical Journal},
   publisher={American Astronomical Society},
   author={Abraham, Roberto G. and van den Bergh, Sidney and Nair, Preethi},
   year={2003},
   month=may, pages={218–229} }

@article{Florian_2016,
	author = {Florian, Michael K. and Li, Nan and Gladders, Michael D.},
	journal = {The Astrophysical Journal},
	month = {nov},
	number = {2},
	pages = {168},
	title = {THE GINI COEFFICIENT AS A MORPHOLOGICAL MEASUREMENT OF STRONGLY LENSED GALAXIES IN THE IMAGE PLANE},
	volume = {832},
	year = {2016}}

@ARTICLE{Lisker_2008,
       author = {{Lisker}, Thorsten},
        title = "{Is the Gini Coefficient a Stable Measure of Galaxy Structure?}",
      journal = {\apjs},
         year = 2008,
        month = dec,
       volume = {179},
       number = {2},
        pages = {319-325},
          doi = {10.1086/591795},
archivePrefix = {arXiv},
       eprint = {0807.1531},
 primaryClass = {astro-ph},
       adsurl = {https://ui.adsabs.harvard.edu/abs/2008ApJS..179..319L}
}

@ARTICLE{Nersesian_2023,
       author = {{Nersesian}, Angelos and {Zibetti}, Stefano and {D'Eugenio}, Francesco and {Baes}, Maarten},
        title = "{Non-parametric galaxy morphology from stellar and nebular emission with the CALIFA sample}",
      journal = {\aap},
         year = 2023,
        month = may,
       volume = {673},
          eid = {A63},
        pages = {A63},
          doi = {10.1051/0004-6361/202345962},
archivePrefix = {arXiv},
       eprint = {2303.01907},
 primaryClass = {astro-ph.GA},
       adsurl = {https://ui.adsabs.harvard.edu/abs/2023A&A...673A..63N}
}

@ARTICLE{Dariush_2007,
       author = {{Dariush}, Ali and {Khosroshahi}, Habib G. and {Ponman}, Trevor J. and {Pearce}, Frazer and {Raychaudhury}, Somak and {Hartley}, Will},
        title = "{The mass assembly of fossil groups of galaxies in the Millennium simulation}",
      journal = {\mnras},
         year = 2007,
        month = nov,
       volume = {382},
       number = {1},
        pages = {433-442},
          doi = {10.1111/j.1365-2966.2007.12385.x},
archivePrefix = {arXiv},
       eprint = {0708.2702},
 primaryClass = {astro-ph},
       adsurl = {https://ui.adsabs.harvard.edu/abs/2007MNRAS.382..433D}
}

@ARTICLE{Chu_2021,
       author = {{Chu}, A. and {Durret}, F. and {M{\'a}rquez}, I.},
        title = "{Physical properties of brightest cluster galaxies up to redshift 1.80 based on HST data}",
      journal = {\aap},
         year = 2021,
        month = may,
       volume = {649},
          eid = {A42},
        pages = {A42},
          doi = {10.1051/0004-6361/202040245},
archivePrefix = {arXiv},
       eprint = {2102.01557},
 primaryClass = {astro-ph.GA},
       adsurl = {https://ui.adsabs.harvard.edu/abs/2021A&A...649A..42C}
}

@article{De_Lucia_2007,
   title={The hierarchical formation of the brightest cluster galaxies},
   volume={375},
   ISSN={1365-2966},
   url={http://dx.doi.org/10.1111/j.1365-2966.2006.11287.x},
   DOI={10.1111/j.1365-2966.2006.11287.x},
   number={1},
   journal={Monthly Notices of the Royal Astronomical Society},
   publisher={Oxford University Press (OUP)},
   author={De Lucia, G. and Blaizot, J.},
   year={2007},
   month=feb, pages={2–14} }

@ARTICLE{Lin_2010,
       author = {{Lin}, Yen-Ting and {Ostriker}, Jeremiah P. and {Miller}, Christopher J.},
        title = "{A New Test of the Statistical Nature of the Brightest Cluster Galaxies}",
      journal = {\apj},
         year = 2010,
        month = jun,
       volume = {715},
       number = {2},
        pages = {1486-1496},
          doi = {10.1088/0004-637X/715/2/1486},
archivePrefix = {arXiv},
       eprint = {0904.3098},
 primaryClass = {astro-ph.CO},
       adsurl = {https://ui.adsabs.harvard.edu/abs/2010ApJ...715.1486L}
}

@article{Diaz-Gimenez_2012,
    author = {Díaz-Giménez, Eugenia and Mamon, Gary A. and Pacheco, Marcela and Mendes de Oliveira, Claudia and Alonso, M. Victoria},
    title = {Compact groups of galaxies selected by stellar mass: the 2MASS compact group catalogue},
    journal = {Monthly Notices of the Royal Astronomical Society},
    volume = {426},
    number = {1},
    pages = {296-316},
    year = {2012},
    month = {10},
    issn = {0035-8711},
    doi = {10.1111/j.1365-2966.2012.21705.x},
    url = {https://doi.org/10.1111/j.1365-2966.2012.21705.x},
    eprint = {https://academic.oup.com/mnras/article-pdf/426/1/296/3264399/426-1-296.pdf},
}

@article{Shen_2014,
	author = {Shen, Shiyin and Yang, Xiaohu and Mo, Houjun and van den Bosch, Frank and More, Surhud},
	doi = {10.1088/0004-637X/782/1/23},
	journal = {The Astrophysical Journal},
	month = {jan},
	number = {1},
	pages = {23},
	publisher = {The American Astronomical Society},
	title = {THE STATISTICAL NATURE OF THE BRIGHTEST GROUP GALAXIES},
	url = {https://doi.org/10.1088/0004-637X/782/1/23},
	volume = {782},
	year = {2014}}

@ARTICLE{Loh_Strauss_2006,
       author = {{Loh}, Yeong-Shang and {Strauss}, Michael A.},
        title = "{The bright end of the luminosity function of red sequence galaxies}",
      journal = {\mnras},
         year = 2006,
        month = feb,
       volume = {366},
       number = {2},
        pages = {373-386},
          doi = {10.1111/j.1365-2966.2005.09714.x},
archivePrefix = {arXiv},
       eprint = {astro-ph/0510500},
 primaryClass = {astro-ph},
       adsurl = {https://ui.adsabs.harvard.edu/abs/2006MNRAS.366..373L}
}

@ARTICLE{Tremaine_Richstone_1977,
       author = {{Tremaine}, Scott D. and {Richstone}, Douglas O.},
        title = "{A test of a statistical model for the luminosities of bright cluster galaxies}",
      journal = {\apj},
         year = 1977,
        month = mar,
       volume = {212},
        pages = {311-316},
          doi = {10.1086/155049},
       adsurl = {https://ui.adsabs.harvard.edu/abs/1977ApJ...212..311T}
}

@ARTICLE{2000AJ....120.2148G,
       author = {{Gladders}, Michael D. and {Yee}, H.~K.~C.},
        title = "{A New Method For Galaxy Cluster Detection. I. The Algorithm}",
      journal = {\aj},
         year = 2000,
        month = oct,
       volume = {120},
       number = {4},
        pages = {2148-2162},
          doi = {10.1086/301557},
archivePrefix = {arXiv},
       eprint = {astro-ph/0004092},
 primaryClass = {astro-ph},
       adsurl = {https://ui.adsabs.harvard.edu/abs/2000AJ....120.2148G}
}

@article{Khullar_2022,
   title={Synthesizing Stellar Populations in South Pole Telescope Galaxy Clusters. I. Ages of Quiescent Member Galaxies at 0.3 &lt; z &lt; 1.4},
   volume={934},
   ISSN={1538-4357},
   url={http://dx.doi.org/10.3847/1538-4357/ac7c0c},
   DOI={10.3847/1538-4357/ac7c0c},
   number={2},
   journal={The Astrophysical Journal},
   publisher={American Astronomical Society},
   author={Khullar, Gourav and Bayliss, Matthew B. and Gladders, Michael D. and Kim, Keunho J. and Calzadilla, Michael S. and Strazzullo, Veronica and Bleem, Lindsey E. and Mahler, Guillaume and McDonald, Michael and Floyd, Benjamin and Reichardt, Christian L. and Ruppin, Florian and Saro, Alexandro and Sharon, Keren and Somboonpanyakul, Taweewat and Stalder, Brian and Stark, Antony A.},
   year={2022},
   month=aug, pages={177} }

@article{Choi_2014,
   title={THE ASSEMBLY HISTORIES OF QUIESCENT GALAXIES SINCEz= 0.7 FROM ABSORPTION LINE SPECTROSCOPY},
   volume={792},
   ISSN={1538-4357},
   url={http://dx.doi.org/10.1088/0004-637X/792/2/95},
   DOI={10.1088/0004-637x/792/2/95},
   number={2},
   journal={The Astrophysical Journal},
   publisher={American Astronomical Society},
   author={Choi, Jieun and Conroy, Charlie and Moustakas, John and Graves, Genevieve J. and Holden, Bradford P. and Brodwin, Mark and Brown, Michael J. I. and van Dokkum, Pieter G.},
   year={2014},
   month=aug, pages={95} }

@article{Gallazzi_2014,
   title={CHARTING THE EVOLUTION OF THE AGES AND METALLICITIES OF MASSIVE GALAXIES SINCEz= 0.7},
   volume={788},
   ISSN={1538-4357},
   url={http://dx.doi.org/10.1088/0004-637X/788/1/72},
   DOI={10.1088/0004-637x/788/1/72},
   number={1},
   journal={The Astrophysical Journal},
   publisher={American Astronomical Society},
   author={Gallazzi, Anna and Bell, Eric F. and Zibetti, Stefano and Brinchmann, Jarle and Kelson, Daniel D.},
   year={2014},
   month=may, pages={72} }

@article{Vaughan_2022,
   title={The SAMI galaxy survey: Galaxy size can explain the offset between star-forming and passive galaxies in the mass–metallicity relationship},
   volume={516},
   ISSN={1365-2966},
   url={http://dx.doi.org/10.1093/mnras/stac2304},
   DOI={10.1093/mnras/stac2304},
   number={2},
   journal={Monthly Notices of the Royal Astronomical Society},
   publisher={Oxford University Press (OUP)},
   author={Vaughan, Sam P and Barone, Tania M and Croom, Scott M and Cortese, Luca and D’Eugenio, Francesco and Brough, Sarah and Colless, Matthew and McDermid, Richard M and van de Sande, Jesse and Scott, Nicholas and Bland-Hawthorn, Joss and Bryant, Julia J and Lawrence, J S and López-Sánchez, Ángel R and Lorente, Nuria P F and Owers, Matt S and Richards, Samuel N},
   year={2022},
   month=aug, pages={2971–2987} }

@article{Bevacqua_2024,
   title={A new perspective on the stellar mass-metallicity relation of quiescent galaxies from the LEGA-C survey},
   volume={690},
   ISSN={1432-0746},
   url={http://dx.doi.org/10.1051/0004-6361/202348831},
   DOI={10.1051/0004-6361/202348831},
   journal={Astronomy \& Astrophysics},
   publisher={EDP Sciences},
   author={Bevacqua, Davide and Saracco, Paolo and Boecker, Alina and D’Ago, Giuseppe and De Lucia, Gabriella and De Propris, Roberto and La Barbera, Francesco and Pasquali, Anna and Spiniello, Chiara and Tortora, Crescenzo},
   year={2024},
   month=oct, pages={A150} }

@ARTICLE{Salpeter_1955,
       author = {{Salpeter}, Edwin E.},
        title = "{The Luminosity Function and Stellar Evolution.}",
      journal = {\apj},
         year = 1955,
        month = jan,
       volume = {121},
        pages = {161},
          doi = {10.1086/145971},
       adsurl = {https://ui.adsabs.harvard.edu/abs/1955ApJ...121..161S}
}

@ARTICLE{2012MNRAS.422..449B,
       author = {{Barden}, Marco and {H{\"a}u{\ss}ler}, Boris and {Peng}, Chien Y. and {McIntosh}, Daniel H. and {Guo}, Yicheng},
        title = "{GALAPAGOS: from pixels to parameters}",
      journal = {\mnras},
         year = 2012,
        month = may,
       volume = {422},
       number = {1},
        pages = {449-468},
          doi = {10.1111/j.1365-2966.2012.20619.x},
archivePrefix = {arXiv},
       eprint = {1203.1831},
 primaryClass = {astro-ph.IM},
       adsurl = {https://ui.adsabs.harvard.edu/abs/2012MNRAS.422..449B}
}

@ARTICLE{2022A&A...664A..92H,
       author = {{H{\"a}u{\ss}ler}, Boris and {Vika}, Marina and {Bamford}, Steven P. and {Johnston}, Evelyn J. and {Brough}, Sarah and {Casura}, Sarah and {Holwerda}, Benne W. and {Kelvin}, Lee S. and {Popescu}, Cristina},
        title = "{GALAPAGOS-2/GALFITM/GAMA - Multi-wavelength measurement of galaxy structure: Separating the properties of spheroid and disk components in modern surveys}",
      journal = {\aap},
         year = 2022,
        month = aug,
       volume = {664},
          eid = {A92},
        pages = {A92},
          doi = {10.1051/0004-6361/202142935},
archivePrefix = {arXiv},
       eprint = {2204.05907},
 primaryClass = {astro-ph.GA},
       adsurl = {https://ui.adsabs.harvard.edu/abs/2022A&A...664A..92H}
}

@ARTICLE{2013MNRAS.430..330H,
       author = {{H{\"a}u{\ss}ler}, Boris and {Bamford}, Steven P. and {Vika}, Marina and {Rojas}, Alex L. and {Barden}, Marco and {Kelvin}, Lee S. and {Alpaslan}, Mehmet and {Robotham}, Aaron S.~G. and {Driver}, Simon P. and {Baldry}, I.~K. and {Brough}, Sarah and {Hopkins}, Andrew M. and {Liske}, Jochen and {Nichol}, Robert C. and {Popescu}, Cristina C. and {Tuffs}, Richard J.},
        title = "{MegaMorph - multiwavelength measurement of galaxy structure: complete S{\'e}rsic profile information from modern surveys}",
      journal = {\mnras},
         year = 2013,
        month = mar,
       volume = {430},
       number = {1},
        pages = {330-369},
          doi = {10.1093/mnras/sts633},
archivePrefix = {arXiv},
       eprint = {1212.3332},
 primaryClass = {astro-ph.CO},
       adsurl = {https://ui.adsabs.harvard.edu/abs/2013MNRAS.430..330H}
}

@ARTICLE{2002AJ....124..266P,
       author = {{Peng}, Chien Y. and {Ho}, Luis C. and {Impey}, Chris D. and {Rix}, Hans-Walter},
        title = "{Detailed Structural Decomposition of Galaxy Images}",
      journal = {\aj},
         year = 2002,
        month = jul,
       volume = {124},
       number = {1},
        pages = {266-293},
          doi = {10.1086/340952},
archivePrefix = {arXiv},
       eprint = {astro-ph/0204182},
 primaryClass = {astro-ph},
       adsurl = {https://ui.adsabs.harvard.edu/abs/2002AJ....124..266P}
}

@ARTICLE{1963BAAA....6...41S,
       author = {{S{\'e}rsic}, J.~L.},
        title = "{Influence of the atmospheric and instrumental dispersion on the brightness distribution in a galaxy}",
      journal = {Boletin de la Asociacion Argentina de Astronomia La Plata Argentina},
         year = 1963,
        month = feb,
       volume = {6},
        pages = {41-43},
       adsurl = {https://ui.adsabs.harvard.edu/abs/1963BAAA....6...41S}
}

@ARTICLE{Ragone-Figueroa_2018,
       author = {{Ragone-Figueroa}, C. and {Granato}, G.~L. and {Ferraro}, M.~E. and {Murante}, G. and {Biffi}, V. and {Borgani}, S. and {Planelles}, S. and {Rasia}, E.},
        title = "{BCG mass evolution in cosmological hydro-simulations}",
      journal = {\mnras},
         year = 2018,
        month = sep,
       volume = {479},
       number = {1},
        pages = {1125-1136},
          doi = {10.1093/mnras/sty1639},
archivePrefix = {arXiv},
       eprint = {1803.08049},
 primaryClass = {astro-ph.GA},
       adsurl = {https://ui.adsabs.harvard.edu/abs/2018MNRAS.479.1125R}
}

@ARTICLE{Best_2007,
       author = {{Best}, P.~N. and {von der Linden}, A. and {Kauffmann}, G. and {Heckman}, T.~M. and {Kaiser}, C.~R.},
        title = "{On the prevalence of radio-loud active galactic nuclei in brightest cluster galaxies: implications for AGN heating of cooling flows}",
      journal = {\mnras},
         year = 2007,
        month = aug,
       volume = {379},
       number = {3},
        pages = {894-908},
          doi = {10.1111/j.1365-2966.2007.11937.x},
archivePrefix = {arXiv},
       eprint = {astro-ph/0611197},
 primaryClass = {astro-ph},
       adsurl = {https://ui.adsabs.harvard.edu/abs/2007MNRAS.379..894B},
}

@article{Von_Der_Linden_2007,
   title={How special are brightest group and cluster galaxies?},
   volume={379},
   ISSN={0035-8711},
   url={http://dx.doi.org/10.1111/j.1365-2966.2007.11940.x},
   DOI={10.1111/j.1365-2966.2007.11940.x},
   number={3},
   journal={Monthly Notices of the Royal Astronomical Society},
   publisher={Oxford University Press (OUP)},
   author={Von Der Linden, Anja and Best, Philip N. and Kauffmann, Guinevere and White, Simon D. M.},
   year={2007},
   month=jul, pages={867–893} }

@ARTICLE{Bautz_Morgan_1970,
       author = {{Bautz}, L.~P. and {Morgan}, W.~W.},
        title = "{On the Classification of the Forms of Clusters of Galaxies}",
      journal = {\apjl},
         year = 1970,
        month = dec,
       volume = {162},
        pages = {L149},
          doi = {10.1086/180643},
       adsurl = {https://ui.adsabs.harvard.edu/abs/1970ApJ...162L.149B}
}

@misc{ben_johnson_2024_12447779,
	author = {Ben Johnson and Dan Foreman-Mackey and Jonathan Sick and Joel Leja and Mike Walmsley and Erik Tollerud and Henry Leung and Spencer Scott and Minjung Park},
	doi = {10.5281/zenodo.12447779},
	month = jun,
	publisher = {Zenodo},
	title = {dfm/python-fsps: v0.4.7},
	url = {https://doi.org/10.5281/zenodo.12447779},
	version = {v0.4.7},
	year = 2024}

@ARTICLE{Conroy_2009_FSPS,
       author = {{Conroy}, Charlie and {Gunn}, James E. and {White}, Martin},
        title = "{The Propagation of Uncertainties in Stellar Population Synthesis Modeling. I. The Relevance of Uncertain Aspects of Stellar Evolution and the Initial Mass Function to the Derived Physical Properties of Galaxies}",
      journal = {\apj},
         year = 2009,
        month = jul,
       volume = {699},
       number = {1},
        pages = {486-506},
          doi = {10.1088/0004-637X/699/1/486},
archivePrefix = {arXiv},
       eprint = {0809.4261},
 primaryClass = {astro-ph},
       adsurl = {https://ui.adsabs.harvard.edu/abs/2009ApJ...699..486C}
}

@ARTICLE{Conroy_2010_FSPS,
       author = {{Conroy}, Charlie and {Gunn}, James E.},
        title = "{The Propagation of Uncertainties in Stellar Population Synthesis Modeling. III. Model Calibration, Comparison, and Evaluation}",
      journal = {\apj},
         year = 2010,
        month = apr,
       volume = {712},
       number = {2},
        pages = {833-857},
          doi = {10.1088/0004-637X/712/2/833},
archivePrefix = {arXiv},
       eprint = {0911.3151},
 primaryClass = {astro-ph.CO},
       adsurl = {https://ui.adsabs.harvard.edu/abs/2010ApJ...712..833C}}

@ARTICLE{Hodges_1958,
       author = {{Hodges}, J.~L.},
        title = "{The significance probability of the smirnov two-sample test}",
      journal = {Arkiv for Matematik},
         year = 1958,
        month = jan,
       volume = {3},
       number = {5},
        pages = {469-486},
          doi = {10.1007/BF02589501},
       adsurl = {https://ui.adsabs.harvard.edu/abs/1958ArM.....3..469H}
}

@article {Smirnoff_1939,
    AUTHOR = {Smirnoff, N.},
     TITLE = {On the estimation of the discrepancy between empirical curves
              of distribution for two independent samples},
   JOURNAL = {Bull. Math. Univ. Moscou},
  FJOURNAL = {Bull. Math. Univ. Moscou},
    VOLUME = {2},
      YEAR = {1939},
    NUMBER = {2},
     PAGES = {16},
   MRCLASS = {62.0X},
  MRNUMBER = {2062},
MRREVIEWER = {M.\ Kac},
}

@ARTICLE{Butcher_Oemler_1984,
       author = {{Butcher}, H. and {Oemler}, Jr., A.},
        title = "{The evolution of galaxies in clusters. V. A study of populations since Z 0.5.}",
      journal = {\apj},
         year = 1984,
        month = oct,
       volume = {285},
        pages = {426-438},
          doi = {10.1086/162519},
       adsurl = {https://ui.adsabs.harvard.edu/abs/1984ApJ...285..426B}
}

@ARTICLE{Thomas_2010,
       author = {{Thomas}, Daniel and {Maraston}, Claudia and {Schawinski}, Kevin and {Sarzi}, Marc and {Silk}, Joseph},
        title = "{Environment and self-regulation in galaxy formation}",
      journal = {\mnras},
         year = 2010,
        month = jun,
       volume = {404},
       number = {4},
        pages = {1775-1789},
          doi = {10.1111/j.1365-2966.2010.16427.x},
archivePrefix = {arXiv},
       eprint = {0912.0259},
 primaryClass = {astro-ph.CO},
       adsurl = {https://ui.adsabs.harvard.edu/abs/2010MNRAS.404.1775T}
}

@ARTICLE{Laporte_2013,
       author = {{Laporte}, Chervin F.~P. and {White}, Simon D.~M. and {Naab}, Thorsten and {Gao}, Liang},
        title = "{The growth in size and mass of cluster galaxies since z = 2}",
      journal = {\mnras},
         year = 2013,
        month = oct,
       volume = {435},
       number = {2},
        pages = {901-909},
          doi = {10.1093/mnras/stt912},
archivePrefix = {arXiv},
       eprint = {1301.5319},
 primaryClass = {astro-ph.CO},
       adsurl = {https://ui.adsabs.harvard.edu/abs/2013MNRAS.435..901L}
}

@ARTICLE{Whiley_2008,
       author = {{Whiley}, I.~M. and {Arag{\'o}n-Salamanca}, A. and {De Lucia}, G. and {von der Linden}, A. and {Bamford}, S.~P. and {Best}, P. and {Bremer}, M.~N. and {Jablonka}, P. and {Johnson}, O. and {Milvang-Jensen}, B. and {Noll}, S. and {Poggianti}, B.~M. and {Rudnick}, G. and {Saglia}, R. and {White}, S. and {Zaritsky}, D.},
        title = "{The evolution of the brightest cluster galaxies since z \raisebox{-0.5ex}\textasciitilde 1 from the ESO Distant Cluster Survey (EDisCS)}",
      journal = {\mnras},
         year = 2008,
        month = jul,
       volume = {387},
       number = {3},
        pages = {1253-1263},
          doi = {10.1111/j.1365-2966.2008.13324.x},
archivePrefix = {arXiv},
       eprint = {0804.2152},
 primaryClass = {astro-ph},

}

@ARTICLE{Zhang_2016,
       author = {{Zhang}, Y. and {Miller}, C. and {McKay}, T. and {Rooney}, P. and {Evrard}, A.~E. and {Romer}, A.~K. and {Perfecto}, R. and {Song}, J. and {Desai}, S. and {Mohr}, J. and {Wilcox}, H. and {Bermeo-Hernandez}, A. and {Jeltema}, T. and {Hollowood}, D. and {Bacon}, D. and {Capozzi}, D. and {Collins}, C. and {Das}, R. and {Gerdes}, D. and {Hennig}, C. and {Hilton}, M. and {Hoyle}, B. and {Kay}, S. and {Liddle}, A. and {Mann}, R.~G. and {Mehrtens}, N. and {Nichol}, R.~C. and {Papovich}, C. and {Sahl{\'e}n}, M. and {Soares-Santos}, M. and {Stott}, J. and {Viana}, P.~T. and {Abbott}, T. and {Abdalla}, F.~B. and {Banerji}, M. and {Bauer}, A.~H. and {Benoit-L{\'e}vy}, A. and {Bertin}, E. and {Brooks}, D. and {Buckley-Geer}, E. and {Burke}, D.~L. and {Carnero Rosell}, A. and {Castander}, F.~J. and {Diehl}, H.~T. and {Doel}, P. and {Cunha}, C.~E. and {Eifler}, T.~F. and {Fausti Neto}, A. and {Fernandez}, E. and {Flaugher}, B. and {Fosalba}, P. and {Frieman}, J. and {Gaztanaga}, E. and {Gruen}, D. and {Gruendl}, R.~A. and {Honscheid}, K. and {James}, D. and {Kuehn}, K. and {Kuropatkin}, N. and {Lahav}, O. and {Maia}, M.~A.~G. and {Makler}, M. and {Marshall}, J.~L. and {Martini}, Paul and {Miquel}, R. and {Ogando}, R. and {Plazas}, A.~A. and {Roodman}, A. and {Rykoff}, E.~S. and {Sako}, M. and {Sanchez}, E. and {Scarpine}, V. and {Schubnell}, M. and {Sevilla}, I. and {Smith}, R.~C. and {Sobreira}, F. and {Suchyta}, E. and {Swanson}, M.~E.~C. and {Tarle}, G. and {Thaler}, J. and {Tucker}, D. and {Vikram}, V. and {da Costa}, L.~N.},
        title = "{Galaxies in X-Ray Selected Clusters and Groups in Dark Energy Survey Data. I. Stellar Mass Growth of Bright Central Galaxies since z\raisebox{-0.5ex}\textasciitilde1.2}",
      journal = {\apj},
         year = 2016,
        month = jan,
       volume = {816},
       number = {2},
          eid = {98},
        pages = {98},
          doi = {10.3847/0004-637X/816/2/98},
archivePrefix = {arXiv},
       eprint = {1504.02983},
 primaryClass = {astro-ph.GA},
       adsurl = {https://ui.adsabs.harvard.edu/abs/2016ApJ...816...98Z}
}

@ARTICLE{Collins_Mann_1998,
       author = {{Collins}, C.~A. and {Mann}, R.~G.},
        title = "{The K-band Hubble diagram for brightest cluster galaxies in X-ray clusters}",
      journal = {\mnras},
         year = 1998,
        month = jun,
       volume = {297},
       number = {1},
        pages = {128-142},
          doi = {10.1046/j.1365-8711.1998.01482.x},
archivePrefix = {arXiv},
       eprint = {astro-ph/9712104},
 primaryClass = {astro-ph},
       adsurl = {https://ui.adsabs.harvard.edu/abs/1998MNRAS.297..128C}
}

@article{Collins_2009,
   title={Early assembly of the most massive galaxies},
   volume={458},
   ISSN={1476-4687},
   url={http://dx.doi.org/10.1038/nature07865},
   DOI={10.1038/nature07865},
   number={7238},
   journal={Nature},
   publisher={Springer Science and Business Media LLC},
   author={Collins, Chris A. and Stott, John P. and Hilton, Matt and Kay, Scott T. and Stanford, S. Adam and Davidson, Michael and Hosmer, Mark and Hoyle, Ben and Liddle, Andrew and Lloyd-Davies, Ed and Mann, Robert G. and Mehrtens, Nicola and Miller, Christopher J. and Nichol, Robert C. and Romer, A. Kathy and Sahlén, Martin and Viana, Pedro T. P. and West, Michael J.},
   year={2009},
   month=apr, pages={603–606} }

@ARTICLE{Stott_2010,
       author = {{Stott}, J.~P. and {Collins}, C.~A. and {Sahl{\'e}n}, M. and {Hilton}, M. and {Lloyd-Davies}, E. and {Capozzi}, D. and {Hosmer}, M. and {Liddle}, A.~R. and {Mehrtens}, N. and {Miller}, C.~J. and {Romer}, A.~K. and {Stanford}, S.~A. and {Viana}, P.~T.~P. and {Davidson}, M. and {Hoyle}, B. and {Kay}, S.~T. and {Nichol}, R.~C.},
        title = "{The XMM Cluster Survey: The Build-up of Stellar Mass in Brightest Cluster Galaxies at High Redshift}",
      journal = {\apj},
         year = 2010,
        month = jul,
       volume = {718},
       number = {1},
        pages = {23-30},
          doi = {10.1088/0004-637X/718/1/23},
archivePrefix = {arXiv},
       eprint = {1005.4681},
 primaryClass = {astro-ph.CO},
       adsurl = {https://ui.adsabs.harvard.edu/abs/2010ApJ...718...23S}
}

@ARTICLE{Lin_2013,
       author = {{Lin}, Yen-Ting and {Brodwin}, Mark and {Gonzalez}, Anthony H. and {Bode}, Paul and {Eisenhardt}, Peter R.~M. and {Stanford}, S.~A. and {Vikhlinin}, Alexey},
        title = "{The Stellar Mass Growth of Brightest Cluster Galaxies in the IRAC Shallow Cluster Survey}",
      journal = {\apj},
         year = 2013,
        month = jul,
       volume = {771},
       number = {1},
          eid = {61},
        pages = {61},
          doi = {10.1088/0004-637X/771/1/61},
archivePrefix = {arXiv},
       eprint = {1305.5254},
 primaryClass = {astro-ph.CO},
       adsurl = {https://ui.adsabs.harvard.edu/abs/2013ApJ...771...61L}
}

@ARTICLE{Cerulo_2019,
       author = {{Cerulo}, P. and {Orellana}, G.~A. and {Covone}, G.},
        title = "{The evolution of brightest cluster galaxies in the nearby Universe - I. Colours and stellar masses from the Sloan Digital Sky Survey and Wide Infrared Survey Explorer}",
      journal = {\mnras},
         year = 2019,
        month = aug,
       volume = {487},
       number = {3},
        pages = {3759-3775},
          doi = {10.1093/mnras/stz1495},
archivePrefix = {arXiv},
       eprint = {1905.12117},
 primaryClass = {astro-ph.GA},
       adsurl = {https://ui.adsabs.harvard.edu/abs/2019MNRAS.487.3759C}
}

@ARTICLE{Moretti_2015,
       author = {{Moretti}, A. and {Bettoni}, D. and {Poggianti}, B.~M. and {Fasano}, G. and {Varela}, J. and {D'Onofrio}, M. and {Vulcani}, B. and {Cava}, A. and {Fritz}, J. and {Couch}, W.~J. and {Moles}, M. and {Kj{\ae}rgaard}, P.},
        title = "{Galaxy luminosity functions in WINGS clusters}",
      journal = {\aap},
         year = 2015,
        month = sep,
       volume = {581},
          eid = {A11},
        pages = {A11},
          doi = {10.1051/0004-6361/201526080},
archivePrefix = {arXiv},
       eprint = {1506.01201},
 primaryClass = {astro-ph.GA},
       adsurl = {https://ui.adsabs.harvard.edu/abs/2015A&A...581A..11M}
}

@INPROCEEDINGS{Popesso_2005_conf,
       author = {{Popesso}, P. and {Biviano}, A. and {B{\"o}hringer}, H. and {Romaniello}, M.},
        title = "{RASS-SDSS galaxy cluster survey. A ubiquitous dwarf galaxy population in clusters}",
    booktitle = {IAU Colloquium 198: Near-fields cosmology with dwarf elliptical galaxies},
         year = 2005,
       editor = {{Jerjen}, H. and {Binggeli}, B.},
        month = jan,
        pages = {346-350},
          doi = {10.1017/S1743921305004035},
archivePrefix = {arXiv},
       eprint = {astro-ph/0506201},
 primaryClass = {astro-ph},
       adsurl = {https://ui.adsabs.harvard.edu/abs/2005nfcd.conf..346P}
}

@ARTICLE{Popesso_2005,
       author = {{Popesso}, P. and {B{\"o}hringer}, H. and {Romaniello}, M. and {Voges}, W.},
        title = "{RASS-SDSS galaxy cluster survey. II. A unified picture of the cluster luminosity function}",
      journal = {\aap},
         year = 2005,
        month = apr,
       volume = {433},
       number = {2},
        pages = {415-429},
          doi = {10.1051/0004-6361:20041870},
archivePrefix = {arXiv},
       eprint = {astro-ph/0410011},
 primaryClass = {astro-ph},
       adsurl = {https://ui.adsabs.harvard.edu/abs/2005A&A...433..415P}
}

@ARTICLE{Wen_Han_2015,
       author = {{Wen}, Z.~L. and {Han}, J.~L.},
        title = "{Dependence of the bright end of composite galaxy luminosity functions on cluster dynamical states}",
      journal = {\mnras},
         year = 2015,
        month = mar,
       volume = {448},
       number = {1},
        pages = {2-8},
          doi = {10.1093/mnras/stu2722},
archivePrefix = {arXiv},
       eprint = {1412.6865},
 primaryClass = {astro-ph.GA},
       adsurl = {https://ui.adsabs.harvard.edu/abs/2015MNRAS.448....2W}
}

@ARTICLE{Goto_2004,
       author = {{Goto}, Tomotsugu and {Yagi}, Masafumi and {Tanaka}, Masayuki and {Okamura}, Sadanori},
        title = "{Evolution of the colour-radius and morphology-radius relations in SDSS galaxy clusters}",
      journal = {\mnras},
         year = 2004,
        month = feb,
       volume = {348},
       number = {2},
        pages = {515-518},
          doi = {10.1111/j.1365-2966.2004.07324.x},
archivePrefix = {arXiv},
       eprint = {astro-ph/0312040},
 primaryClass = {astro-ph},
       adsurl = {https://ui.adsabs.harvard.edu/abs/2004MNRAS.348..515G}
}

@ARTICLE{Andreon_Berg_2012,
       author = {{Andreon}, S. and {Berg{\'e}}, J.},
        title = "{Richness-mass relation self-calibration for galaxy clusters}",
      journal = {\aap},
         year = 2012,
        month = nov,
       volume = {547},
          eid = {A117},
        pages = {A117},
          doi = {10.1051/0004-6361/201220115},
archivePrefix = {arXiv},
       eprint = {1209.5938},
 primaryClass = {astro-ph.CO},
       adsurl = {https://ui.adsabs.harvard.edu/abs/2012A&A...547A.117A}
}

@ARTICLE{Chiu_2022,
       author = {{Chiu}, I.-Non and {Ghirardini}, Vittorio and {Liu}, Ang and {Grandis}, Sebastian and {Bulbul}, Esra and {Bahar}, Y. Emre and {Comparat}, Johan and {Bocquet}, Sebastian and {Clerc}, Nicolas and {Klein}, Matthias and {Liu}, Teng and {Li}, Xiangchong and {Miyatake}, Hironao and {Mohr}, Joseph and {More}, Surhud and {Oguri}, Masamune and {Okabe}, Nobuhiro and {Pacaud}, Florian and {Ramos-Ceja}, Miriam E. and {Reiprich}, Thomas H. and {Schrabback}, Tim and {Umetsu}, Keiichi},
        title = "{The eROSITA Final Equatorial-Depth Survey (eFEDS). X-ray observable-to-mass-and-redshift relations of galaxy clusters and groups with weak-lensing mass calibration from the Hyper Suprime-Cam Subaru Strategic Program survey}",
      journal = {\aap},
         year = 2022,
        month = may,
       volume = {661},
          eid = {A11},
        pages = {A11},
          doi = {10.1051/0004-6361/202141755},
archivePrefix = {arXiv},
       eprint = {2107.05652},
 primaryClass = {astro-ph.CO},
       adsurl = {https://ui.adsabs.harvard.edu/abs/2022A&A...661A..11C}
}

@article{Brown_2007,
   title={The Evolving Luminosity Function of Red Galaxies},
   volume={654},
   ISSN={1538-4357},
   url={http://dx.doi.org/10.1086/509652},
   DOI={10.1086/509652},
   number={2},
   journal={The Astrophysical Journal},
   publisher={American Astronomical Society},
   author={Brown, Michael J. I. and Dey, Arjun and Jannuzi, Buell T. and Brand, Kate and Benson, Andrew J. and Brodwin, Mark and Croton, Darren J. and Eisenhardt, Peter R.},
   year={2007},
   month=jan, pages={858–877} }

@ARTICLE{Bell_2004,
       author = {{Bell}, Eric F. and {Wolf}, Christian and {Meisenheimer}, Klaus and {Rix}, Hans-Walter and {Borch}, Andrea and {Dye}, Simon and {Kleinheinrich}, Martina and {Wisotzki}, Lutz and {McIntosh}, Daniel H.},
        title = "{Nearly 5000 Distant Early-Type Galaxies in COMBO-17: A Red Sequence and Its Evolution since z\raisebox{-0.5ex}\textasciitilde1}",
      journal = {\apj},
         year = 2004,
        month = jun,
       volume = {608},
       number = {2},
        pages = {752-767},
          doi = {10.1086/420778},
archivePrefix = {arXiv},
       eprint = {astro-ph/0303394},
 primaryClass = {astro-ph},
       adsurl = {https://ui.adsabs.harvard.edu/abs/2004ApJ...608..752B}
}

@ARTICLE{Bell_2008,
       author = {{Brown}, Michael J.~I. and {Zheng}, Zheng and {White}, Martin and {Dey}, Arjun and {Jannuzi}, Buell T. and {Benson}, Andrew J. and {Brand}, Kate and {Brodwin}, Mark and {Croton}, Darren J.},
        title = "{Red Galaxy Growth and the Halo Occupation Distribution}",
      journal = {\apj},
         year = 2008,
        month = aug,
       volume = {682},
       number = {2},
        pages = {937-963},
          doi = {10.1086/589538},
archivePrefix = {arXiv},
       eprint = {0804.2293},
 primaryClass = {astro-ph},
       adsurl = {https://ui.adsabs.harvard.edu/abs/2008ApJ...682..937B}
}

@ARTICLE{Rudnick_2009,
       author = {{Rudnick}, Gregory and {von der Linden}, Anja and {Pell{\'o}}, Roser and {Arag{\'o}n-Salamanca}, Alfonso and {Marchesini}, Danilo and {Clowe}, Douglas and {De Lucia}, Gabriella and {Halliday}, Claire and {Jablonka}, Pascale and {Milvang-Jensen}, Bo and {Poggianti}, Bianca and {Saglia}, Roberto and {Simard}, Luc and {White}, Simon and {Zaritsky}, Dennis},
        title = "{The Rest-frame Optical Luminosity Function of Cluster Galaxies at z < 0.8 and the Assembly of the Cluster Red Sequence}",
      journal = {\apj},
         year = 2009,
        month = aug,
       volume = {700},
       number = {2},
        pages = {1559-1588},
          doi = {10.1088/0004-637X/700/2/1559},
archivePrefix = {arXiv},
       eprint = {0905.4503},
 primaryClass = {astro-ph.CO},
       adsurl = {https://ui.adsabs.harvard.edu/abs/2009ApJ...700.1559R}
}

@article{Stott_2009,
   title={The evolution of the red sequence slope in massive galaxy clusters},
   volume={394},
   ISSN={1365-2966},
   url={http://dx.doi.org/10.1111/j.1365-2966.2009.14477.x},
   DOI={10.1111/j.1365-2966.2009.14477.x},
   number={4},
   journal={Monthly Notices of the Royal Astronomical Society},
   publisher={Oxford University Press (OUP)},
   author={Stott, J. P. and Pimbblet, K. A. and Edge, A. C. and Smith, G. P. and Wardlow, J. L.},
   year={2009},
   month=apr, pages={2098–2108} }

@article{De_Lucia_2004,
   title={Observing the build–up of the colour-magnitude relation at redshift $\sim0.8$},
   volume={2004},
   ISSN={1743-9221},
   url={http://dx.doi.org/10.1017/S1743921304001024},
   DOI={10.1017/s1743921304001024},
   number={IAUC195},
   journal={Proceedings of the International Astronomical Union},
   publisher={Cambridge University Press (CUP)},
   author={De Lucia, G. and Poggianti, B. M. and Aragón-Salamanca, A. and Clowe, D. and Halliday, C. and Jablonka, P. and Milvang-Jensen, B. and Pelló, R. and Poirier, S. and Rudnick, G. and Saglia, R. and Simard, L. and White, S. D. M.},
   year={2004},
   month=mar }

@article{De_Lucia_2007_RS,
   title={The build-up of the colour-magnitude relation in galaxy clusters since z  0.8},
   volume={374},
   ISSN={1365-2966},
   url={http://dx.doi.org/10.1111/j.1365-2966.2006.11199.x},
   DOI={10.1111/j.1365-2966.2006.11199.x},
   number={3},
   journal={Monthly Notices of the Royal Astronomical Society},
   publisher={Oxford University Press (OUP)},
   author={De Lucia, G. and Poggianti, B. M. and Aragon-Salamanca, A. and White, S. D. M. and Zaritsky, D. and Clowe, D. and Halliday, C. and Jablonka, P. and Von Der Linden, A. and Milvang-Jensen, B. and Pello, R. and Rudnick, G. and Saglia, R. P. and Simard, L.},
   year={2007},
   month=jan, pages={809–822} }

@ARTICLE{Strazzullo_2019,
       author = {{Strazzullo}, V. and {Pannella}, M. and {Mohr}, J.~J. and {Saro}, A. and {Ashby}, M.~L.~N. and {Bayliss}, M.~B. and {Bocquet}, S. and {Bulbul}, E. and {Khullar}, G. and {Mantz}, A.~B. and {Stanford}, S.~A. and {Benson}, B.~A. and {Bleem}, L.~E. and {Brodwin}, M. and {Canning}, R.~E.~A. and {Capasso}, R. and {Chiu}, I. and {Gonzalez}, A.~H. and {Gupta}, N. and {Hlavacek-Larrondo}, J. and {Klein}, M. and {McDonald}, M. and {Noordeh}, E. and {Rapetti}, D. and {Reichardt}, C.~L. and {Schrabback}, T. and {Sharon}, K. and {Stalder}, B.},
        title = "{Galaxy populations in the most distant SPT-SZ clusters. I. Environmental quenching in massive clusters at 1.4 {\ensuremath{\lesssim}} z {\ensuremath{\lesssim}} 1.7}",
      journal = {\aap},
         year = 2019,
        month = feb,
       volume = {622},
          eid = {A117},
        pages = {A117},
          doi = {10.1051/0004-6361/201833944},
archivePrefix = {arXiv},
       eprint = {1807.09768},
 primaryClass = {astro-ph.GA},
       adsurl = {https://ui.adsabs.harvard.edu/abs/2019A&A...622A.117S}
}

@article{Baldry_2004,
   title={Quantifying the Bimodal Color‐Magnitude Distribution of Galaxies},
   volume={600},
   ISSN={1538-4357},
   url={http://dx.doi.org/10.1086/380092},
   DOI={10.1086/380092},
   number={2},
   journal={The Astrophysical Journal},
   publisher={American Astronomical Society},
   author={Baldry, Ivan K. and Glazebrook, Karl and Brinkmann, Jon and Ivezić, Željko and Lupton, Robert H. and Nichol, Robert C. and Szalay, Alexander S.},
   year={2004},
   month=jan, pages={681–694} }

@article{Gavazzi_2010,
   title={A snapshot on galaxy evolution occurring in the Great Wall: the role of Nurture atz = 0},
   volume={517},
   ISSN={1432-0746},
   url={http://dx.doi.org/10.1051/0004-6361/201014153},
   DOI={10.1051/0004-6361/201014153},
   journal={Astronomy and Astrophysics},
   publisher={EDP Sciences},
   author={Gavazzi, G. and Fumagalli, M. and Cucciati, O. and Boselli, A.},
   year={2010},
   month=jul, pages={A73} }

@ARTICLE{Kodama_Arimoto_1997,
       author = {{Kodama}, T. and {Arimoto}, N.},
        title = "{Origin of the colour-magnitude relation of elliptical galaxies.}",
      journal = {\aap},
         year = 1997,
        month = apr,
       volume = {320},
        pages = {41-53},
          doi = {10.48550/arXiv.astro-ph/9609160},
archivePrefix = {arXiv},
       eprint = {astro-ph/9609160},
 primaryClass = {astro-ph},
       adsurl = {https://ui.adsabs.harvard.edu/abs/1997A&A...320...41K}
}

@article{Stott_2011,
   title={Little change in the sizes of the most massive galaxies since z = 1: The sizes of brightest cluster galaxies},
   volume={414},
   ISSN={0035-8711},
   url={http://dx.doi.org/10.1111/j.1365-2966.2011.18404.x},
   DOI={10.1111/j.1365-2966.2011.18404.x},
   number={1},
   journal={Monthly Notices of the Royal Astronomical Society},
   publisher={Oxford University Press (OUP)},
   author={Stott, J. P. and Collins, C. A. and Burke, C. and Hamilton-Morris, V. and Smith, G. P.},
   year={2011},
   month=mar, pages={445–457} }

@article{Aragon_Salamanca_1998,
   title={The K-band Hubble diagram for the brightest cluster galaxies: a test of hierarchical galaxy formation models},
   volume={297},
   ISSN={1365-2966},
   url={http://dx.doi.org/10.1046/j.1365-8711.1998.01495.x},
   DOI={10.1046/j.1365-8711.1998.01495.x},
   number={2},
   journal={Monthly Notices of the Royal Astronomical Society},
   publisher={Oxford University Press (OUP)},
   author={Aragon-Salamanca, A. and Baugh, C. M. and Kauffmann, G.},
   year={1998},
   month=jun, pages={427–434} }

@article{Santos_2012,
   title={DeepChandraobservation of the galaxy cluster WARPJ1415.1+3612 atz=1: An evolved cool-core cluster at high redshift},
   volume={539},
   ISSN={1432-0746},
   url={http://dx.doi.org/10.1051/0004-6361/201118162},
   DOI={10.1051/0004-6361/201118162},
   journal={Astronomy \& Astrophysics},
   publisher={EDP Sciences},
   author={Santos, J. S. and Tozzi, P. and Rosati, P. and Nonino, M. and Giovannini, G.},
   year={2012},
   month=mar, pages={A105} }

@article{Ding_2024,
   title={Miscentring of optical galaxy clusters based on Sunyaev–Zeldovich counterparts},
   volume={536},
   ISSN={1365-2966},
   url={http://dx.doi.org/10.1093/mnras/stae2601},
   DOI={10.1093/mnras/stae2601},
   number={1},
   journal={Monthly Notices of the Royal Astronomical Society},
   publisher={Oxford University Press (OUP)},
   author={Ding, Jupiter and Dalal, Roohi and Sunayama, Tomomi and Strauss, Michael A and Oguri, Masamune and Okabe, Nobuhiro and Hilton, Matt and Monteiro-Oliveira, Rogério and Sifón, Cristóbal and Staggs, Suzanne T},
   year={2024},
   month=nov, pages={572–591} }

@ARTICLE{2025arXiv250707093S,
       author = {{SDSS Collaboration} and {Adamane Pallathadka}, Gautham and {Aghakhanloo}, Mojgan and {Aird}, James and {Almeida}, Andr{\'e}s and {Amrita}, Singh and {Anders}, Friedrich and {Anderson}, Scott F. and {Arseneau}, Stefan and {Gonz{\'a}lez Avila}, Consuelo and {Aviram}, Shir and {Aydar}, Catarina and {Badenes}, Carles and {Barrera-Ballesteros}, Jorge K. and {Bauer}, Franz E. and {Behmard}, Aida and {Berg}, Michelle and {Besser}, F. and {Moni Bidin}, Christian and {Bizyaev}, Dmitry and {Blanc}, Guillermo and {Blanton}, Michael R. and {Bovy}, Jo and {Brandt}, William Nielsen and {Brownstein}, Joel R. and {Buchner}, Johannes and {Bulbul}, Esra and {Burchett}, Joseph N. and {Carigi}, Leticia and {Carlberg}, Joleen K. and {Casey}, Andrew R. and {Chakraborty}, Priyanka and {Chanam{\'e}}, Julio and {Chandra}, Vedant and {Chiappini}, Cristina and {Chilingarian}, Igor and {Comparat}, Johan and {Covey}, Kevin and {Crumpler}, Nicole and {Cunha}, Katia and {D'Onghia}, Elena and {Dai}, Xinyu and {Darling}, Jeremy and {Davis}, Megan and {De Lee}, Nathan and {Deacon}, Niall and {M{\'e}ndez Delgado}, Jos{\'e} Eduardo and {Demasi}, Sebastian and {Demianenko}, Mariia and {Demke}, Delvin and {Donor}, John and {Drory}, Niv and {Villa Durango}, Monica Alejandra and {Dwelly}, Tom and {Egorov}, Oleg and {Egorova}, Evgeniya and {El-Badry}, Kareem and {Eracleous}, Mike and {Fan}, Xiaohui and {Farr}, Emily and {Finkbeiner}, Douglas P. and {Fries}, Logan and {Frinchaboy}, Peter and {Gentile Fusillo}, Nicola Pietro and {Serrano F{\'e}lix}, Luis Daniel and {Gaensicke}, Boris and {Galligan}, Emma and {Garc{\'\i}a}, Pablo and {Gelfand}, Joseph and {Grabowski}, Katie and {Grebel}, Eva and {Green}, Paul J and {Greve}, Hannah and {Grier}, Catherine and {Griffith}, Emily and {Guetzoyan}, Paloma and {Gupta}, Pramod and {Hackshaw}, Zoe and {Hall}, Patrick B. and {Hawkins}, Keith and {Heged{\H{u}}s}, Viola and {Hekker}, Saskia and {Herbst}, T.~M. and {Hermes}, J.~J. and {Hern{\'a}ndez-Garc{\'\i}a}, Lorena and {Hiremath}, Pranavi and {Hogg}, David W and {Holtzman}, Jon and {Horne}, Keith and {Horta}, Danny and {Huang}, Yang and {Hutchinson}, Brian and {H{\"a}berle}, Maximilian and {Ibarra-Medel}, Hector Javier and {Ji}, Alexander P. and {Jofre}, Paula and {Johnson}, James W. and {Johnson}, Jennifer and {Johnston}, Evelyn J. and {Kaldor}, Mary and {Katkov}, Ivan and {Khalatyan}, Arman and {Khoperskov}, Sergey and {Klessen}, Ralf and {Kluge}, Matthias and {Koekemoer}, Anton M. and {Kollmeier}, Juna A. and {Kounkel}, Marina and {Kreckel}, Kathryn and {Krishnarao}, Dhanesh and {Krumpe}, Mirko and {Lacerna}, Ivan and {Laporte}, Chervin and {Lepine}, Sebastien and {Li}, Jing and {Liang}, Fu-Heng and {Limberg}, Guilherme and {Liu}, Xin and {Loebman}, Sarah and {Long}, Knox and {Lu}, Yuxi and {Lucey}, Madeline and {Lugo-Aranda}, Alejandra Z. and {Mart{\'\i}nez Martinez-Aldama}, Mary Loli and {McKinnon}, Kevin and {Medan}, Ilija and {Merloni}, Andrea and {Morrison}, Sean and {Myers}, Natalie and {M{\'e}sz{\'a}ros}, Szabolcs and {M{\"u}ller-Horn}, Johanna and {Nepal}, Samir and {Ness}, Melissa and {Nidever}, David and {Nitschelm}, Christian and {Oravetz}, Audrey and {Otto}, Jonah and {Pan}, Kaike and {P{\'e}rez Paolino}, Facundo and {Negrete Pe{\~n}aloza}, Castalia Alenka and {Pinsonneault}, Marc and {Taghizadeh Popp}, Manuchehr and {Price-Whelan}, Adrian and {Pulatova}, Nadiia and {Queiroz}, Anna Barbara and {Raddick}, Jordan and {Rankine}, Amy and {Rix}, Hans-Walter and {Rom{\'a}n-Z{\'u}{\~n}iga}, Carlos and {Fern{\'a}ndez Rosso}, Daniela and {Runnoe}, Jessie and {Mahmud Saad}, Serat and {Salvato}, Mara and {Sanchez}, Sebastian F. and {Sattler}, Natascha and {Saydjari}, Andrew and {Sayres}, Conor and {Schlaufman}, Kevin and {Schneider}, Donald P. and {Schwope}, Axel and {Seaton}, Lucas M. and {Seeburger}, Rhys and {Serna}, Javier and {Sharma}, Sanjib and {Shen}, Yue and {Sinha}, Amaya and {Sizemore}, Brian and {Sniegowska}, Marzena and {Song}, Yingyi and {Souto}, Diogo and {Stassun}, Keivan and {Steinmetz}, Matthias and {Stone}, Zachary and {Stone-Martinez}, Alexander and {Stringfellow}, Guy S. and {Mata S{\'a}nchez}, Aurora and {S{\'a}nchez-Gallego}, Jos{\'e} and {Tan}, Jonathan and {Tayar}, Jamie and {Thai}, Riley and {Thakar}, Ani and {Thibodeaux}, Pierre and {Ting}, Yuan-Sen and {Tkachenko}, Andrew and {Trakhtenbrot}, Benny and {Fernandez Trincado}, Jose G. and {Troup}, Nicholas and {Trump}, Jonathan R. and {Ulloa}, Natalie and {Van der Marel}, Roeland P. and {Vera}, Pablo and {Villanova}, Sandro and {Villase{\~n}or}, Jaime and {Wang}, Ji and {Way}, Zachary and {Weijmans}, Anne-Marie and {Wheeler}, Adam and {Wilson}, John C. and {Wofford}, Aida and {Wong}, Tony},
        title = "{The Nineteenth Data Release of the Sloan Digital Sky Survey}",
      journal = {arXiv e-prints},
         year = 2025,
        month = jul,
          eid = {arXiv:2507.07093},
        pages = {arXiv:2507.07093},
          doi = {10.48550/arXiv.2507.07093},
archivePrefix = {arXiv},
       eprint = {2507.07093},
 primaryClass = {astro-ph.GA},
       adsurl = {https://ui.adsabs.harvard.edu/abs/2025arXiv250707093S}
}

@article{Donzelli_2011,
   title={THE LUMINOSITY PROFILES OF BRIGHTEST CLUSTER GALAXIES},
   volume={195},
   ISSN={1538-4365},
   url={http://dx.doi.org/10.1088/0067-0049/195/2/15},
   DOI={10.1088/0067-0049/195/2/15},
   number={2},
   journal={The Astrophysical Journal Supplement Series},
   publisher={American Astronomical Society},
   author={Donzelli, C. J. and Muriel, H. and Madrid, J. P.},
   year={2011},
   month=jul, pages={15} }

@article{Calzadilla_2023,
   title={SPT-CL J2215−3537: A Massive Starburst at the Center of the Most Distant Relaxed Galaxy Cluster},
   volume={947},
   ISSN={1538-4357},
   url={http://dx.doi.org/10.3847/1538-4357/acc6c2},
   DOI={10.3847/1538-4357/acc6c2},
   number={2},
   journal={The Astrophysical Journal},
   publisher={American Astronomical Society},
   author={Calzadilla, Michael S. and Bleem, Lindsey E. and McDonald, Michael and Gladders, Michael D. and Mantz, Adam B. and Allen, Steven W. and Bayliss, Matthew B. and Eilers, Anna-Christina and Floyd, Benjamin and Hlavacek-Larrondo, Julie and Khullar, Gourav and Kim, Keunho J. and Mahler, Guillaume and Sharon, Keren and Somboonpanyakul, Taweewat and Stalder, Brian and Stark, Antony A.},
   year={2023},
   month=apr, pages={44} }

@ARTICLE{1982ApJ...255..382H,
       author = {{Hickson}, P.},
        title = "{Systematic properties of compact groups of galaxies.}",
      journal = {\apj},
         year = 1982,
        month = apr,
       volume = {255},
        pages = {382-391},
          doi = {10.1086/159838},
       adsurl = {https://ui.adsabs.harvard.edu/abs/1982ApJ...255..382H}
}

@ARTICLE{1989ApJS...70..687H,
       author = {{Hickson}, Paul and {Kindl}, Enrico and {Auman}, Jason R.},
        title = "{A Photometric Catalog of Compact Groups of Galaxies}",
      journal = {\apjs},
         year = 1989,
        month = aug,
       volume = {70},
        pages = {687},
          doi = {10.1086/191354},
       adsurl = {https://ui.adsabs.harvard.edu/abs/1989ApJS...70..687H}
}

@article{Lisenfeld_2017,
   title={The role of molecular gas in galaxy transition in compact groups},
   volume={607},
   ISSN={1432-0746},
   url={http://dx.doi.org/10.1051/0004-6361/201730898},
   DOI={10.1051/0004-6361/201730898},
   journal={Astronomy \& Astrophysics},
   publisher={EDP Sciences},
   author={Lisenfeld, U. and Alatalo, K. and Zucker, C. and Appleton, P. N. and Gallagher, S. and Guillard, P. and Johnson, K.},
   year={2017},
   month=nov, pages={A110} }

@ARTICLE{2011A&A...533A.142B,
       author = {{Bitsakis}, T. and {Charmandaris}, V. and {da Cunha}, E. and {D{\'\i}az-Santos}, T. and {Le Floc'h}, E. and {Magdis}, G.},
        title = "{A mid-IR study of Hickson compact groups. II. Multiwavelength analysis of the complete GALEX-Spitzer sample}",
      journal = {\aap},
         year = 2011,
        month = sep,
       volume = {533},
          eid = {A142},
        pages = {A142},
          doi = {10.1051/0004-6361/201117355},
archivePrefix = {arXiv},
       eprint = {1107.3418},
 primaryClass = {astro-ph.CO},
       adsurl = {https://ui.adsabs.harvard.edu/abs/2011A&A...533A.142B}
}

@ARTICLE{2012ApJ...752...12H,
       author = {{Harrison}, Craig D. and {Miller}, Christopher J. and {Richards}, Joseph W. and {Lloyd-Davies}, E.~J. and {Hoyle}, Ben and {Romer}, A. Kathy and {Mehrtens}, Nicola and {Hilton}, Matt and {Stott}, John P. and {Capozzi}, Diego and {Collins}, Chris A. and {Deadman}, Paul-James and {Liddle}, Andrew R. and {Sahl{\'e}n}, Martin and {Stanford}, S. Adam and {Viana}, Pedro T.~P.},
        title = "{The XMM Cluster Survey: The Stellar Mass Assembly of Fossil Galaxies}",
      journal = {\apj},
         year = 2012,
        month = jun,
       volume = {752},
       number = {1},
          eid = {12},
        pages = {12},
          doi = {10.1088/0004-637X/752/1/12},
archivePrefix = {arXiv},
       eprint = {1202.4450},
 primaryClass = {astro-ph.CO},
       adsurl = {https://ui.adsabs.harvard.edu/abs/2012ApJ...752...12H}
}

@article{Golden_Marx_2022,
   title={The Observed Evolution of the Stellar Mass–Halo Mass Relation for Brightest Central Galaxies},
   volume={928},
   ISSN={1538-4357},
   url={http://dx.doi.org/10.3847/1538-4357/ac4cb4},
   DOI={10.3847/1538-4357/ac4cb4},
   number={1},
   journal={The Astrophysical Journal},
   publisher={American Astronomical Society},
   author={Golden-Marx, Jesse B. and Miller, C. J. and Zhang, Y. and Ogando, R. L. C. and Palmese, A. and Abbott, T. M. C. and Aguena, M. and Allam, S. and Andrade-Oliveira, F. and Annis, J. and Bacon, D. and Bertin, E. and Brooks, D. and Buckley-Geer, E. and Carnero Rosell, A. and Carrasco Kind, M. and Castander, F. J. and Costanzi, M. and Crocce, M. and da Costa, L. N. and Pereira, M. E. S. and De Vicente, J. and Desai, S. and Diehl, H. T. and Doel, P. and Drlica-Wagner, A. and Everett, S. and Evrard, A. E. and Ferrero, I. and Flaugher, B. and Fosalba, P. and Frieman, J. and García-Bellido, J. and Gaztanaga, E. and Gerdes, D. W. and Gruen, D. and Gruendl, R. A. and Gschwend, J. and Gutierrez, G. and Hartley, W. G. and Hinton, S. R. and Hollowood, D. L. and Honscheid, K. and Hoyle, B. and James, D. J. and Jeltema, T. and Kim, A. G. and Krause, E. and Kuehn, K. and Kuropatkin, N. and Lahav, O. and Lima, M. and Maia, M. A. G. and Marshall, J. L. and Melchior, P. and Menanteau, F. and Miquel, R. and Mohr, J. J. and Morgan, R. and Paz-Chinchón, F. and Petravick, D. and Pieres, A. and Plazas Malagón, A. A. and Prat, J. and Romer, A. K. and Sanchez, E. and Santiago, B. and Scarpine, V. and Schubnell, M. and Serrano, S. and Sevilla-Noarbe, I. and Smith, M. and Soares-Santos, M. and Suchyta, E. and Tarle, G. and Varga, T. N.},
   year={2022},
   month=mar, pages={28} }

@ARTICLE{Golden_Marx_2025,
       author = {{Golden-Marx}, Jesse B. and {Zhang}, Y. and {Ogando}, R.~L.~C. and {Yanny}, B. and {da Silva Pereira}, M.~E. and {Hilton}, M. and {Aguena}, M. and {Allam}, S. and {Andrade-Oliveira}, F. and {Bacon}, D. and {Brooks}, D. and {Carnero Rosell}, A. and {Carretero}, J. and {Cheng}, T.-Y. and {da Costa}, L.~N. and {De Vicente}, J. and {Desai}, S. and {Doel}, P. and {Everett}, S. and {Ferrero}, I. and {Frieman}, J. and {Garc{\'\i}a-Bellido}, J. and {Gatti}, M. and {Giannini}, G. and {Gruen}, D. and {Gruendl}, R.~A. and {Gutierrez}, G. and {Hinton}, S.~R. and {Hollowood}, D.~L. and {Honscheid}, K. and {James}, D.~J. and {Kuehn}, K. and {Lee}, S. and {Mena-Fern{\'a}ndez}, J. and {Menanteau}, F. and {Miquel}, R. and {Mohr}, J. and {Palmese}, A. and {Pieres}, A. and {Plazas Malag{\'o}n}, A.~A. and {Samuroff}, S. and {Sanchez}, E. and {Schubnell}, M. and {Sevilla-Noarbe}, I. and {Smith}, M. and {Suchyta}, E. and {Tarle}, G. and {Vikram}, V. and {Walker}, A.~R. and {Weaverdyck}, N. and {Wiseman}, P.},
        title = "{The hierarchical growth of bright central galaxies and intracluster light as traced by the magnitude gap}",
      journal = {\mnras},
         year = 2025,
        month = apr,
        volume = {538},
       number = {2},
        pages = {622-638},
          doi = {10.1093/mnras/staf277},
archivePrefix = {arXiv},
       eprint = {2409.02184},
 primaryClass = {astro-ph.GA},
       adsurl = {https://ui.adsabs.harvard.edu/abs/2025MNRAS.538..622G}
}

@article{De_Propris_2003,
   title={The 2dF Galaxy Redshift Survey: the luminosity function of cluster galaxies},
   volume={342},
   ISSN={1365-2966},
   url={http://dx.doi.org/10.1046/j.1365-8711.2003.06510.x},
   DOI={10.1046/j.1365-8711.2003.06510.x},
   number={3},
   journal={Monthly Notices of the Royal Astronomical Society},
   publisher={Oxford University Press (OUP)},
   author={De Propris, Roberto and Colless, Matthew and Driver, Simon P. and Couch, Warrick and Peacock, John A. and Baldry, Ivan K. and Baugh, Carlton M. and Bland-Hawthorn, Joss and Bridges, Terry and Cannon, Russell and Cole, Shaun and Collins, Chris and Cross, Nicholas and Dalton, Gavin B. and Efstathiou, George and Ellis, Richard S. and Frenk, Carlos S. and Glazebrook, Karl and Hawkins, Edward and Jackson, Carole and Lahav, Ofer and Lewis, Ian and Lumsden, Stuart and Maddox, Steve and Madgwick, Darren S. and Norberg, Peder and Percival, Will and Peterson, Bruce and Sutherland, Will and Taylor, Keith},
   year={2003},
   month=jul, pages={725–737} }

@article{McNaught_Roberts_2014,
   title={Galaxy And Mass Assembly (GAMA): the dependence of the galaxy luminosity function on environment, redshift and colour},
   volume={445},
   ISSN={0035-8711},
   url={http://dx.doi.org/10.1093/mnras/stu1886},
   DOI={10.1093/mnras/stu1886},
   number={2},
   journal={Monthly Notices of the Royal Astronomical Society},
   publisher={Oxford University Press (OUP)},
   author={McNaught-Roberts, Tamsyn and Norberg, Peder and Baugh, Carlton and Lacey, Cedric and Loveday, J. and Peacock, J. and Baldry, I. and Bland-Hawthorn, J. and Brough, S. and Driver, Simon P. and Robotham, A. S. G. and Vázquez-Mata, J. A.},
   year={2014},
   month=oct, pages={2125–2145} }

@ARTICLE{2013A&A...550A..58V,
       author = {{Vulcani}, B. and {Poggianti}, B.~M. and {Oemler}, A. and {Dressler}, A. and {Arag{\'o}n-Salamanca}, A. and {De Lucia}, G. and {Moretti}, A. and {Gladders}, M. and {Abramson}, L. and {Halliday}, C.},
        title = "{The galaxy stellar mass function and its evolution with time show no dependence on global environment}",
      journal = {\aap},
         year = 2013,
        month = feb,
       volume = {550},
          eid = {A58},
        pages = {A58},
          doi = {10.1051/0004-6361/201118388},
archivePrefix = {arXiv},
       eprint = {1111.0830},
 primaryClass = {astro-ph.CO},
       adsurl = {https://ui.adsabs.harvard.edu/abs/2013A&A...550A..58V}
}

@article{Hatch_2009,
   title={The growth and assembly of a massive galaxy at},
   volume={395},
   ISSN={1365-2966},
   url={http://dx.doi.org/10.1111/j.1365-2966.2009.14525.x},
   DOI={10.1111/j.1365-2966.2009.14525.x},
   number={1},
   journal={Monthly Notices of the Royal Astronomical Society},
   publisher={Oxford University Press (OUP)},
   author={Hatch, N. A. and Overzier, R. A. and Kurk, J. D. and Miley, G. K. and Röttgering, H. J. A. and Zirm, A. W.},
   year={2009},
   month=may, pages={114–125} }

@article{Koyama_2012,
   title={Massive starburst galaxies in a z = 2.16 proto-cluster unveiled by panoramic Hα mapping},
   volume={428},
   ISSN={0035-8711},
   url={http://dx.doi.org/10.1093/mnras/sts133},
   DOI={10.1093/mnras/sts133},
   number={2},
   journal={Monthly Notices of the Royal Astronomical Society},
   publisher={Oxford University Press (OUP)},
   author={Koyama, Yusei and Kodama, Tadayuki and Tadaki, Ken-ichi and Hayashi, Masao and Tanaka, Masayuki and Smail, Ian and Tanaka, Ichi and Kurk, Jaron},
   year={2012},
   month=oct, pages={1551–1564} }

@ARTICLE{2018MNRAS.473.1977S,
       author = {{Shimakawa}, Rhythm and {Kodama}, Tadayuki and {Hayashi}, Masao and {Prochaska}, J. Xavier and {Tanaka}, Ichi and {Cai}, Zheng and {Suzuki}, Tomoko L. and {Tadaki}, Ken-ichi and {Koyama}, Yusei},
        title = "{MAHALO Deep Cluster Survey I. Accelerated and enhanced galaxy formation in the densest regions of a protocluster at z = 2.5}",
      journal = {\mnras},
         year = 2018,
        month = jan,
       volume = {473},
       number = {2},
        pages = {1977-1999},
          doi = {10.1093/mnras/stx2494},
archivePrefix = {arXiv},
       eprint = {1708.06369},
 primaryClass = {astro-ph.GA},
       adsurl = {https://ui.adsabs.harvard.edu/abs/2018MNRAS.473.1977S}
}

@BOOK{2012drzp.book.....G,
       author = {{Gonzaga}, S. and {Hack}, W. and {Fruchter}, A. and {Mack}, J.},
        title = "{The DrizzlePac Handbook}",
         year = 2012,
    publisher = {Space Telescope Science Institute},
       adsurl = {https://ui.adsabs.harvard.edu/abs/2012drzp.book.....G}
}

@INPROCEEDINGS{2021AAS...23821602H,
       author = {{Hoffmann}, S.~L. and {Mack}, J. and {Avila}, R. and {Martlin}, C. and {Cohen}, Y. and {Bajaj}, V.},
        title = "{New Drizzlepac Handbook Version 2.0 Released In Hdox: Updated Documentation For HST Image Analysis}",
    booktitle = {American Astronomical Society Meeting Abstracts},
         year = 2021,
       series = {American Astronomical Society Meeting Abstracts},
       volume = {53},
        month = jun,
          eid = {216.02},
        pages = {216.02},
       adsurl = {https://ui.adsabs.harvard.edu/abs/2021AAS...23821602H}
}

@MISC{2010wfc..rept....6P,
       author = {{Pirzkal}, N. and {Viana}, A. and {Rajan}, A.},
        title = "{The WFC3 IR 'Blobs''}",
 howpublished = {Instrument Science Report WFC3 2010-06, 16 pages},
         year = 2010,
        month = apr,
        pages = {6},
       adsurl = {https://ui.adsabs.harvard.edu/abs/2010wfc..rept....6P}
}

@ARTICLE{2024ApJ...976..169C,
       author = {{Calzadilla}, Michael S. and {McDonald}, Michael and {Benson}, Bradford A. and {Bleem}, Lindsey E. and {Croston}, Judith H. and {Donahue}, Megan and {Edge}, Alastair C. and {Floyd}, Benjamin and {Garmire}, Gordon P. and {Hlavacek-Larrondo}, Julie and {Huynh}, Minh T. and {Khullar}, Gourav and {Kraft}, Ralph P. and {McNamara}, Brian R. and {Noble}, Allison G. and {Romero}, Charles E. and {Ruppin}, Florian and {Somboonpanyakul}, Taweewat and {Voit}, G. Mark},
        title = "{The SPT-Chandra BCG Spectroscopic Survey. I. Evolution of the Entropy Threshold for ICM Cooling and AGN Feedback in Galaxy Clusters over the Last 10 Gyr}",
      journal = {\apj},
         year = 2024,
        month = dec,
       volume = {976},
       number = {2},
          eid = {169},
        pages = {169},
          doi = {10.3847/1538-4357/ad8916},
archivePrefix = {arXiv},
       eprint = {2311.00396},
 primaryClass = {astro-ph.GA},
       adsurl = {https://ui.adsabs.harvard.edu/abs/2024ApJ...976..169C}
}

@ARTICLE{2002AJ....124.1810S,
       author = {{Strauss}, Michael A. and {Weinberg}, David H. and {Lupton}, Robert H. and {Narayanan}, Vijay K. and {Annis}, James and {Bernardi}, Mariangela and {Blanton}, Michael and {Burles}, Scott and {Connolly}, A.~J. and {Dalcanton}, Julianne and {Doi}, Mamoru and {Eisenstein}, Daniel and {Frieman}, Joshua A. and {Fukugita}, Masataka and {Gunn}, James E. and {Ivezi{\'c}}, {\v{Z}}eljko and {Kent}, Stephen and {Kim}, Rita S.~J. and {Knapp}, G.~R. and {Kron}, Richard G. and {Munn}, Jeffrey A. and {Newberg}, Heidi Jo and {Nichol}, R.~C. and {Okamura}, Sadanori and {Quinn}, Thomas R. and {Richmond}, Michael W. and {Schlegel}, David J. and {Shimasaku}, Kazuhiro and {SubbaRao}, Mark and {Szalay}, Alexander S. and {Vanden Berk}, Dan and {Vogeley}, Michael S. and {Yanny}, Brian and {Yasuda}, Naoki and {York}, Donald G. and {Zehavi}, Idit},
        title = "{Spectroscopic Target Selection in the Sloan Digital Sky Survey: The Main Galaxy Sample}",
      journal = {\aj},
         year = 2002,
        month = sep,
       volume = {124},
       number = {3},
        pages = {1810-1824},
          doi = {10.1086/342343},
archivePrefix = {arXiv},
       eprint = {astro-ph/0206225},
 primaryClass = {astro-ph},
       adsurl = {https://ui.adsabs.harvard.edu/abs/2002AJ....124.1810S}
}

@ARTICLE{2005AJ....130.1535G,
       author = {{Graham}, Alister W. and {Driver}, Simon P. and {Petrosian}, Vah{\'e} and {Conselice}, Christopher J. and {Bershady}, Matthew A. and {Crawford}, Steven M. and {Goto}, Tomotsugu},
        title = "{Total Galaxy Magnitudes and Effective Radii from Petrosian Magnitudes and Radii}",
      journal = {\aj},
         year = 2005,
        month = oct,
       volume = {130},
       number = {4},
        pages = {1535-1544},
          doi = {10.1086/444475},
archivePrefix = {arXiv},
       eprint = {astro-ph/0504287},
 primaryClass = {astro-ph},
       adsurl = {https://ui.adsabs.harvard.edu/abs/2005AJ....130.1535G}
}

@ARTICLE{2022A&A...661A..12G,
       author = {{Ghirardini}, Vittorio and {Bahar}, Y. Emre and {Bulbul}, Esra and {Liu}, Ang and {Clerc}, Nicolas and {Pacaud}, Florian and {Comparat}, Johan and {Liu}, Teng and {Ramos-Ceja}, Miriam E. and {Hoang}, Duy and {Ider-Chitham}, Jacob and {Klein}, Matthias and {Merloni}, Andrea and {Nandra}, Kirpal and {Ota}, Naomi and {Predehl}, Peter and {Reiprich}, Thomas H. and {Sanders}, Jeremy and {Schrabback}, Tim},
        title = "{The eROSITA Final Equatorial-Depth Survey (eFEDS). Characterization of morphological properties of galaxy groups and clusters}",
      journal = {\aap},
         year = 2022,
        month = may,
       volume = {661},
          eid = {A12},
        pages = {A12},
          doi = {10.1051/0004-6361/202141639},
archivePrefix = {arXiv},
       eprint = {2106.15086},
 primaryClass = {astro-ph.CO},
       adsurl = {https://ui.adsabs.harvard.edu/abs/2022A&A...661A..12G}
}

@ARTICLE{2004AJ....128..163L,
       author = {{Lotz}, Jennifer M. and {Primack}, Joel and {Madau}, Piero},
        title = "{A New Nonparametric Approach to Galaxy Morphological Classification}",
      journal = {\aj},
         year = 2004,
        month = jul,
       volume = {128},
       number = {1},
        pages = {163-182},
          doi = {10.1086/421849},
archivePrefix = {arXiv},
       eprint = {astro-ph/0311352},
 primaryClass = {astro-ph},
       adsurl = {https://ui.adsabs.harvard.edu/abs/2004AJ....128..163L}
}

@ARTICLE{2019ApJ...876...40P,
       author = {{Pintos-Castro}, I. and {Yee}, H.~K.~C. and {Muzzin}, A. and {Old}, L. and {Wilson}, G.},
        title = "{The Evolution of the Quenching of Star Formation in Cluster Galaxies since z {\ensuremath{\sim}} 1}",
      journal = {\apj},
         year = 2019,
        month = may,
       volume = {876},
       number = {1},
          eid = {40},
        pages = {40},
          doi = {10.3847/1538-4357/ab14ee},
archivePrefix = {arXiv},
       eprint = {1904.00023},
 primaryClass = {astro-ph.GA},
       adsurl = {https://ui.adsabs.harvard.edu/abs/2019ApJ...876...40P}
}

@ARTICLE{2026ApJ..1000..103V,
       author = {{Vicentin}, Marcelo C. and {Strauss}, Michael A. and {Sodr{\'e}}, Jr., Laerte and {Yates}, Robert M. and {Araya-Araya}, Pablo and {Stoppacher}, Doris},
        title = "{The Journey to Dominance: How Brightest Cluster Galaxies Evolve Differently from Other Massive Galaxies}",
      journal = {\apj},
         year = 2026,
        month = mar,
       volume = {1000},
       number = {1},
          eid = {103},
        pages = {103},
          doi = {10.3847/1538-4357/ae45fc},
archivePrefix = {arXiv},
       eprint = {2603.19521},
 primaryClass = {astro-ph.GA},
       adsurl = {https://ui.adsabs.harvard.edu/abs/2026ApJ..1000..103V}
}

@ARTICLE{2004ApJ...617..879L,
       author = {{Lin}, Yen-Ting and {Mohr}, Joseph J.},
        title = "{K-band Properties of Galaxy Clusters and Groups: Brightest Cluster Galaxies and Intracluster Light}",
      journal = {\apj},
         year = 2004,
        month = dec,
       volume = {617},
       number = {2},
        pages = {879-895},
          doi = {10.1086/425412},
archivePrefix = {arXiv},
       eprint = {astro-ph/0408557},
 primaryClass = {astro-ph},
       adsurl = {https://ui.adsabs.harvard.edu/abs/2004ApJ...617..879L}
}

@article{Lauer_2014,
	author = {Lauer, Tod R. and Postman, Marc and Strauss, Michael A. and Graves, Genevieve J. and Chisari, Nora E.},
	doi = {10.1088/0004-637X/797/2/82},
	journal = {The Astrophysical Journal},
	month = {dec},
	number = {2},
	pages = {82},
	publisher = {The American Astronomical Society},
	title = {BRIGHTEST CLUSTER GALAXIES AT THE PRESENT EPOCH},
	url = {https://doi.org/10.1088/0004-637X/797/2/82},
	volume = {797},
	year = {2014}}

@article{10.1093/mnras/staa3286,
    author = {De Propris, Roberto and West, Michael J and Andrade-Santos, Felipe and Ragone-Figueroa, Cinthia and Rasia, Elena and Forman, William and Jones, Christine and Kipper, Rain and Borgani, Stefano and Lambas, Diego García and Romashkova, Elena A and Patra, Kishore C},
    title = {Brightest cluster galaxies: the centre can(not?) hold},
    journal = {Monthly Notices of the Royal Astronomical Society},
    volume = {500},
    number = {1},
    pages = {310-318},
    year = {2021},
    month = {01},
    issn = {0035-8711},
    doi = {10.1093/mnras/staa3286},
    url = {https://doi.org/10.1093/mnras/staa3286},
    eprint = {https://academic.oup.com/mnras/article-pdf/500/1/310/34819542/staa3286.pdf},
}

@ARTICLE{2012MNRAS.420.2120M,author = {{Mann}, Andrew W. and {Ebeling}, Harald},
        title = "{X-ray-optical classification of cluster mergers and the evolution of the cluster merger fraction}",
      journal = {\mnras},
         year = 2012,
        month = mar,
       volume = {420},
       number = {3},
        pages = {2120-2138},
          doi = {10.1111/j.1365-2966.2011.20170.x},
archivePrefix = {arXiv},
       eprint = {1111.2396},
 primaryClass = {astro-ph.CO},
       adsurl = {https://ui.adsabs.harvard.edu/abs/2012MNRAS.420.2120M}
}

@article{Gini_normalization_paper,
	author = {Mu{\~n}oz, Juan F. and Pav{\'\i}a, Jose M. and {\'A}lvarez-Verdejo, Encarnaci{\'o}n},
	journal = {Social Indicators Research},
	number = {1},
	pages = {2},
	title = {A Practical Guide to Proper Estimation and Inference of the Gini Index by Avoiding often Encountered Methodological Pitfalls},
	volume = {183},
	year = {2026}}
\bibliographystyle{aasjournalv7}

\appendix
%At high z -- are we probing more or less of the luminosity distribution? Think about how this scales with z. Less of the avaliable luminosity distribution here. 
\section{Flat Richness Cuts}
\label{Appendix: Richness}

We advise against making a flat richness cut (like that discussed in Section \ref{max_min_gini} as an indirect method of indirectly limiting depth) when applying G to observed galaxy clusters, and provide this appendix as an illustration of this recommendation. As discussed in sections \ref{max_min_gini} and \ref{application_benefits_G}, the Gini index is sensitive to which limiting magnitude the luminosity function is truncated. Applying a flat richness cut of $\rm N \in [15,25]$ to our sample of clusters, such as that shown in row 3 of Figure \ref{fig:G_M12_N_richness_cut}, is effectively equivalent to limiting depth with individual values of $\rm M_{lim}$ determined by the the magnitude of the Nth brightest galaxy. This cut drives notable non-physical scatter in G, with an average $\rm \sim 33\%$ increase in fit scatter and larger variation in G for individual clusters. Here, G and $\rm M_{1,2}$ are still non-evolving within $\rm 2 \sigma$. 

%This section is an illustration of why. 

\begin{deluxetable*}{ccccccc}[h!]
%\digitalasset
\tablewidth{\textwidth}
\tablecaption{Limiting Richness Parametric Evolution Results \label{tab:evolutionary_parameters_richness}}
\tablehead{
\colhead{Indicator}& \colhead{All BCGs?}& \colhead{Center Type}& \colhead{m} & \colhead{b} & \colhead{$\rm \sigma$}
}
\startdata
G & Yes & BCG & $\rm 0.07\pm 0.04$ & $\rm 0.69\pm 0.01$ & $\rm 0.09\pm 0.01$   \\
G & Yes & Algorithm & $\rm 0.07\pm0.04$ & $\rm 0.68\pm 0.01$ & $\rm 0.09\pm 0.01$  \\
G & No & Algorithm &  $\rm 0.08\pm 0.04$ & $\rm 0.67\pm 0.01$ & $\rm 0.1\pm 0.01$  \\
$\rm M_{1,2}$ & Yes & BCG &  $\rm 0.2\pm 0.3$ & $\rm 1.18\pm0.08$ & $\rm 0.69\pm 0.06$ \\
$\rm M_{1,2}$ & Yes & Algorithm & $\rm 0\pm 0.3$ & $\rm 1.07\pm 0.08$ & $\rm 0.71\pm 0.06$
\\
$\rm M_{1,2}$ & No & Algorithm & $\rm -0.04\pm 0.3$ & $\rm 1.01\pm0.08$ & $\rm 0.69\pm 0.06$  \\
\enddata
\tablecomments{Summary of parameters fit to the G and $\rm M_{1,2}$ (``Indicator") data for the flat richness cut. Columns 2-3 specify the choice of BCG or light-weighted (``Algorithm") cluster centers (``Center Type") and if non-RS BCGs are included (``Yes"/``No")  in the calculation of G and $\rm M_{1,2}$ (``All BCGs?"). The fit slope (m), intercept (b) and intrinsic scatter ($\sigma$) are given in columns 4-6 with $\rm 1\sigma$ normally-distributed errors.    }
\end{deluxetable*}

\begin{deluxetable*}{ccccccccccc}[h]
%\digitalasset
\tablewidth{\columnwidth}
\tablecaption{Limiting Richness Empirical Evolution Results \label{tab:K-S_test_richness}}
\tablehead{\colhead{}&\colhead{}&\twocolhead{Full Sample} & \twocolhead{HST-SPT} & \twocolhead{High-Low z} \\
\colhead{All BCGs?}& \colhead{Center Type} & \colhead{$\rm p_{G}$} & \colhead{$\rm p_{M_{1,2}}$} & \colhead{$\rm p_{G}$} & \colhead{$\rm p_{M_{1,2}}$}&\colhead{$\rm p_{G}$} & \colhead{$\rm p_{M_{1,2}}$}
}
\startdata
Yes & BCG & 0.40 & 0.67 & 0.46 & 0.91 & 0.14 & 0.99  \\
Yes & Algorithm & 0.54 & 0.71 & 0.57 & 0.80 & 0.26 & 0.90\\
No & Algorithm & 0.50 & 0.59 & 0.83 & 0.57 & 0.14 & 0.69\\
\enddata
\tablecomments{Summary of p-values returned from the application of the KS-test to G ($\rm p_{G}$) and $\rm M_{1,2}$ ($\rm p_{M_{1,2}}$) data for the flat richness cut. Columns 1-2 specify the choice of BCG or light-weighted (``Algorithm") cluster centers (``Center Type") and if non-RS BCGs are included (``Yes"/``No")  in the calculation of G and $\rm M_{1,2}$ (``All BCGs?"). The KS-test is applied to the distributions of three sub-samples: ``Full Sample" (Column 3) compares the combined HST-SPT and SDSS-Planck sample split across $\rm \bar{z} = 0.56$, ``HST-SPT" (Column 4) compares the HST-SPT sample split across $\rm \bar{z} = 0.56$, and ``High-Low z" (Column 5) compares the lowest ($\rm z < 0.32 $) and highest ($\rm z > 0.79 $)  z-quartiles of the combined HST-SPT and SDSS-Planck sample.}
\end{deluxetable*}

 \begin{figure*}[ht!]
\centering
\includegraphics[width=\textwidth]{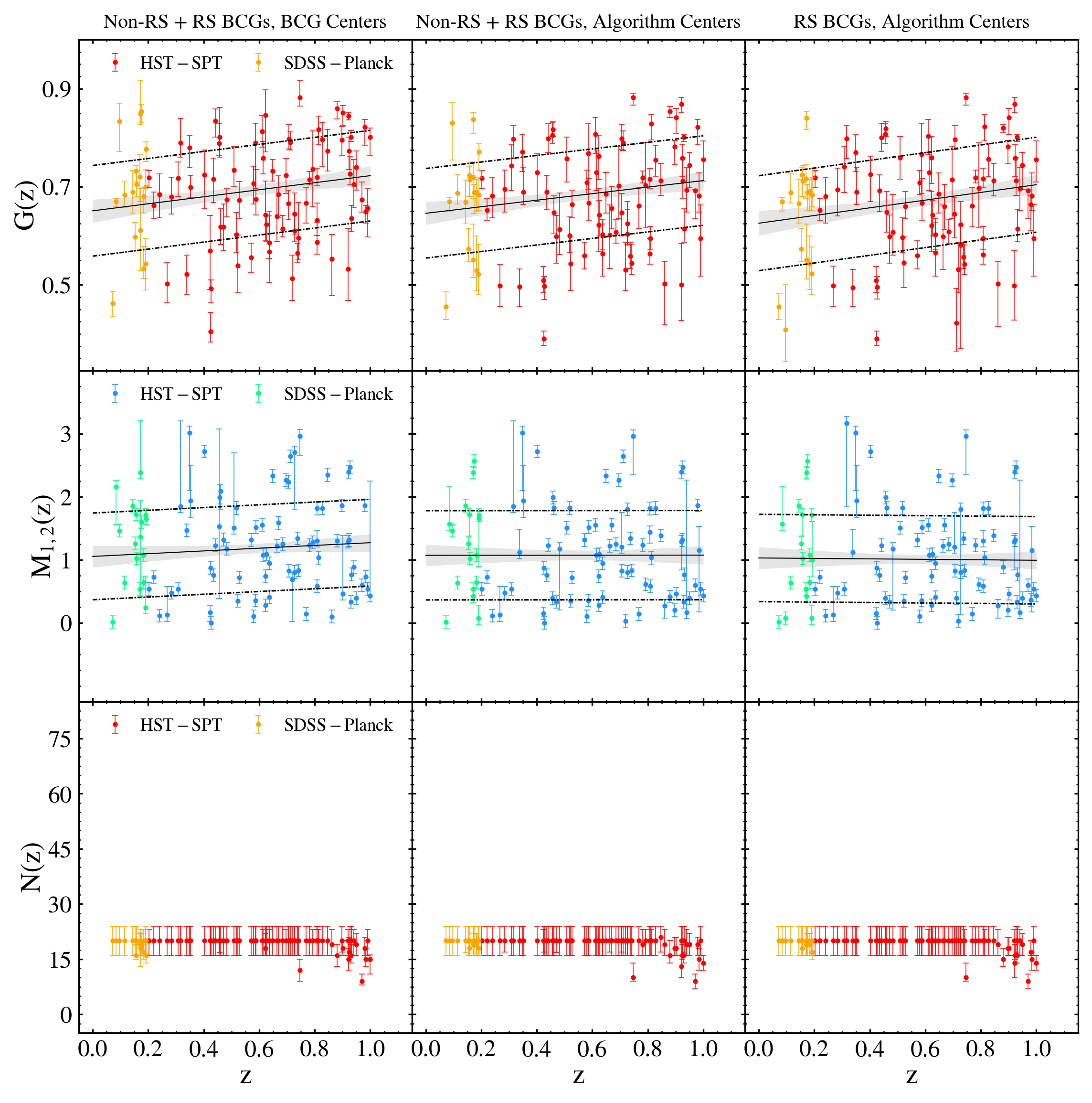}
\caption{Reproduction of figures \ref{fig:G_M12_v_z_systematics} and \ref{fig:N_v_z_systematics} for a constant richness cut of $\rm N\in [15,25]$ to indirectly limit depth. This cut drives notable non-physical scatter in G, both for individual clusters and globally with an average $\rm \sim 33\%$ increase in fit scatter when compared to that from the limiting magnitude cut. G and $\rm M_{1,2}$ are still non-evolving within $\rm 2\sigma$.
\label{fig:G_M12_N_richness_cut}}
\end{figure*}

%Fitting parameter table Still inconsistent with evolution within two sigma

%\section{HST-SPT Sample Images and Gini Indices}

%We present this appendix as a tool for building intuition with the Gini index. We order the HST-images of HST-SPT clusters, labeled with their redshifts, SPT IDs, and median $\rm M_{1,2}$ and G values with $\rm 1\sigma$ uncertainties from run A, from low to high G. Our identified BCG is marked with a diamond. All core galaxies that survive our median $2\sigma$ RS, $\rm 225~kpc$ size, and $\rm M^*_{F110W} + 2$ limiting-magnitude cuts are marked with points. The $\rm 225~kpc$ boundary of the cluster core from the BCG center is circled. 



\end{document}

% End of file `sample7.tex'.